\documentclass{article}
\usepackage{latexsym}
\usepackage{amsmath,amsthm}
\usepackage{amssymb}
\usepackage{graphics}
\usepackage{color}
\newtheorem{theorem}{Theorem}[section]
\newtheorem{proposition}{Proposition}[section]
\newtheorem{conjecture}{Conjecture}[section]
\newtheorem{question}{Question}[section]
\newtheorem{assumption}{Assumption}[section]
\newtheorem{lemma}{Lemma}[section]

\newtheorem{corollary}{Corollary}[section]
\begin{document}
\title{Strong cosmic censorship for surface-symmetric 
cosmological spacetimes with
collisionless matter}
\author{Mihalis Dafermos\thanks{University of Cambridge,
Department of Pure Mathematics and Mathematical Statistics,
Wilberforce Road, Cambridge CB3 0WB United Kingdom} \and Alan D.~Rendall\thanks{
Max-Planck-Institut f\"ur Gravitationsphysik
(Albert-Einstein-Institut), 
Am M\"uhlenberg 1,
D-14476 Potsdam}}
\maketitle
\begin{abstract}
This paper addresses
strong cosmic censorship for spacetimes 
with self-gravitating collisionless matter, evolving from 
surface-symmetric compact initial data. The global dynamics
exhibit qualitatively different features according
to the sign of the curvature $k$ of the symmetric surfaces
and the cosmological constant $\Lambda$.
With a suitable formulation, the question of strong cosmic censorship
is settled in the affirmative if $\Lambda=0$ or
$k\le0$, $\Lambda>0$. In the case
$\Lambda>0$, $k=1$, we give a detailed geometric characterization of possible
``boundary'' components of spacetime; 
the remaining obstruction to showing strong cosmic censorship in this case
has to do with the possible formation of extremal
Schwarzschild-de Sitter-type black holes. In the special case
that the initial symmetric surfaces are all expanding,
strong cosmic censorship is shown in the past for
all  $k,\Lambda$. Finally, our results also 
lead to a geometric characterization of the future boundary of black
hole interiors for the collapse of asymptotically
flat data: in particular, in the case of small perturbations of Schwarzschild data,
it is shown that these solutions do \emph{not} exhibit Cauchy
horizons emanating from $i^+$ with strictly positive limiting area radius.
\end{abstract}
\tableofcontents
\section{Introduction}
\emph{Strong cosmic censorship} is the conjecture that 
the classical fate of all observers in general relativity
be uniquely predictable from initial data, assuming the initial
data to be sufficiently generic. In other words,
it is the conjecture that, generically,
classical general relativity is a deterministic theory,
in the same sense that classical
mechanics is.

The above conjecture finds a rigorous formulation as follows:
recall that the proper mathematical description of general relativity
is the initial value problem for the Einstein equations
\begin{equation}
\label{eeqs}
R_{\mu\nu}-\frac12g_{\mu\nu}R=8\pi T_{\mu\nu}-\Lambda g_{\mu\nu}
\end{equation}
coupled to appropriate matter equations.
Given an initial data set for a suitable Einstein-matter system, it
is a classical theorem~\cite{chge:givp} that there exists a 
unique maximal globally hyperbolic spacetime $(\mathcal{M},g)$, 
the so-called \emph{maximal Cauchy development}, 
together with matter fields defined on $\mathcal{M}$,
solving this initial value problem.
\emph{Strong cosmic censorship} is 
then the conjecture that for generic initial
data, $(\mathcal{M},g)$ be inextendible. To make
this precise, a specific definition of generic and inextendible must
be chosen.

In this paper, strong cosmic censorship will be addressed
for cosmological solutions to the Einstein-Vlasov system with surface symmetry. 
That is to say, we consider the initial value problem for the system $(\ref{eeqs})$
coupled to the Vlasov equation\footnote{See Section~\ref{vlasovsec}.},
with
initial Riemannian $3$-manifold given by
a doubly warped product ${\mathbb S}^1\times \Sigma$, where $\Sigma$
is a compact $2$-surface of constant curvature\footnote{The cases where
the curvature $k$ of $\Sigma$ is $1$, $0$ and $-1$ are
known as \emph{spherical}, \emph{plane} and 
\emph{hyperbolic} symmetry, respectively.}, and initial
second fundamental form and Vlasov distribution function 
invariant with respect to the local isometries
of $\Sigma$. 
In the case of spherical symmetry, 
our results will also apply
to the interiors
of black holes forming from the collapse
of non-compact asymptotically flat data, with topology 
${\mathbb R}^3$.

Physically, solutions to the Einstein-Vlasov system describe
self-gravitating collisionless matter. The motivation for this 
system has been discussed at length in~\cite{ar:vlasov}. 
Suffice it to say here that it is the simplest matter model in which 
the issue of singularities can be reasonably posed, and thus, provides
a natural starting point for the study of strong cosmic censorship in
general relativity.

\subsection{The main results}
\subsubsection{$k\le 0$,  $\Lambda\ge0$}
The first main result of this paper
characterizes the global geometry of solutions with $k\le0$,
$\Lambda\ge0$, and, in particular, 
 resolves a suitable
formulation of strong cosmic censorship in the affirmative.
\begin{theorem}
\label{introthe}
Let $(\mathcal{M},g)$ denote the maximal development of surface symmetric
data as described above, with
$k\le 0$, $\Lambda\ge0$. The spacetime $(\mathcal{M},g)$ is
surface symmetric with natural projection
map $\pi_1:\mathcal{M}\to\mathcal{Q}$.
If $\Lambda>0$, then,
with a suitable choice of time orientation,
the universal cover $\tilde{\mathcal{Q}}$ of the future development quotient
has Penrose diagram\footnote{See Section~\ref{sssec}. $\tilde{\mathcal{S}}$ denotes the lift of the quotient of
the initial hypersurface
to the universal cover.} given by:
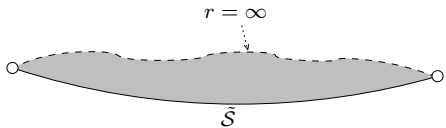
\[
\input{futcas2.pstex_t}
\]
where the future boundary is acausal, to which the area radius
function $r$ extends continuously to $\infty$.
In the case $\Lambda = 0 $,
then, either
\begin{equation}
\label{special}
k=\Lambda=0, \qquad f=0, \qquad R_{\alpha\beta\gamma\delta}=0,
\end{equation}
or,
$\tilde{\mathcal{Q}}$ has Penrose diagram:
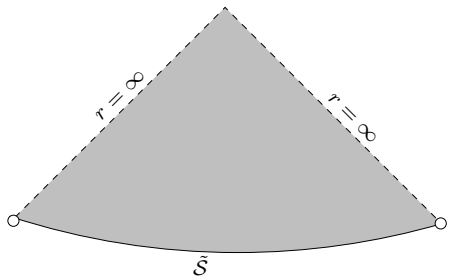
\[
\input{futcas1.pstex_t}
\]
where, again, $r$ extends continuously to $\infty$
on the future boundary.
In general, for $\Lambda\ge0$, if $(\ref{special})$ does not hold,
then the universal cover of the past development has
Penrose diagram given by either:
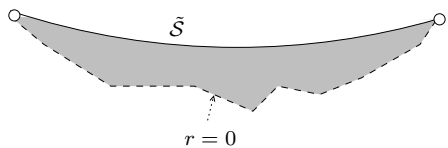
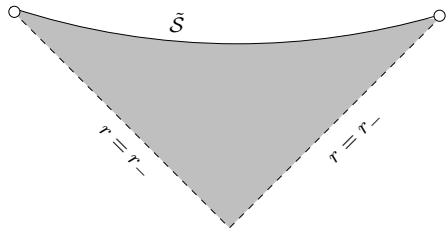
\[
\input{pastcas2.pstex_t}
\]
or:
\[
\input{pastcas1.pstex_t}
\]
where $r$ extends continuously on the past boundary to $0$, or 
a constant $r_-\ge0$, respectively. 

The spacetime $(\mathcal{M},g)$
is future inextendible as a $C^2$ Lorentzian metric. In the case $k=0$, $\Lambda=0$,
it is past inextendible as a $C^2$ Lorentzian metric for all initial data 
where 
the Vlasov distribution function $f$
is not identically $0$, and moreover, if the second Penrose diagram applies, then $r_-=0$,
whereas in the case $k= -1$,  or the case
$k=0, \Lambda>0$, it is past
inextendible for a suitable notion of generic\footnote{The 
genericity statement is 
that there should exist a fixed constant such that
on initial data, the Vlasov distribution function $f$
should not vanish identically on any open set 
of the mass shell intersecting the set where angular momentum is less than
this constant.
} initial data. In particular, strong cosmic censorship (in the sense 
of $C^2$-inextendibility) holds in this symmetry class for all cases considered here.
\end{theorem}

From the above Penrose diagrams, one sees for instance that there are
timelike curves ending on the future boundary which have future event
horizons in the case $\Lambda>0$ while there are no timelike curves
with this property in the case $\Lambda=0$ (cf.~\cite{he:lssst}, p.~129
for the terminology).

\subsubsection{The past evolution of antitrapped data}
The second result concerns strong cosmic censorship only
in the past direction. It applies independently of the signs
of $k$ and $\Lambda$, in the special case where the initial
data are antitrapped, i.e.~the symmetric surfaces are all 
initially expanding in both future null directions.
\begin{theorem}
\label{introthe15}
Let $(\mathcal{M},g)$ denote the maximal development of surface
symmetric data as described above, with $\Lambda$ and $k$
arbitrary, and such
that the initial data are antitrapped. The universal cover $\tilde{\mathcal{Q}}$
of the quotient of the
past Cauchy development has Penrose diagram given by one of the second two
Penrose diagrams of Theorem~\ref{introthe}, where $r$ extends continuously to the boundary
as indicated there.

In the case $k=1$ or the case $k=0$, $\Lambda\le 0$, then $(\mathcal{M},g)$ is past inextendible as
a $C^2$ Lorentzian metric for all data
where $f$ is not identically $0$. In the case of $k<0$, or the case
$k=0$, $\Lambda>0$,  $(\mathcal{M},g)$
is past inextendible for data satisfying
a suitable generic condition. In particular, strong
cosmic censorship holds in the past for all cases considered here.
\end{theorem}

\subsubsection{$k=1$, $\Lambda\ge  0$}
\label{???}
The case $k=1$, $\Lambda\ge0$ is qualitatively different from the $k\le0$ case.
If $\Lambda=0$, solutions will not expand forever. If $\Lambda>0$, one can
have the formation of interesting small-scale structure. Indeed, this
is clear already from the special Schwarzschild-de 
Sitter class of solutions. The Penrose diagram of the non-extremal
case is depicted below:
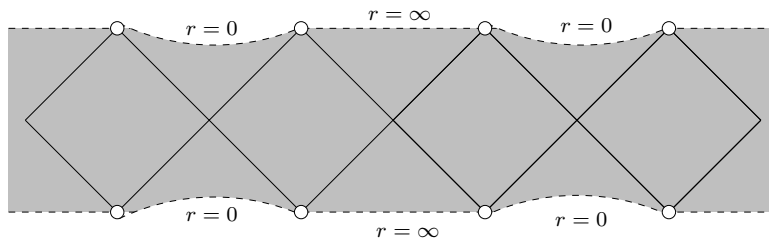
\[
\input{desit.pstex_t}
\]
Cosmological solutions with an arbitrary number of black holes
can be constructed by passing to quotients.
It turns out that the above solution is key to understanding
the evolution of general initial data. An essential 
difficulty, however, arises from the so-called
\emph{extremal} case, depicted here (cf.~\cite{podolsky}):
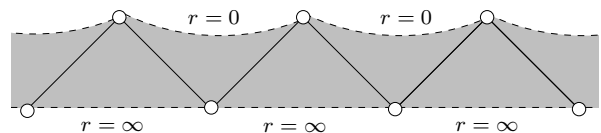
\[
\input{desitext.pstex_t}
\]
These solutions indicate that in the extremal case,
the behaviour on the horizon does not determine
the behaviour of its future. In our dynamical setting,
the possibility of the formation
of (a generalised\footnote{See~$(\ref{case2})$ in the statement of Theorem~\ref{introthe2}.} notion of) asymptotically extremal horizons will in fact limit our ability to understand 
the singular behaviour of certain components ($\mathcal{N}^i_x$ in the statement below)
of the boundary or spacetime.
Modulo the presence of such components, strong cosmic censorship is resolved.
We have
\begin{theorem}
\label{introthe2}
Let $(\mathcal{M},g)$ denote the maximal development of surface symmetric
data as described above, with
$k=1$ and $\Lambda\ge 0$.  

If $\Lambda=0$, then the future and past evolution of initial data
have quotient with universal cover $\tilde{Q}$ with Penrose diagram
as depicted in one of the second two diagrams of Theorem~\ref{introthe}, with
$r$ extending continuously to the boundary as indicated. Moreover, 
in the case of the latter diagram, then either
$f$ vanishes identically or $r_{\pm}=0$. If $f$ does not vanish identically,
then 
the spacetime is both future and past inextendible as a $C^2$ metric. In particular,
strong cosmic censorship holds. 

In the case $\Lambda>0$, then
a future (resp.~past) boundary $\mathcal{B}^{\pm}$ can be attached 
to $\tilde{Q}$
such that either
$\mathcal{B}^\pm$
is as depicted in the last Penrose diagram
of Theorem~\ref{introthe}, 
where $r$ extends continuously to a constant $0\le r_+<\infty$,
or else
\begin{equation}
\label{evwsis}
\mathcal{B}^{\pm}=\mathcal{B}^{\pm}_s\cup\mathcal{B}^{\pm}_\infty\cup \mathcal{B}^{\pm}_h
\cup\left(\bigcup_{x\in\mathcal{B}^{\pm}_h}(\mathcal{N}^1_x\cup\mathcal{N}^2_x)\right)
\end{equation}
where $\mathcal{B}^{\pm}_s$, $\mathcal{B}^{\pm}_\infty$ 
are open subsets of $\mathcal{B}^\pm$,
such that $r$ extends continuously to $0$ along $\mathcal{B}^{\pm}_s$, and $r$ extends
continuously to $\infty$ along $\mathcal{B}^{\pm}_\infty$, $\mathcal{B}^\pm_\infty$ is
acausal, and where $\mathcal{B}^{\pm}_h$ is
characterized by the facts that $\mathcal{B}_h^\pm\cap\mathcal{B}_\infty^\pm=\emptyset$,
$\mathcal{B}_h^\pm$ are future endpoints in the topology of ${\mathbb R}^{1+1}$
of two null rays $\mathcal{H}^i_x\subset \mathcal{Q}$, where $i=1,2$, 
such that $\mathcal{H}^i_x$ are future affine complete and $r$
has a (possibly infinite) limiting final value $r^i_x>0$ along $\mathcal{H}^i_x$. The 
$\mathcal{N}^i_x$ are (possibly empty) half-open null segments emanating from (but not containing)
$x$ on whose interior points in the limit $0<r<\infty$.\footnote{The union $(\ref{evwsis})$
is thus by definition disjoint, except for possible coinciding future (resp.~past)
endpoints of $\mathcal{N}^1_x$ and $\mathcal{N}^2_y$ for points $x\ne y$.}

If 
\[
r^i_x\ne \frac{1}{\sqrt\Lambda}
\]
then we say $\mathcal{H}^i_x$ is non-extremal.
In this case, if either 
\begin{equation}
\label{case1}
r^i_x>\frac{1}{\sqrt\Lambda}
\end{equation}
or, defining regular null coordinates $u$, $v$ along $\mathcal{H}^i_x$,
with $\mathcal{H}^i_x$ corresponding to $u=u_0$,
\begin{equation}
\label{case2}
r^i_x<\frac{1}{\sqrt\Lambda}, \qquad 1-\frac{2m}r(u_0,v) \ge 0, \qquad -\partial_u 
r(u_0,v) \ge e^{\alpha \int_{v_0}^v \Omega^2 (-\partial_u r)^{-1} (u_0,\bar{v})d\bar{v}} 
\end{equation}
for some constant $\alpha>0$ and for all $v\ge v_0$, where $m$ denotes
the Hawking mass, and $\Omega^2$ is such that the metric of $\mathcal{Q}$
takes the form $-\Omega^2dudv$, and where 
$v_0$ is a sufficiently late affine time along $\mathcal{H}^i_x$,
then we have 
$\mathcal{N}^i_x=\emptyset$.  In case ($\ref{case1})$, 
$x\in \overline{{ \mathcal{B}}^\pm_\infty\cap \overline{J^\pm(\mathcal{H}^i_x)}}$, 
while in case ($\ref{case2})$, 
$x\in \overline{{ \mathcal{B}}^\pm_s\cap \overline{J^\pm(\mathcal{H}^i_x)}}$.

Suppose $(\mathcal{M}',g')$ is a non-trivial
future (resp.~past)
extension.
In the case of the latter diagram of Theorem~\ref{introthe}, 
then either $f$ must vanish identically, or $r_\pm>0$ with
$r<r_\pm$.
 Otherwise, in the case $(\ref{evwsis})$,
  if $\gamma$ is a future (resp.~past) directed 
geodesic leaving $\mathcal{M}$, 
and $\widetilde{\pi_1(\gamma)}$ denotes a lift of $\pi_1(\gamma)$
to $\tilde{\mathcal{Q}}$,
then 
$\overline{\widetilde{\pi_1(\gamma)}}\cap\overline{\mathcal{N}^i_x}
\ne\emptyset$, for some $x$, $i$.
\end{theorem}
As an illustration of possible structure for $\mathcal{B}^+$,
see the Penrose diagram below:
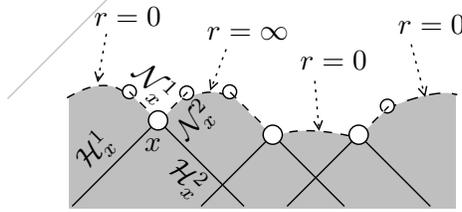
\[
\input{kpos3.pstex_t}
\]

In the case $(\ref{evwsis})$, if the set of $x\in \mathcal{B}^\pm_h$
for which $(\ref{case1})$ or $(\ref{case2})$ does not hold is empty,
then it follows from Theorem~\ref{introthe2} that $\mathcal{M}$ is inextendible.
Intuition would have it that the case where this set is non-empty should be
in some sense exceptional, and that strong cosmic censorship
should thus still hold for $\Lambda>0$. 
This remains, however, an open 
problem. (See Section~\ref{ops}.)

Exclusion of such horizons would have other applications. For instance
we also have the following
\begin{theorem}
\label{finitetheorem}
Let $(\mathcal{M},g)$ denote the maximal development of surface symmetric
data as described above, with
$k=1$ and $\Lambda> 0$.  
Suppose for all $x\in \mathcal{B}^{\pm}_h$,
either $(\ref{case1})$ or $(\ref{case2})$ is satisfied. Let $\mathcal{F}$ denote
a fundamental domain for $\mathcal{Q}$ in $\tilde{\mathcal{Q}}$.
Then $\overline{\mathcal{F}}\cap \mathcal{B}^{\pm}_h$ is finite.
\end{theorem}
The above theorem says that if all horizons satisfy $(\ref{case1})$ or $(\ref{case2})$, then 
there can only be finitely many black (resp.~white) holes
and finitely many cosmological regions.

One should note, that, despite the picture above,
in general the set $\mathcal{B}^\pm_h$ can fail to be discrete,
in fact, in principle it can have non-empty interior in $\mathcal{B}^\pm$. 
See also the discussion in Appendix~\ref{homog}.

\subsubsection{The asymptotically flat case}
The ideas of the proof of Theorem~\ref{introthe2} 
can be adapted to the asymptotically flat setting, where more
can in fact be said.
\begin{theorem}
\label{afth}
Let $(\mathcal{M},g)$ denote the maximal development of 
spherically symmetric asymptotically flat data for
the Einstein-Vlasov system, with no anti-trapped surfaces
present initially. Suppose
$\mathcal{Q}\setminus J^-(\mathcal{I}^+)\ne\emptyset$.\footnote{Here $\mathcal{Q}$
as before denotes the $2$-dimensional Lorentzian quotient. See~\cite{trapped} for
an explanation of $\mathcal{I}^+$. A sufficient
condition for $\mathcal{Q}\setminus J^-(\mathcal{I}^+)\ne \emptyset$ is 
the existence of a single trapped
or marginally trapped surface in $\mathcal{Q}$.}
Then the Penrose diagram of the solution is as below:
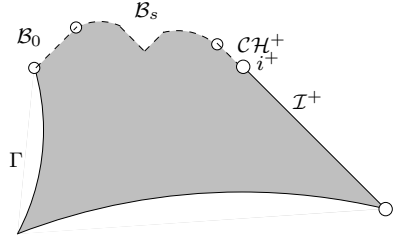
\[
\input{asymptflat3.pstex_t}
\]
where the null segment $\mathcal{B}_0$
possibly consists
of a single point, $r$ extends continuously to $0$
on $\mathcal{B}_s$, a nonempty achronal curve, and finally,
$\mathcal{CH}^+$ is a possibly empty null half-open segment characterized by 
$r\ne 0$ in the limit at its interior points.

Let $M_f$ denote the final Bondi mass of the black hole, and let $r_+$ denote
the asymptotic area radius of $\mathcal{H}^+$. By the results of~\cite{dr:ep, trapped},
it follows that $\mathcal{I}^+$ is complete and
$r_+\le 2M_f$. There exists a universal constant $\delta_0>1$
such that if 
\begin{equation}
\label{veasuv9nkn}
2M_f r_+^{-1}\le \delta_0,
\end{equation}
then
\begin{equation}
\label{kevo}
\mathcal{CH}^+=\emptyset.
\end{equation}
More generally, $(\ref{kevo})$ holds whenever $(\ref{case2})$
is satisfied.

Moreover, in the case $(\ref{kevo})$,
there is a non-empty
set $\mathcal{A}\subset\mathcal{Q}$ representing marginally trapped surfaces in $\mathcal{M}$,
such that $D^-(\mathcal{A})$ has past boundary $\mathcal{H}^+$, and
\begin{equation}
\label{property}
i^+=\overline{\mathcal{A}}\setminus
\mathcal{A}.
\end{equation}
 
Finally, if $(\mathcal{M}',g')$
is a $C^2$ extension, and $\gamma$ is a causal
geodesic leaving $\mathcal{M}$,
then 
\[
\overline{\pi_1(\gamma)}\cap\left(\overline{\mathcal{CH}^+}
\cup \overline{\mathcal{B}_0\setminus\overline{\Gamma}}\right)
\neq\emptyset.
\]
\end{theorem}
In particular, strong cosmic censorship would follow
if it can be shown that for generic initial data,
$\mathcal{CH}^+\cup \mathcal{B}_0\setminus\overline{\Gamma}=
\emptyset$.

The condition $(\ref{veasuv9nkn})$ can be interpreted as the statement
that the portion of the final Bondi mass
due to the persistent ``atmosphere'' of the black hole has to be small
in relation to the portion due to the black hole itself. By simple monotonicity arguments,
it is immediate that $(\ref{veasuv9nkn})$ holds for data containing a trapped
surface such that, outside the trapped region, the data are suitably close
to Schwarzschild
data, specifically such that an inequality $(\ref{veasuv9nkn})$ holds where $r_+$ is replaced
by the area radius of the outermost marginally trapped surface, and $M_f$ is replaced
by the ADM mass. For open questions surrounding $(\ref{veasuv9nkn})$, see Section~\ref{ops}.

In the case $(\ref{veasuv9nkn})$ at least, 
we see that the picture of the interior of these black holes is
analogous to that in the collapse
of a self-gravitating scalar field in the
absence of charge, studied by Christodoulou~\cite{chr:bhf}.
In particular, there is no null component of the boundary of
$\mathcal{Q}$ emanating from $i^+$
for which $r$ is bounded below by a positive constant near $i^+$.\footnote{Note, however,
that it is \emph{not} here
claimed that $\mathcal{B}_s$ is acausal. $\mathcal{B}_s$ may well contain
a null segment emanating from $i^+$ or elsewhere.} This
is in contrast to the case of a scalar field in the presence of (even arbitrarily
small) charge,
studied in~\cite{md:si,md:cbh}.\footnote{It is not a priori obvious
that Vlasov matter should not exhibit similar properties
with the scalar field in the presence of charge, as, both
these models share a non-zero $T_{uv}$ term.} See, however, the discussion
in Section~\ref{ops}.

Questions about the structure of $\mathcal{A}$ in a neighborhood of $i^+$ often
are considered under the heading ``dynamical horizons''. See Section~\ref{ops}
for conjectures which refine $(\ref{property})$.

\subsection{Previous results}
\label{prevressec}
The study of the surface-symmetric cosmological
solutions of the Einstein-Vlasov system considered here was
initiated in~\cite{crushing, gr:cosmo}. In the case of plane and hyperbolic
symmetries, it was shown in~\cite{hrr:onthe} that if $\Lambda= 0$,
then the maximal development is foliated by so-called
constant areal hypersurfaces, such that the areal function is
a time function taking the values $(R_0,\infty)$ for some $R_0\ge 0$.
Future inextendibility (and detailed asymptotic behaviour)
has been shown~\cite{tr} for the case $\Lambda>0$. 
These results have been extended~\cite{tsb} to the spherical case under the
assumption that $r>1/\sqrt{\Lambda}$ initially. 
A global foliation of the maximal development by constant mean curvature 
surfaces ranging
in $(-\infty,\infty)$ follows from the results of~\cite{br:emh,oh:gpmc}
in the spherical case for $\Lambda =0$.
Past inextendibility has been shown in~\cite{hrr:onthe}
for $\Lambda=0$, in the case of plane  symmetry.
Future inextendibility under additional small-data assumptions has 
been shown in~\cite{gr:fc} in the case of hyperbolic symmetry.
Past inextendibility for small data under various assumptions on $k$ and
$\Lambda$ has been shown in~\cite{gr:cosmo, tsb:cqg}.

Study of the spherically symmetric Einstein-Vlasov
system for asymptotically flat data was initiated in~\cite{rr} where it was 
shown that for sufficiently small data, the solution disperses. Certain results
for large data are proven in~\cite{dr:ep} and have been described already in the statement
of Theorem~\ref{afth}.

For more details, we refer the reader to the survey 
article~\cite{ha:living}.

\subsection{Overview}
This paper is essentially self-contained and can be read linearly. In particular,
it is independent of the previous work discussed in Section~\ref{prevressec}.
We only appeal to the local existence proven in~\cite{dr:ep}, the results 
of~\cite{trapped} for the asymptotically flat case, and,  for the
cosmological $k=1$, $\Lambda=0$ case,
to certain results from~\cite{oh:gpmc}.
For readers wishing to see in advance some of the main new ideas present here,
we give in this section a relatively complete overview.

\subsubsection{Preliminaries: null coordinates, Raychaudhuri,
the Hawking mass, conservation laws}
In Sections~\ref{sssec}--\ref{consec}, we formulate the Einstein-Vlasov system 
under surface symmetry and introduce its most basic features.

The problem of coordinates is resolved once and for all by adopting null coordinates,
which are easily shown to 
cover the entire maximal development in all cases considered here.
(See Sections~\ref{sssec} and~\ref{Eenc}.)
In particular, one dispenses entirely with spacelike foliations and
the well-known problems 
that occur when such foliations break down (e.g.~Schwarzschild-type coordinates as
trapped regions form, or
area radial coordinates in the $k=1$, $\Lambda\ge 0$ case as one
crosses the cosmological horizon of Schwarzschild-de Sitter).

The structure of the Einstein part of the system is discussed in
Section~\ref{Eenc}. Here, important monotonicity properties 
follow from the Raychaudhuri equation applied to the natural
$2$-dimensional foliation of the surface-symmetric null cones by the surfaces of
symmetry. In null coordinates,
this Raychaudhuri equation appears as
the set of null constraint equations $(\ref{const1})$ and $(\ref{const2})$.
Monotonicity arises since the right hand side
has a sign, in view of the dominant energy condition~$(\ref{dec})$.

Another source for monotonicity is the so-called Hawking mass,
introduced in Section~\ref{HMS}.
This quantity is most powerful in regions where the future pointing
null derivatives of $r$ have opposite sign, in which case it gives rise to an energy
estimate for the Vlasov matter in characteristic rectangles, via the inequalities
$(\ref{pum})$ and $(\ref{pvm})$. This
estimate was exploited in~\cite{dr:ep}. We make heavy use of this monotonicity
in Section~\ref{bhs}. We will also make use of monotonicity properties of $m$ in 
a special timelike direction
in the proof of Proposition~\ref{otherprop} of Section~\ref{pastev}.

The Vlasov matter itself is discussed in Section~\ref{vlasovsec}. 
As we shall see below, of utmost importance for our analysis at several levels is the existence of a conservation law,
namely, conservation of particle current. This conservation law
is described in Section~\ref{consec}.

\subsubsection{Spacetime integral estimates}
\label{SIE}
In Section~\ref{localest}, we introduce new
semi-global \emph{a priori} 
estimates for the Einstein-Vlasov system 
tied directly to the causal structure. The null decomposition of
the energy momentum tensor is essential.\footnote{The reader should
note that $T_{uv}$ cannot be replaced with $T_{vv}$, say,  in the estimate
below. Compare with
the well-known null condition for non-linear wave equations.}

The estimate is non-standard. One considers a characteristic 
rectangle $J^-(p)\cap J^+(q)\setminus \{p\}$\footnote{Causal relations here and in what
follows are to be understood in $\mathbb R^{1+1}$.} in the quotient $\mathcal{Q}$ spacetime,
such that one of its future boundary segments has
finite affine length on which the area radius $r$ is uniformly bounded above and below.
One first obtains an estimate for the spacetime integral
\[
\int T_{uv} du dv
\]
(see $(\ref{comput})$). In this estimate, conservation of particle current
is exploited to bound terms containing the Vlasov field, and the remaining terms
containing only metric
quantities are bounded with the help of the auxiliary 
assumptions.\footnote{Compare these with the
spacetime estimates introduced in~\cite{chr:bv}
for a self-gravitating scalar field.} Pointwise estimates can then be
derived in a standard manner. 

In Section~\ref{alternatif}, the a priori assumptions necessary for the above
estimate are retrieved from an alternative set of assumptions, namely,
that
$r$ be bounded above and below in $J^-(p)\cap J^+(q)\setminus \{p\}$,
and the spacetime volume of this region be finite.

The above-mentioned estimates 
lead naturally to an extension theorem, Theorem~\ref{extenthe}, 
which easily leads to a general characterization (given in
Propositions~\ref{reform} and~\ref{alternatifprop}) for possible ``boundary points''
of the quotient spacetime. At this stage, the proposition is equally applicable 
in
the cosmological case for arbitrary values of
$k$ and $\Lambda$. More refined characterizations,
giving the Penrose diagrams of the main theorems, can be then obtained
by exploiting different manifestations of monotonicity 
in each of the separate cases.   
We turn to this now.

\subsubsection{The Penrose diagram for $k\le 0$}
In the $k\le0$ case,
simple application of the Raychaudhuri inequalities $(\ref{const1})$ and $(\ref{const2})$
is sufficient to refine Proposition~\ref{reform} and obtain
the Penrose diagrams of Theorem~\ref{introthe}. This is accomplished
in Section~\ref{easiersec}. Similarly, this monotonicity is used in Section~\ref{pastev}
to obtain the Penrose diagram of Theorem~\ref{introthe15}.

\subsubsection{The Penrose diagram for $k=1$}
In the cosmological $k=1$, $\Lambda=0$ case, a more subtle monotonicity
is required which cannot be seen when restricting to a single null direction.
Such a monotonicity has been exploited in~\cite{oh:gpmc} 
to show that the spacetime can be covered by
a constant mean curvature foliation where the lapse is controlled.
In Section~\ref{k1L0}, we use the estimate of~\cite{oh:gpmc} to 
bound \emph{a priori} the
volume of spacetime (cf.~\cite{gerhardt}). In view of Propositions~\ref{reform}
and~\ref{alternatifprop}, this allow us to refine our characterization
of the boundary of spacetime, in particular to obtain the Penrose diagram
of  Theorem~\ref{introthe2} for this particular case.

For the case $k=1$, $\Lambda>0$, or the asymptotically flat $k=1$ case,
the situation is more complicated, as horizons
can indeed form. If $\mathcal{H}^1_x$ is a non-extremal cosmological horizon, 
i.e.~$(\ref{case1})$ holds, then
the Raychaudhuri equation alone is enough to show that $\mathcal{N}^1_x=\emptyset$,
where $\mathcal{N}^1_x$ is as defined in the statement of Theorem~\ref{introthe2}.
See Section~\ref{cosmohor}.
For black (resp.~white) hole horizons satisfying $(\ref{case2})$, global arguments 
must be used at the spacetime
integral level. We turn to this now.

\subsubsection{Black hole interior boundary and apparent horizon}
\label{BHexp}
First some background: The difficulties in understanding the nature of the boundary of spacetime
in a spherically symmetric
black hole interior arise from the competition of the mass ratio
times the volume form
and the $T_{uv}$ component of the energy momentum tensor. See equation $(\ref{newevol1})$. 
In the case of a spherically symmetric massless scalar field, the $T_{uv}$
component vanishes, and the boundary near $i^+$ can be understood with
relative ease~\cite{chr:bv}. In the case of a scalar field coupled (only gravitationally) to
the Maxwell equations, the situation is much more complicated, and,
as proven in~\cite{md:si, md:cbh},
this competition leads generically to the mass-inflation scenario with its weak
null singularities, first conjectured
in~\cite{pi:ih}. The results of~\cite{md:si, md:cbh} rest on very precise pointwise
control of the solution in a series of regions (red-shift, blue-shift, etc.)~where different physical effects
dominate.

For the present case, in Section~\ref{bhs},  we adapt our spacetime integral estimates,
together with monotonicity arising as before from Raychaudhuri, to obtain that
if $\mathcal{N}^1_x\ne\emptyset$, corresponding to a horizon
$\mathcal{H}^1_x$ satisfying $(\ref{case2})$,  then
\[
\int_{\mathcal{U}} T_{uv} dudv \le \epsilon {\rm Vol} (\mathcal{U})
\]
holds for $\mathcal{U}$ a subset of a sufficiently small neighborhood (in the topology
of the closure) of $x$ to the future $\mathcal{H}^+_x$. (The non-extremality
condition is essential for the monotonicity to apply.) Moreover, by requiring
$\mathcal{U}$ to lie in a more restricted neighborhood, $\epsilon$ can
be made arbitrarily small.
Thus, at the spacetime integral level, we have shown that
the $T_{uv}$ term is dominated,
and this leads then in  a straightforward manner to a contradiction.
Consequently, $\mathcal{N}^1_x=\emptyset$, and this leads to the remaining
conclusions on the structure of the Penrose diagram in Theorem~\ref{introthe2}.

In the above, spacetime integral estimates 
circumvent the need for detailed pointwise control. There is, however,
a price to pay: Much less is understood about the black hole interior. In particular,
we cannot obtain detailed information about the apparent horizon $\mathcal{A}$, 
its eventual achronality say, as was shown in~\cite{md:cbh} for scalar fields. 
See the comments in  Section~\ref{ops}.

On the other hand, the fact that, in constrast to the situation in~\cite{md:cbh},
we here have $\mathcal{N}^1_x=\emptyset$, allows us to easily obtain
the statement about $\mathcal{A}$ of Theorem~\ref{afth}, solely by applying Raychaudhuri.
No such simple \emph{a posteriori} argument is available in~\cite{md:cbh}!

\subsubsection{The asymptotically flat case}
The Penrose diagram of the statement of Theorem~\ref{afth} follows
easily from monotonicity provided by Raychaudhuri. 
Given the condition $(\ref{case2})$, the statement $\mathcal{CH}^+=\emptyset$
and the information about $\mathcal{A}$ 
follow as a special case of the result of
 the previous section.

The aspect special to the asymptotically flat case is that
we can now give a sufficient condition for $(\ref{case2})$
to hold, namely $(\ref{veasuv9nkn})$,
and this can be related to initial data in view of the monotonicity properties of $m$.

The argument that $(\ref{veasuv9nkn})$ implies $(\ref{case2})$ proceeds again through
spacetime integral estimates, and exploits monotonicity properties of $m$:
Integrating equation $(\ref{pum})$ over a spacetime region in a sufficiently
small neighborhood of $i^+$ in the topology of the boundary yields
(in view of the nonpositivity of each of the terms on the right hand side) an 
estimate for a spacetime integral of $T_{uv}$ in terms of the $v$-length of the region
and a measure of the change in $m$. The condition $(\ref{veasuv9nkn})$ ensures that,
restricting to a sufficiently small neighborhood, the latter is small. 
This allows one to dominate this integral
by the integral of the middle term of the right hand side of
 $(\ref{newevol2})$, and this easily leads to $(\ref{case2})$
upon integration of $(\ref{newevol2})$.

\subsubsection{Radial null extendibility across $r=\infty$ and horizon points}
\label{eari}
Having discussed how one obtains
the complete characterization of the Penrose diagram,
we now turn to discuss the issue of strong cosmic censorship, i.e.~generic
inextendibility. 

The first issue is inextendibility across boundary portions corresponding
to $r=\infty$. For this, one could apply the method of~\cite{dr2} using extendibility
of Killing fields. This method will  in fact be used in Section~\ref{HARD} 
below to understand inextendibility across $r=0$.

An alternative approach,
followed here, is to consider geodesics crossing into an extension.
Conservation of angular momentum leads easily to the following statement:
The set of boundary points $p$ for which there exits a radial null geodesic exiting 
the spacetime at $p$ is dense in the set of all boundary points $p$ such that there
does not exist a geodesic crossing at $p$ for which $r\to0$.
 (This is the statement $(\ref{withgeo})$ of Proposition~\ref{scc1}.)

Radial null extendibility across $r=\infty$ and horizon points is
is easily shown with
Raychaudhuri. 
Thus, there must be more boundary points available in $\mathcal{B}^\pm$
if the spacetime is to be extendible!

\subsubsection{The ``easy'' $r=0$ cases}
Inextendibility across $r=0$ is immediate in the $k=1$ case, in view
of a local curvature computation showing that the Kretschmann scalar
must blow up as this boundary is approached. This is discussed as part of the proof
of Theorem~\ref{scc15}, and
the relevant computation is in Appendix~\ref{curvexp}. 

In the $k=0$, $\Lambda\le 0$ case, the same argument applies
as long as the assumption of Proposition~\ref{otherprop} holds. 
For $\Lambda<0$, this in turn follows from the assumptions of Theorem~\ref{introthe15},
(which is the only Theorem which applies to $\Lambda<0$), while, in the case $\Lambda=0$, it 
follows in view of Proposition~\ref{eitherf0}  
as long as $f$ does not vanish identically.

\subsubsection{The ``hard'' $r=0$ cases: locally-induced Cauchy horizon rigidity}
\label{HARD}
Showing generic inextendibilty across $r=0$ boundaries in
 the case $k<0$, or the case $k=0$, $\Lambda>0$, is more tricky.
 The results will depend on a general rigidity theorem for hyperbolic
 symmetric spacetimes (Theorem~\ref{rigid}) for
 the portion of the Cauchy horizon for which $r=0$.
 This theorem is described below. In the body of the paper, the 
 rigidity theorem has been separated out in
 Section~\ref{Rigidsection}, as it is completely independent of the Vlasov equation.

Let $\mathcal{H}^+$ here denote a portion of a (without loss of generality future)
Cauchy horizon in a hyperbolic
symmetric spacetime for
which $r=0$. (In the Vlasov case, we have shown that if this is non-empty, then
it must represent the entire Cauchy horizon.)
The (locally defined) Killing vector fields
can be extended $C^2$ in a neighbourhood of any $p\in\mathcal{H}^+$,
and their integral curves through points of $\mathcal{H}^+$ must stay
on $\mathcal{H}^+$. In Lemma~\ref{Nprevlem}, it is proven that
we can write $\mathcal{H}^+$ as $\overline{\mathcal{H}^+_1\cup\mathcal{H}^+_2}$,
where $\mathcal{H}^+_1$ are points $p$ where the span of the Killing
fields is $1$-dimensional in a neighborhood of $p$ on $\mathcal{H}^+$, and
$\mathcal{H}^+_2$ is the set where this span is $2$-dimensional.
Moreover, it is proven that $\mathcal{H}^+_1\cup\mathcal{H}^+_2$
is a $C^3$ null hypersurface whose null generator lies in the span of
the Killing vectors. 

A particularly simple way in which the above could happen is if $\mathcal{H}^+$
is what is known as a \emph{Killing horizon}, i.e.~when there is a single Killing field $K$
everywhere tangent to the null generator of
$\mathcal{H}^+$. (This is the case for Gowdy symmetry, by the results of~\cite{carter}. It is 
also the case locally around a $p\in \mathcal{H}^+_1$.)
If this is the case, then, it has been shown
in~\cite{unique} that
\begin{equation}
\label{identR}
{\rm Ric}(K,K)=0.
\end{equation}

In the case of $\mathcal{H}^+_2$, however, the situation is in general
considerably more complicated.
As the example of the standard future light
cone in Minkowski space (thought of as the past Cauchy horizon of
its hyperbolically symmetric future) reveals,
the Killing field generating the null direction of the horizon can vary from point
to point. Thus the Cauchy horizon is no longer necessarily a Killing horizon.
Nonetheless, we still recover the identity $(\ref{identR})$ on $\mathcal{H}^+_2$, at least
in the nondegenerate case. 
Essentially, for this we compute
both sides of the well-known identity $(\ref{theid})$ for Killing fields by using a 
well-chosen frame. The properties of this frame are derived in Section~\ref{framesec}.
For this, heavy use is made of the Lie algebra of hyperbolic
symmetry, and the twist-free condition for the
Killing fields. Various cases must be considered separately.

In the highly degenerate case of vanishing surface gravity, we 
can deduce from a local calculation the inequality ${\rm Ric}(K,K)\le 0$ on
$\mathcal{H}^+_2$.
Thus $(\ref{identR})$ holds on all of $\mathcal{H}^+_1\cup \mathcal{H}^+_2$
as long as the spacetime satisfies
the null convergence condition. In particular, $(\ref{identR})$ holds
on all of $\mathcal{H}^+_1\cup\mathcal{H}^+_2$ when Theorem~\ref{rigid} is specialised
to our Vlasov case.

Finally, for the $k=0$ case, $(\ref{identR})$ follows from the rigidity
theorem proven in our~\cite{dr3} for general $T^2$-symmetric spacetimes.

These inferences for our Vlasov case are stated explicitly as Proposition~\ref{Riccivanish}.

Armed with this rigidity, we may now prove generic inextendibility.
For, given a suitable genericity assumption on initial data,
we show in Proposition~\ref{positRicci}   that
\begin{equation}
\label{ineqR}
{\rm Ric}(K,K)>0
\end{equation}
for a dense open subset of $\mathcal{H}^+$.
This contradicts $(\ref{identR})$, showing thus generic inextendibility.
The proof of Proposition~\ref{positRicci}  reveals
the nature of the necessary genericity assumption: Assuming $\mathcal{H}^+\ne\emptyset$,
we construct a
transverse timelike geodesic crossing the Cauchy horizon into the original spacetime,
with arbitrary small angular momentum. By global hyperbolicity, this must intersect the Cauchy
surface somewhere. By a continuity argument, if the lift of this geodesic to the tangent bundle
intersects the support of
$f$, then one can show $(\ref{ineqR})$. Thus, a sufficient genericity assumption
would be that 
the Vlasov field is supported for all points in phase space with angular momentum
less than a fixed constant.
By varying slightly, one weakens this to the condition that the Vlasov field not
vanish identically in any open set intersecting the set where angular momentum
is less than a fixed constant.

Note that this class of initial data is more general than that
studied in previous analysis of the Einstein-Vlasov system; it is easily handled here,
however, in view of the nature of our estimates. See also the comments in
Section~\ref{mgd}.

\subsubsection{The case $0<r_{\pm}<\infty$: Globally-induced Cauchy horizon rigidity} 
\label{globind}
The final case which must be understood is when the Penrose diagram is as in
the last diagram of Theorem~\ref{introthe}, and $0<r_{\pm}<\infty$.

Suppose $\mathcal{M}$ were extendible at every point corresponding to $\mathcal{B}^\pm$.
Then $\mathcal{B}^\pm$ would correspond to
a \emph{compact} Cauchy horizon
 foliated by closed null curves. The results 
of~\cite{frw}
would apply to yield an additional Killing field in the direction of
the null generator. One could then apply $(\ref{identR})$ and argue as in
the previous section, or, better, argue that the flux of matter through 
the Cauchy horizon must vanish,
and then, by conservation of particle current (see below), 
that the initial matter must vanish, i.e.~$f=0$ identically in the spacetime.
  
Without assuming regularity everywhere for $\mathcal{B}^+$ or $\mathcal{B}^-$, one
cannot argue as above. Nonetheless, we are still able
to carry out global arguments, at least in all cases except the case where
$k=1$, $\Lambda>0$, $r< r_{\pm}$.

We first note that we have previously reduced the problem to null radial geodesic
extendibility. Supposing that there exists a null radial geodesic crossing to 
an extension, without loss of generality let this be $v=0$,
 we first show that we can bound the particle flux along this geodesic.
 This bound is enunciated in $(\ref{yeniyeniyIldIz})$. It follows because
 we can bound $N^v$ pointwise in suitable coordinates 
 from the null components of energy momentum,
 which themselves are bounded as they can be related to curvature
 in a parallely propagated frame along a geodesic passing to the extension.
 By conservation of matter, from $(\ref{yeniyeniyIldIz})$ we obtain uniform
bounds on the flux of matter through the boundaries $v=0$ and $v=V$
of a region $\mathcal{F}$ defined as the union of
$3$ copies of a fundamental domain. Refer to the diagram in Section~\ref{thelastcases}. 
For, using the discrete transformation of the universal cover $\tilde{\mathcal{Q}}$ 
of $\mathcal{Q}$ (corresponding to the ${\mathbb S}^1$ factor), it follows that these
fluxes must coincide. Thus, considering a sequence of constant $u$ curves
in $\mathcal{F}$ approaching $\mathcal{B}^-$, the flux through these curves
must approach the initial flux.

The problem is thus reduced to showing that the flux through these curves approaches
zero. 

First, via Raychaudhuri, our bounds for the flux, and spacetime integral estimates
in the style of Section~\ref{localest} (compare $(\ref{comput})$ and $(\ref{computALT})$), 
we are able to derive upper and lower bounds for $\Omega^2$
with respect to the null coordinate system employed.
Here, essential use is made of $(\ref{newconst})$ (or the bound on the total volume
of spacetime in the case $k=1$, $\Lambda=0$)
and it is for this monotonicity to be in the right direction that we must exclude the exceptional case.
 Applying again Raychaudhuri,
this allows us to deduce
$(\ref{claim})$. This states that an ``energy'' flux corresponding to the Vlasov matter
approaches zero.  
The above relation can be thought of as the source of globally-induced rigidity
from the monotonicity on the Einstein side.

In view of the bounds obtained, it is only a small step (via inequality $(\ref{ieqhelp})$)
to pass from
the statement of the energy flux to that of the particle flux.
We may then again deduce by a limiting argument and conservation of
particle number, that the matter must vanish identically.

Though not essential for showing cosmic censorship, we can exclude
$\infty>r_\pm>0$ if $f$ does not vanish identically in various cases, including
the case $k\ge 0$, $\Lambda=0$, for arbitrary data, and $k\ge 0$, $\Lambda<0$
for antitrapped data, where the latter result concerns, however, only the past. 
See Propositions~\ref{stoparel9ov} and~\ref{stomellov} and their corollaries.

The results of this section demonstrate the power of the essentially geometrically invariant
approach of this paper, and of spacetime integral estimates: 
Were the analysis ``married'' to, say, area radial coordinates,
then it would be very difficult to obtain appropriate estimates
near $\mathcal{B}^+$, where these coordinates break down.

\subsubsection{The class of initial data}
\label{mgd}
As discussed in Section~\ref{HARD}, to infer generic inextendibility,
one must work in a class of initial data more general than $f$
of compact support in momentum space, considered in previous work. 
The nature of the fundamental estimates (Section~\ref{localest}) make it irrelevant
whether one works in the class of data of compact support or whether
$f$ is allowed
to decay suitably fast in momentum. In view of the comments in the next
section, it would be useful to extend this analysis to other symmetry classes.

\subsection{Extensions to $T^2$ symmetry}
The locally-induced rigidity argument for Cauchy horizons can
be adapted to general $T^2$-symmetric spacetimes.  (The $k=0$ surface
symmetric case is a subset of these.)
In view of previous work~\cite{mw}, it follows that 
strong cosmic censorship holds for this model, as
the arguments of~\cite{mw} can be easily adapted to the class of data discussed in
Section~\ref{mgd}.\footnote{Inextendibility to the future has already
been shown in our~\cite{dr2}.} These issues are discussed in a separate paper~\cite{dr3}.

A special case of $T^2$ is 
Gowdy symmetry, where H.~Ringstr\"om~\cite{hr} has proven strong cosmic
censorship in 
the vacuum case by a deep study of the generic asymptotic profile
of the solution in the approach to $r=0$. Generalisation of this method
to general $T^2$ spacetimes would appear quite difficult, as the
expected asymptotic profiles are much more complicated.

One sees then that the inclusion of Vlasov matter allows one to prove 
strong cosmic censorship for a class of spacetimes for which otherwise
it would appear out of reach!  We believe that this is yet another reason that
this matter model has an important role to play in mathematical relativity.

\subsection{Acknowledgements}
M.D.~thanks the Albert Einstein Institute and A.D.R.~thanks the 
University of Cambridge for hospitality during visits. Some of this research was carried
out during the progamme ``Global Problems in Mathematical Relativity'' at the Isaac Newton
Institute of the University of Cambridge. The authors thank Piotr Chru\'sciel for useful
discussions. M.D.~also thanks Dan Pollack and Igor Rodnianski.
M.D.~is supported in part by NSF grant
DMS-0302748 and a Marie Curie International Re-integration grant from the European 
Commission.

\section{Surface symmetry}
\label{sssec}
We use the term \emph{surface symmetry} to
describe spherical, hyperbolic, or plane symmetry. 
We require the geometric component of the 
initial data to be of
the form $(\mathcal{S}, g, K)$ with
\[
\mathcal{S} = {\mathbb S}^1\times \Sigma,
\]
with doubly warped product metric $a(\theta)d\theta^2+ r^2(\theta)\gamma_{\Sigma}$, where $\gamma_{\Sigma}$ is a metric of constant curvature.
The matter component will be described in Section~\ref{vlasovsec}.
It can be easily shown that the maximal development $(\mathcal{M},g)$ of initial data
is of the form
\[
\mathcal{Q}\times \Sigma ,
\]
with doubly warped product metric
\[
-\Omega^2dudv+r^2\gamma_{\Sigma}.
\]
Topologically, we have $\mathcal{Q}={\mathbb R}\times {\mathbb S}^1$.
We may lift the Lorentzian $2$-manifold $\mathcal{Q}$ to its universal cover
$\tilde{\mathcal{Q}}$. Standard arguments show
that $\tilde{\mathcal{Q}}$ can be causally represented as a bounded subset of
${\mathbb R}^{1+1}$ as depicted below,
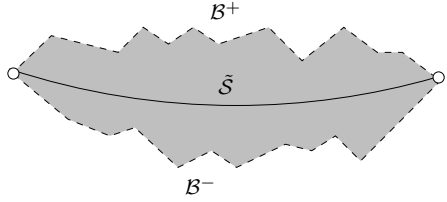
\[
\input{global.pstex_t}
\]
i.e. such that
\[
\tilde{\mathcal{Q}}=\bigcup_{p,q\in\tilde{\mathcal{Q}}}J^-(p)\cap J^+(q),
\]
and the lift of the projection of $\mathcal{S}$ to $\mathcal{Q}$, denoted
$\tilde{\mathcal{S}}$, is a Cauchy surface for $\tilde{\mathcal{Q}}$.
Here and in what follows, $J^-(p)$, and the notion of a \emph{Cauchy surface}, etc.,
refer to the topology and causal structure of ${\mathbb R}^{1+1}$.
In particular, the future
and past boundaries $\mathcal{B}^+$ and $\mathcal{B}^-$
of $\mathcal{Q}$ in ${\mathbb R}^{1+1}$ are achronal.
The above representation is often called a \emph{Penrose diagram}.
Clearly, such a representation defines a bounded system of global null coordinates
on $\tilde{\mathcal{Q}}$.

\section{The Einstein equations in null coordinates}
\label{Eenc}
Let $(u,v)$ denote null coordinates on $\tilde{\mathcal{Q}}$.
The Einstein equations
\begin{equation}
\label{Eeq}
R_{\mu\nu}-\frac12g_{\mu\nu}R=8\pi T_{\mu\nu}-\Lambda g_{\mu\nu}
\end{equation}
give rise to  the system
\begin{equation}
\label{evol1}
\partial_u\partial_v r=-\frac{k\Omega^2}{4r}-\frac {1}{r}\partial_ur\partial_v r
+4\pi r T_{uv} + \frac14r \Omega^2\Lambda,
\end{equation}
\begin{equation}
\label{evol2}
\partial_u\partial_v \log \Omega =
-4\pi T_{uv} +\frac{k\Omega^2}{4r^2} +\frac{1}{r^2}\partial_ur\partial_vr
-{\pi\Omega^2}g^{AB}T_{AB},
\end{equation}
\begin{equation}
\label{const1}
\partial_v(\Omega^{-2}\partial_vr)=-4\pi rT_{vv}\Omega^{-2},
\end{equation}
\begin{equation}
\label{const2}
\partial_u(\Omega^{-2}\partial_ur)=-4\pi rT_{uu}\Omega^{-2}.
\end{equation}
Here the constant
$k$ denotes the curvature of $\gamma_{\Sigma}$, and $x^A$ denote
coordinates on $\Sigma$.

Although from the point of view of the Penrose diagram, it is natural
to consider bounded null coordinates, we will often consider coordinate
systems with unbounded range. This will necessarily be the case for
the coordinates respecting the periodicity. 

\section{The Hawking mass}
\label{HMS}
Let us introduce the notation
\[
\nu=\partial_ur, \lambda=\partial_vr.
\]
We define the
\emph{Hawking mass} by
\begin{equation}
\label{Hawkmassdef}
m=\frac r2(k+4\Omega^{-2}\nu\lambda),
\end{equation}
and the \emph{mass ratio}
\[
\mu =\frac{2m}r.
\]
In the region where $\nu\ne0$, we may define
\[
\kappa=-\frac14\Omega^{2}\nu^{-1}.
\]
We have
\[
\kappa(k-\mu)=\lambda.
\]
The constraint equation $(\ref{const2})$
can be rewritten
\begin{equation}
\label{newconst}
\partial_u \kappa = \kappa (4\pi r\nu^{-1}T_{uu}).
\end{equation}
In terms of $\kappa$, $m$, we may rewrite the evolution equations
$(\ref{evol1})$, $(\ref{evol2})$ as
\begin{equation}
\label{newevol1}
\partial_u\lambda=\partial_v\nu
=2r^{-2}m \kappa\nu +4\pi rT_{uv}-r\kappa\nu\Lambda,
\end{equation}
\begin{equation}
\label{newevol2}
\partial_u\partial_v \log \Omega
=-4\pi T_{uv} -2r^{-3}\kappa\nu m +4\pi \kappa\nu
g^{AB}T_{AB}.
\end{equation}
We finally compute the identities
\begin{equation}
\label{pum}
\partial_um= r^2\Omega^{-2}(8\pi T_{uv}\nu -\Lambda g_{uv}\nu - 8\pi  T_{uu}\lambda),
\end{equation}
\begin{equation}
\label{pvm}
\partial_vm= r^2 \Omega^{-2}(8\pi T_{uv}\lambda -\Lambda g_{uv}\lambda -8\pi T_{vv}\nu).
\end{equation}

\section{The Vlasov equation}
\label{vlasovsec}
Let $P\subset T\mathcal{M}$ denote the set of all future directed
timelike vectors of length $-1$. We will call $P$ the mass shell. 
Vlasov matter is completely described by a nonnegative function
$f:P\to{\mathbb R}$. The equations of motion for $f$ are simply that $f$ be preserved
along geodesic flow on $P$. In coordinates we have
\begin{equation}
\label{Vlasoveq}
p^\alpha \partial_{x^\alpha}f -\Gamma^\alpha_{\beta\gamma}p^\beta p^\gamma \partial_{p^\alpha}
f=0
\end{equation}
where $p^\alpha$ define the momentum coordinates on the tangent bundle
conjugate to $x^\alpha$.
The fact that $f$ is supported on $P$ yields
the relation 
\begin{equation}
\label{MSR}
-\Omega^2p^up^v+r^2\gamma_{AB}p^Ap^B = -1
\end{equation}
on the support of $f$.
We call $(\ref{MSR})$ the mass-shell relation.

 The energy momentum tensor is defined by
\begin{equation}
\label{EMdef}
T_{\alpha\beta}(x)=\int_{\pi^{-1}(x)} p_{\alpha}p_{\beta} f,
\end{equation}
where $\pi:P\to \mathcal{M}$, and the integral is with respect to the natural
volume form on $\pi^{-1}(x)$.
For the correct formulation 
of the symmetry assumption for the matter ensuring the results of
Section~\ref{sssec}, see~\cite{dr:ep}.
It will follow that $T_{\alpha\beta}$ 
will be of the form $T_{ab}dx^a dx^b+T_{AB} dy^A dy^B$, where $T_{ab}$
is a $2$-tensor on $\mathcal{Q}$ with components
\begin{equation}
\label{defuu}
T_{uu}=\int_0^\infty\int_{-\infty}^{\infty}\int_{-\infty}^{\infty}
r^2p_up_uf(p^u)^{-1}\sqrt{\gamma}dp^u dp^Adp^B,
\end{equation}
\begin{equation}
\label{defuv}
T_{uv}=\int_0^\infty\int_{-\infty}^{\infty}\int_{-\infty}^{\infty}
r^2p_up_vf(p^u)^{-1}\sqrt{\gamma}dp^u dp^Adp^B,
\end{equation}
\begin{equation}
\label{defvv}
T_{vv}=\int_0^\infty\int_{-\infty}^{\infty}\int_{-\infty}^{\infty}
r^2p_vp_vf(p^u)^{-1}\sqrt{\gamma}dp^u dp^Adp^B.
\end{equation}

The Vlasov equation $(\ref{Vlasoveq})$, 
equations $(\ref{evol1})$--$(\ref{const2})$, together with
the definitions $(\ref{defuu})$--$(\ref{defvv})$ yield a closed system, in view of the fact that
the Christoffel symbols can be computed
from $\Omega$, $r$ by the formulas of~\cite{dr:ep}.

Note the inequalities
\begin{equation}
\label{dec}
T_{uu}\ge 0, \qquad T_{vv}\ge 0, \qquad T_{uv}\ge0;
\end{equation}
These follow from the \emph{dominant energy condition}. Both the dominant
and \emph{strong} energy conditions for Vlasov matter follow directly from
the definitions of the energy-momentum tensor independently of
symmetry assumptions.

The quantity $r^4\gamma_{AB}p^A p^B$ is the squared modulus of the angular
momentum and is a conserved quantity along particle trajectories. It is
convenient to have an alternative form of this quantity which is 
independent of local coordinates. In~\cite{dr2}, Killing vectors $X$, $Y$ and $Z$
were introduced in the case of spherical and hyperbolic symmetry. It is
possible to define Killing tensors by 
$K^{\alpha\beta}=X^\alpha X^\beta+Y^\alpha Y^\beta+k Z^\alpha Z^\beta$
where $k$ is the parameter in the definition of surface symmetry. Then
a computation shows that 
$r^4\gamma_{AB}p^A p^B=K_{\alpha\beta}p^\alpha p^\beta$. The conservation
law is seen to follow from a general property of Killing tensors (see~\cite{Wald}, 
p.~444).

We will assume on initial data that 
\begin{equation}
\label{newa}
\sup_{(\pi_1\circ\pi)^{-1}(\tilde{\mathcal{S}})}
f(1+(p^\alpha)^3)<\infty,
\end{equation}
for $\alpha=u,v$,
for some global system of null coordinates on $\tilde{\mathcal{S}}$ respecting
periodicity.\footnote{By compactness, the finiteness of the left hand
side of $(\ref{newa})$, though not its value,
is coordinate-independent.}
This  allows more general data than in~\cite{dr:ep}, where it was required  that
$f$ be compactly supported on $\pi^{-1}(x)$.
Allowing such more 
general data will be important
for the proof of strong cosmic censorship in certain
cases.
In particular
\begin{equation}
\label{tofmovo}
F\doteq \sup_{(\pi_1\circ\pi)^{-1}(\tilde{\mathcal{S}})}
f<\infty.
\end{equation}
From the Vlasov equation, we have that 
\begin{equation}
\label{fromvlasov}
0\le f\le F
\end{equation}
 on $P$ over all of spacetime.
The constant $F$ does not depend on choice of coordinates.

Finally, we assume initially
\[
X=\sup_{(\pi_1\circ\pi)^{-1}(\mathcal{S})\cap
{\rm Supp}(f)} 
r^4\gamma_{AB}p^Ap^B<\infty,
\]
i.e.~the particles are
of bounded angular momentum.
By conservation of angular momentum, we have that
\begin{equation}
\label{am0}
r^4\gamma_{AB}p^Ap^B \le
X
\end{equation}
on $P\cap({\rm Supp})(f)$ over all points of spacetime.

\section{Conservation of particle current}
\label{consec}
Define the particle current vector field $N$
by 
\[
N^\alpha=\int_{-\infty}^{\infty}\int_{-\infty}^{\infty}\int_0^\infty
r^2 p^\alpha f\frac{dp^u}{p^u}\sqrt{\gamma} dp^A dp^B.
\]
This vector field is divergence free. We obtain the conservation law
\begin{eqnarray}
\nonumber
\label{cpc}
\int_{u_1}^{u_2} N^v \Omega^2 r^2(u,v_1) du
&+&\int_{v_1}^{v_2} N^u\Omega^2 r^2(u_1,v)dv
=
\int_{u_1}^{u_2} N^v\Omega^2 r^2(u,v_2) du\\
&+&\int_{v_1}^{v_2} N^u\Omega^2 r^2(u_2,v) dv.
\end{eqnarray}
This conservation law will be crucial for obtaining \emph{a priori}
estimates.
 
\section{Local estimates}
\label{localest}
Let $\mathcal{D}$ be a region
$[0,U]\times[0,V]\setminus \{(U,V)\}$, and consider a sufficiently regular
solution of $(\ref{evol1})$--$(\ref{const2})$, $(\ref{Vlasoveq})$,
$(\ref{defuu})$--$(\ref{defvv})$ in an open set containing $\mathcal{D}$.

Let us assume that
\begin{equation}
\label{thirdset}
\int_0^V\Omega^2(U,v)dv<\infty
\end{equation}
together with 
\begin{equation}
\label{lower-r}
r(U,v)\ge r_0>0
\end{equation}
\begin{equation}
\label{upper-r}
r(U,v)\le R<\infty.
\end{equation}
For the purposes
of this section only, set
\[
F=\sup_{(\pi_1\circ\pi)^{-1}(\{0\}\times[0,V]\cup[0,U]\times\{0\})}
f,
\]
\[
X=\sup_{(\pi_1\circ\pi)^{-1}((\{0\}\times[0,V]\cup[0,U]\times\{0\}))\cap
{\rm Supp}(f)} 
r^4\gamma_{AB}p^Ap^B.
\]
We assume that $F$, $X$ are finite. Finally
we assume that in any regular coordinate system defined
on an open set containing $\mathcal{D}$,
\begin{equation}
\label{newway}
\sup_{(\pi_1\circ\pi)^{-1}(\{0\}\times[0,V]\cup[0,U]\times\{0\})}
f(1+(p^\alpha)^3) <\infty
\end{equation}
for $\alpha=u,v$.
We will derive \emph{a priori} estimates on $\mathcal{D}$ which will
lead in the next section to an extension principle.

\subsection{Estimates on Christoffel symbols and curvature}
\label{Ccsec}
We first derive \emph{a priori}
estimates for the Christoffel symbols and curvature.

Note that we have the bounds 
\begin{equation}
\label{am}
r^4\gamma_{AB}p^A p^B\le X
\end{equation}
and 
\begin{equation}
\label{trivial}
0\le f\le F
\end{equation}
throughout $\pi^{-1}(\mathcal{D})$.

It is convenient to choose a new set of coordinates on a subset of
the original $[0,U]\times [0,V]\setminus (U,V)$ as follows:

Set $\Omega^2=1$ along $\{U\}\times [0,V)$, and 
choose a new $v$-coordinate $\tilde{v}$
such that the old $(U,V)$ has sufficiently small value $\tilde{V}>0$.
Now choose a new $u$-coordinate $\tilde{u}$ such that $(U,V)$ has sufficiently small value $\tilde{U}>0$, and define $\Omega^2=1$ along $[0,\tilde{U}]\times\{0\}$. We thus have a new coordinate system in 
$[0,\tilde{U}]\times[0,\tilde{V})$, which is a (small) subset of the original region, 
for which the old $(U,V)$ is still a limit point.

Clearly, by compactness it suffices to obtain bounds in 
this new region. In what follows we drop the tildes and let us call
\[
\mathcal{D}=[0,U]\times[0,V).
\]
Define now the region
\begin{eqnarray*}
\tilde{\mathcal{D}}&=&
\{(u,v)\in \mathcal{D}: r(u^*,v^*)> r_0/2, \forall u^*\ge u, v^*\le v,
(u^*,v^*)\in \mathcal{D}\}\\
&&\hbox{}\cap\{(u,v)\in \mathcal{D}: r(u^*,v^*)< 2R, \forall
 u^*\ge u, v^*\le v, (u^*,v^*)\in\mathcal{D}
\}.
\end{eqnarray*}

First we compute:
\begin{eqnarray}
\label{comput}
T_{uv}	& = &	\int_0^\infty\int_{-\infty}^{\infty}
\int_{-\infty}^{\infty}r^2 p_up_v f\frac{dp^u}{p^u}\sqrt{\gamma} dp^A dp^B\\
	\nonumber
	& = &	(g_{uv})^2
			\int_0^\infty\int_{-\infty}^{\infty}
			\int_{-\infty}^{\infty}
				r^2 p^up^v 
			f\frac{dp^u}{p^u}\sqrt{\gamma} dp^A dp^B\\
	\nonumber
	& = &	-\frac12 g_{uv}\int_0^\infty\int_{-\infty}^{\infty}
			\int_{-\infty}^{\infty}
				r^2 \Omega^2 p^u p^v  f\frac{dp^u}{p^u}\sqrt{\gamma} dp^A dp^B\\
	\nonumber
	& = &	\frac14\Omega^2 \int_0^\infty\int_{-\infty}^{\infty}\int_{-\infty}^{\infty}
				r^2 (1+r^2\gamma_{AB}p^A p^B)  f\frac{dp^u}{p^u}\sqrt{\gamma} dp^A dp^B\\
	\nonumber
	&\le&	\frac14\Omega^2 (1+Xr_0^{-2}) \int_0^\infty\int_{-\infty}^{\infty}\int_{-\infty}^{\infty}
			r^2 (p^u+p^v) f \frac{dp^u}{p^u}\sqrt{\gamma} dp^A dp^B\\
\nonumber
			&&\hbox{}+\frac14\Omega^2(1+Xr_0^{-2}) 
			 \int_{\min\{\Omega^{-2},1\}}^1\int_{-\infty}^{\infty}\int_{-\infty}^{\infty} 
			r^2 f \frac{dp^u}{p^u}\sqrt{\gamma} dp^A dp^B \\
	\nonumber
	&\le&	(1+Xr_0^{-2})\frac14\Omega^2 N^u+
	(1+Xr_0^{-2})\frac14\Omega^2N^v\\
\nonumber
&&\hbox{} +\frac14\Omega^2 Xr_0^{-2}F(1+Xr_0^{-2})|\log \Omega^2|.
\end{eqnarray}

From the computation $(\ref{comput})$, and the conservation of particle current, we have that, for $(u_1,v_1)\in\mathcal{D}$,
\begin{equation}
\label{frcomput}
\int_{u_1}^U\int_0^{v_1} T_{uv}du dv \le A\sup_{\tilde{\mathcal{D}}}
|\log\Omega^2| \int_{u_1}^U\int_0^{v_1}\Omega^2dudv
+ UB_1+VB_2.
\end{equation}
The constants above (and those that will appear below!)
only depend on $r_0$, $R$, $F$, $X$ and the initial mass particle current.\footnote{Here,
initial means in the original $\mathcal{D}$, before the change of coordinates and the restriction to the tip.} 
On the other hand, integrating $(\ref{evol1})$ in $u$ and $v$ we obtain
\begin{eqnarray}
\label{suplambdabd}
\nonumber
\int_0^{v_1} \left(\sup_{u\in [u_1,U]}
|\lambda|(u,v)\right) dv &\le& \tilde{A}_1\left(1+\sup_{\tilde{\mathcal{D}}}
|\log\Omega^2|\right)\int_{u_1}^U\int_0^{v_1}\Omega^2dudv\\
&&\hbox{}+U\tilde{B}_1
+V\tilde{B}_2,
\end{eqnarray}
and similarly
\begin{eqnarray}
\label{supvubd}
\nonumber
\int_{u_1}^U \left(\sup_{v\in [0,v_1]}|\nu|(u,v)\right) 
du&\le& \tilde{A}_2\left(1+\sup_{\tilde{\mathcal{D}}}
|\log\Omega^2|\right) \int_{u_1}^U\int_0^{v_1}\Omega^2dudv\\
&&\hbox{}+U\tilde{B}_1' +V\tilde{B}_2'.
\end{eqnarray}
Putting the above two computations together and using $(\ref{evol2})$,
we obtain finally,
\begin{eqnarray*}
|\log \Omega^2(u_1,v_1)|
&\le& \tilde{A}'\left(1+\sup_{\tilde{\mathcal{D}}}|\log \Omega^2|\right)
\int_{u_1}^U\int_0^{v_1}\Omega^2dudv\\
&&\hbox{}+
\hat{A}\left(1+\sup_{\tilde{\mathcal{D}}}
|\log\Omega^2|\right)^2\left(\int_{u_1}^U\int_0^{v_1}\Omega^2dudv\right)^2\\
&&\hbox{}+UB'_1+VB'_2+UVB_3,
\end{eqnarray*}
and this gives, for small enough $U$, $V$, an a priori bound
\[
|\log \Omega^2|\le C
\]
\[
\int_{u_1}^U\int_0^{v_1}\Omega^2dudv \le C,
\]
for a $C$ which moreover can be made arbitary small, by appropriate
choice of $U$, $V$.

Since 
\[
|r(u_1,v_1)-r(U,v)|\le \int_{u_1}^U \left(\max_{v\in [0,v_1]}|\nu|(u,v)\right) 
du,
\]
we have from
$(\ref{supvubd})$ that
\[
|r(u_1,v_1)-r(U,v)| \le \tilde{A}_2C+ \tilde{B}_2U
\]
and thus, for appropriate choice of $U$, $V$, we can show that
$\tilde{\mathcal{D}}$ is open and closed in the topology of $\mathcal{D}$,
and thus, by connectedness
\[
\tilde{\mathcal{D}}=\mathcal{D},
\]
i.e.~we have the bounds
\begin{equation}
\label{ur}
r(u,v)\ge r_0/2
\end{equation}
\begin{equation}
\label{lr}
r(u,v)\le 2R
\end{equation}
in $\mathcal{D}$.

We thus have 
\begin{equation}
\label{logbound}
|\log\Omega^2|\le \tilde{C}
\end{equation}
and
\begin{equation}
\label{cvb}
\int_0^U\int_0^V\Omega^2 du dv\le \tilde{C}.
\end{equation}
Note that this latter bound is coordinate invariant.
By compactness, we have that
$(\ref{cvb})$ holds in our original null coordinates, where we
can now return to the original larger set $\mathcal{D}$.
We then easily show that $(\ref{logbound})$ holds
in the original null coordinates, after renaming $\bar{C}$.
Also, we can rename $r_0$, and $R$ such that $(\ref{lr})$
and $(\ref{ur})$ hold in the original $\mathcal{D}$.

Using the bounds $(\ref{logbound})$, 
$(\ref{cvb})$, 
we obtain from $(\ref{frcomput})$
the bound
\[
\int_0^U\int_0^V T_{uv}dudv \le C,
\]
from $(\ref{supvubd})$
\[
\int_0^U\sup_{0\le v\le V}|\nu(u,v)|du \le C
\]
and from $(\ref{suplambdabd})$
\[
\int_0^V\sup_{0\le u\le U}|\lambda(u,v)|dv \le C.
\]

Integrating $(\ref{evol1})$,
let us note that we have easy one-sided pointwise bounds
\begin{eqnarray}
\nonumber
\label{nubound}
-\nu(u,v) &\le&  \left(\max\{0,-\nu(u,0)\}+
\int_0^v(4|k|r^{-1}\Omega^2+\frac14r\Omega^2|\Lambda|)dv
\right)e^{\int_0^v{r^{-1}|\lambda| dv}}\\
&\le&\bar{N}
\end{eqnarray}
and similarly
\begin{equation}
\label{lambdabound}
-\lambda \le \bar{L}.
\end{equation}

Let us denote by 
\begin{equation}
\label{W}
W=\max(\sup |\log \Omega^2(0,v)|,\sup |\log \Omega^2(u,0)|).
\end{equation}
Integrating the geodesic equation we have
\[
p^v(s)= p^v(s')e^{-\int_{v(s')}^{v(s)}\Gamma_{vv}^vdv}+
	\int_{v(s')}^{v(s)} 2(-\nu)\Omega^{-2}r\gamma_{AB}p^Ap^Be^{-\int_{v(\tilde{s})}^{v(s)}\Gamma^v_{vv}dv}
			(p^v)^{-1}dv.
\]
From $(\ref{newevol2})$, we have the inequality
\[
\partial_u\Gamma^v_{vv} \ge -16\pi T_{uv}+\frac{k\Omega^2}{2r^2}
+\frac{2}{r^2}\partial_ur\partial_vr,
\]
and thus
\begin{eqnarray*}
\int_{v_1}^{v_2} \Gamma^v_{vv}(u(v),v)dv &\ge& -16\pi \int_{v_1}^{v_2}
\int_0^{u(v)}
T_{uv}  +\int_{v_1}^{v_2}
\int_0^{u(v)}{\left(\frac{k\Omega^2}{2r^2}
+\frac{2}{r^2}\partial_ur\partial_vr\right) d\bar{u}}dv\\
&&\hbox{}+\int_{v_1}^{v_2}\Gamma^v_{vv}(0,v)dv
\end{eqnarray*}
in view of $(\ref{W})$. Here, $u(v)$ can be 
any continuous function of $v$, in particular,
the projection of the path of a geodesic. 
From $(\ref{frcomput})$, $(\ref{suplambdabd})$,
and $(\ref{supvubd})$, we immediately obtain
\[
\int_{v_1}^{v_2}\Gamma^v_{vv}dv \ge -G.
\]

We wish to prove that 
\begin{equation}
\label{toprove}
p^v\le \bar C (p^v(0)+1)
\end{equation}
throughout this geodesic, for some appropriately defined $\bar{C}$. 
Suppose not, at point $s_1$. 
Then there is a point $s'$ for which $p^v(s')= \max (1, p^v(0)) $, 
and $p^v(s)\ge p^v(s')$ for all $s_1\ge s'\ge s$.
We have then, in view of our bounds 
\begin{eqnarray*}
p^v(s_1)	&\le&	(1+p^v(0))e^{G}+ 2\bar{N}
e^{2C}r_0^{-3}XV\\
		&\le&	(\bar {C}/2)(p^v(0)+1),
\end{eqnarray*}
where the above inequality constrains the choice of $\bar C$. 
But this is a contradiction. We thus indeed have $(\ref{toprove})$.

We turn now to prove an upper bound for $p^u$.
We argue exactly as before: From the inequality
\[
\partial_v\Gamma^u_{uu} \ge -16\pi T_{uv}+\frac{k\Omega^2}{2r^2}
+\frac{2}{r^2}\partial_ur\partial_vr,
\]
we obtain a bound for
\[
\int_{u_1}^{u_2}\Gamma^u_{uu}(u,v(u))du.
\]
Integrating the geodesic equation between parameters $s'$ and $s_1$ as 
described before, we
obtain from
\[
p^u(s)= p^u(s')e^{-\int_{u(s')}^{u(s)}\Gamma_{uu}^udu}+
	\int_{u(s')}^{u(s)} 2(-\lambda)\Omega^{-2}r\gamma_{AB}p^Ap^Be^{-\int_{u(\tilde{s})}^{u(s)}\Gamma^u_{uu}du}
			(p^u)^{-1}du,
\]
in view also of our bound $(\ref{lambdabound})$,
the estimate
\begin{eqnarray*}
p^u(s_1)
		&\le&	(\bar C/2)(p^u(0)+1),
\end{eqnarray*}
for appropriate choice of $\bar C$.
The contradiction proves
\begin{equation}
\label{toprove2}
p^u\le \bar C(p^u(0)+1).
\end{equation}

From $(\ref{newway})$, $(\ref{toprove})$, $(\ref{toprove2})$ and the fact 
that $f$ is constant along geodesics,
we have that
\[
\sup_{(\pi_1\circ\pi)^{-1}([0,U]\times [0,V]\setminus (U,V))}
f(1+(p^\alpha)^3) <\infty,
\]
for $\alpha=u,v$.
From this together with conservation of angular momentum $(\ref{am})$,
one obtains uniform pointwise
bounds for $T_{uu}$, $T_{vv}$, $T_{uv}$ in 
$[0,U]\times [0,V]\setminus (U,V)$.
From these one obtains uniform pointwise bounds 
for all Christoffel symbols,
and all components of the curvature tensor.

\subsection{Higher-order estimates}
We may prove higher order estimates following~\cite{dr:ep}.

\subsection{Alternative assumptions}
\label{alternatif}
Finally, we retrieve here the assumptions  of the beginning
of Section~\ref{localest} from a variation
of the basic set of assumptions.

Consider $\mathcal{D}$ as in the beginning of Section~\ref{localest}. 
Instead of assuming $(\ref{thirdset})$, 
let us assume \emph{a priori} that 
\begin{equation}
\label{altome}
B\doteq \int_0^U\int_0^V \Omega^2 du dv<\infty.
\end{equation}
Let us also now assume that $(\ref{lower-r})$ and $(\ref{upper-r})$ hold in all
of $\mathcal{D}$, and assume bounds on $f$ as before.

We have that
\begin{equation}
\label{bd-cd}
\log \Omega^2 (0, v)  \le C,
\qquad
\log \Omega^2(u,0) \le C,
\end{equation}
for $0\le v\le V$, $0\le u\le U$, respectively, 
by compactness, for some constant $C$.
To retrieve $(\ref{thirdset})$,
it clearly suffices to show that there exists a $\tilde{C}$ such that
\[
\log \Omega^2 (u,v)\le \tilde{C}
\]
throughout $\mathcal{D}$.
In view of $(\ref{altome})$ and $(\ref{bd-cd})$, integrating $(\ref{evol2})$,
it suffices to bound
\[
\int_0^u\int_0^v r^{-2}\nu\lambda d\bar{u}d\bar{v}
\]
uniformly in $u$, $v$.

For this, note first the inequality:
\begin{equation}
\label{allowehave}
\sup_v |\nu(u,v) |\le \tilde{C}\left(|\nu(u,0)| +|\nu (u,V)|+ \int_0^V \Omega^2(u,\bar{v})
 d\bar{v}\right).
\end{equation}
This follows from the inequalities
\begin{eqnarray*}
\nu(u,v) &=&e^{\int_v^V \lambda r^{-1}d\bar{v}}
\left(\nu(u,V) -\int_v^V (-k\Omega^2{4r}^{-1}+4\pi rT_{uv} +\frac14r\Omega^2
\Lambda) e^{-\int_{\bar{v}}^{\bar{V}}\lambda r^{-1} }\right)\\
		&\le & \tilde{C} \left(\max\{0, \partial_u r(u,V )\} +\int_0^V \Omega^2\right),
\end{eqnarray*}
and 
\begin{eqnarray*}
\nu (u,v) &=&e^{\int_0^v \lambda r^{-1}d\bar{v}}\left( \nu(u,0) +
\int_0^v (-k\Omega^2{4r}^{-1}+4\pi rT_{uv} +\frac14r\Omega^2
\Lambda) e^{-\int_0^{\bar{v}}\lambda r^{-1}}\right)\\
		&\ge & \tilde{C} \left(\min\{0, \partial_u r(u,0)\} - \int_0^V \Omega^2\right).
\end{eqnarray*}

From $(\ref{allowehave})$, we bound
\begin{eqnarray*}
\left|\int_0^u\int_0^v r^{-2}\nu\lambda \right|&\le&  \int_{0}^u \sup_v |\nu|\left( \int_0^v |\lambda| dv\right) du\\
							&\le &2(R-r_0)   \int_0^u \sup_v |\nu| du\\
							&\le& 2(R-r_0) \cdot\\
							&&\hbox{}\cdot\tilde{C}\int_0^u
							 \left(|\nu(\bar{u},0)| +|\nu(\bar{u},V)|+ 
							 \int_0^V \Omega^2(\bar{u},\bar{v}) d\bar{v}\right)d\bar{u}\\
							&\le & 2(R-r_0)\tilde{C}\left(4(R-r_0) +
							\int_0^u\int_0^V \Omega^2\right)\\
							&\le & 2(R-r_0)\tilde{C}(4(R-r_0) +B),
\end{eqnarray*}
where in the second and fourth inequality above we use the fact that,
by Raychaudhuri, $\lambda$ changes sign at most
once on a constant-$u$ ray, and similarly $\nu$ changes sign at most
once on a constant-$v$ ray.

We have thus retrieved $(\ref{thirdset})$, as required.

\section{The global theory}
\label{globalsec}

We return to the setup of Section~\ref{sssec}.
We will consider maximal developments of initial
surfaces $\tilde{\mathcal{S}}$.

\subsection{The extension theorem} 
We have the following extension theorem:
\begin{theorem}
\label{extenthe}
Let $\mathcal{D}\subset \tilde{\mathcal{Q}}$ satisfy
the hypotheseis of the beginning of Section~\ref{localest}, or alternatively,
the hypotheseis of Section~\ref{alternatif}.
Then $\mathcal{D}\subset J^-(p)$,
for a  $p\in \tilde{\mathcal{Q}}$.
\end{theorem}
The proof of this theorem follows from the estimates of the previous
section, the local existence theorem of~\cite{dr:ep}\footnote{suitably modified
so as to allow for non-compact support of $f$ on the mass-shell}, and 
the maximality of $\tilde{\mathcal{Q}}$.

\subsection{General characterization of $\mathcal{B}^{\pm}$}
\label{GCS}
We consider the cosmological case here. Let $(\mathcal{M},g)$ be as in the
statement of Theorems~\ref{introthe}--\ref{introthe2}, with quotient $\mathcal{Q}$
and universal cover $\tilde{\mathcal{Q}}$. 

Let $\mathcal{B}^{\pm}$ denote the future (resp.~past) boundary
of $\tilde{\mathcal{Q}}$ in the topology of the Penrose diagram.
Let $\mathcal{B}_1^{\pm}$ consist of the subset of ``first
singularities'', i.e.~the subset of $\mathcal{B}^{\pm}$ 
which are ``preceded'' by a $\mathcal{D}\subset\tilde{\mathcal{Q}}$,
in the following sense.
For $p\in\mathcal{B}^{\pm}$, and $q\in J^\mp(p)\cap\tilde{\mathcal{Q}}$,
we can define the set
$\mathcal{D}_{p,q}=(J^\mp(p)\cap J^\pm(q))\cap \tilde{\mathcal{Q}}$.
We say $p\in \mathcal{B}_1^{\pm}$ if $p\in \mathcal{B}^{\pm}$,
and there exists a $q\in J^\mp(p)\cap\tilde{\mathcal{Q}}$ such
that $\mathcal{D}_{p,q}\cup \{p\}= J^\mp(p)\cap J^\pm(q)$.

Define $r_{\inf}(p)=\lim_{q\to p}\inf_{\mathcal{D}_{p,q}} r$
and let $r_{\sup}(p)=\lim_{q\to p}\sup_{\mathcal{D}_{p,q}} r$. 

Let $\mathcal{N}^{\pm}$ denote the union of two null segments
forming the boundary of the future (resp. past) of $\Sigma$
as a subset of ${\mathbb R}^{1+1}$:
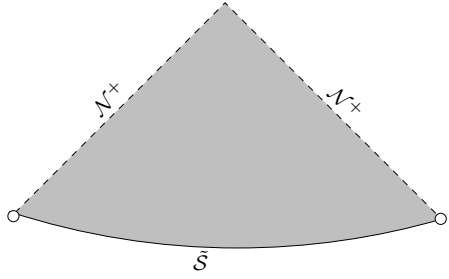
\[
\input{case20.pstex_t} 
\]

We have
\begin{proposition}
\label{reform}
Either $\mathcal{B}_1^{\pm}=\emptyset$ and
$\mathcal{B}^{\pm}=\mathcal{N}^{\pm}$, or
\[
\mathcal{B}^{\pm}=\cup_{x\in\mathcal{B}_1^{\pm}}
(\{x\}\cup\hat{\mathcal{N}}^1_x\cup\hat{\mathcal{N}}^2_x),
\]
where $\hat{\mathcal{N}}^1_x$ and $\hat{\mathcal{N}}^2_x$
are (possibly empty) null segments emanating from $x$.

If $\mathcal{B}^\pm=\mathcal{N}^\pm$, then if $r_{\inf}(x)=0$ for some $x\in\mathcal{N}^\pm$,
it follows that $r_{\sup}=0$ identically on $\mathcal{N}^\pm$. In this case, define
$\mathcal{B}^\pm_s=\mathcal{N}^\pm$. 

Otherwise, if $\mathcal{B}^\pm\ne\mathcal{N}^\pm$,
define
 \[
 \mathcal{N}^i_x= \overline{\{y\in \hat{\mathcal{N}^i_x}:r_{\inf}(y)\ne 0\}}\cap{\hat{\mathcal{N}}}^i_x,
 \]
\[
\mathcal{B}^{\pm}_{s,1}=\{x\in \mathcal{B}^{\pm}_1: r_{\inf}(x)=0\},
\]
\[
\mathcal{B}^{\pm}_s=
\mathcal{B}^{\pm}_{s,1}\cup\big(\bigcup_{x\in\mathcal{B}^\pm_1}\hat{\mathcal{N}}^i_x\big)
\setminus
\big(\bigcup_{x\in\mathcal{B}^\pm_1}\mathcal{N}^i_x\big).
\]
The set $\mathcal{B}^\pm_s$ is an open subset
of $\mathcal{B}^\pm$, $y\in\mathcal{B}^\pm_s$ satisfies
$r_{\sup(y)}=0$, and $\mathcal{N}^{i}_x$ is a connected
(possibly empty) half-open segment for all $x\in\mathcal{B}^\pm_1$.

If $x\in\mathcal{B}^{\pm}_1$
with $0<r_{\inf}(x)\le \infty$,
then
$J^{\mp}(x)\setminus (I^{\mp}\cup\{x\})$ consists of two
null rays $\mathcal{H}^1_x$ and $\mathcal{H}^2_x$ 
of infinite affine length:
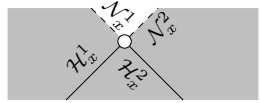
\[
\input{hor.pstex_t}
\]
Let $\mathcal{B}^{\pm}_H\subset
\mathcal{B}_1^{\pm}$ denote the set of points $x$ with this property. 

If $\mathcal{B}^{\pm}=\mathcal{N}^\pm$ then let us
define $\mathcal{B}^\pm_{\infty}=\mathcal{B}^{\pm}=\mathcal{N}^\pm$
if $r_{\inf}=\infty$ at any point of $\mathcal{N}^{\pm}$.
It follows that $r_{\inf}=\infty$
for all points of $\mathcal{N}^\pm$.

Otherwise, if $\mathcal{B}^\pm\ne\mathcal{N}^\pm$,
define $\mathcal{B}^{\pm}_\infty$
by
\[
\mathcal{B}^{\pm}_{\infty}=\{x\in\mathcal{B}^\pm_1:r_{\inf}(x)=\infty\}.
\]
It follows that
\[
\mathcal{B}^\pm_{\infty}\subset \mathcal{B}^\pm_H.
\]
Finally, setting $\mathcal{B}^\pm_h=\mathcal{B}^\pm_H\setminus \mathcal{B}^\pm_{\infty}$,
we have
\[
\mathcal{B}^\pm=\mathcal{B}^\pm_s\cup \mathcal{B}^\pm_\infty\cup
\mathcal{B}^\pm_h\cup
\bigcup_{x\in\mathcal{B}^\pm_H}{(\mathcal{N}^1_x\cup\mathcal{N}^2_x)}
\]
where this union is disjoint, except for possible coinciding future (resp.~past)
endpoints of $\mathcal{N}^1_x$ and $\mathcal{N}^2_y$ for points $x\ne y$.
\end{proposition}

\begin{proof}
The first paragraph of the statement of the Proposition follows from simple causality, and
the remark that, by the periodicity, either $\mathcal{B}^\pm\cap \mathcal{N}^\pm=\emptyset$
or $\mathcal{B}^\pm=\mathcal{N}^\pm$.

Now, let us be in the latter case, and
without loss of generality, suppose $\mathcal{B}^+=\mathcal{N}^+$, and
there exists an $x\in\mathcal{N}^+$
such that $r_{\inf}(x)=0$. 
Let $x_i\to x$ be a sequence such that $r(x_i)\to 0$. 
We may choose now $\tilde{x}_i=(u_i,v_i)$ to lie in a single fundamental domain $\mathcal{F}$
for $\mathcal{Q}^+$, defined by $U_1 >u \ge U_2$, say,
with $\pi (x_i)=\pi(\tilde{x}_i)$. 
Let $y_i\in\tilde{\mathcal{S}}$ be such 
that $v_i\doteq v(\tilde{x}_i)=v(y_i)$.
Since by compactness $r(y_i)>c>0$, we have by Raychaudhuri that $\nu(z_i)<0$ for some
$z_i$, with $u(z_i)< U_1$, $v(z_i)=v_i$, for all sufficiently large $i$. 
It follows again by Raychaudhuri that $\nu(u,v_i)<0$, for $u\ge U_1$,
and thus $r(u,v_i) \le r(\tilde{x}_i)$. In particular, $r_{\inf}=0$ for $\mathcal{B}^+\cap \{v=V\}
\cap\{ u\ge U_1\}$, where $V=\lim v_i$,
and thus by periodicity, $r_{\inf}=0$ identically on $\mathcal{B}^+$.

But now, by considering constant $u$-curves, for $\tilde{u}\ge U_1$, one obtains
that, for any such curve, there exists a $\tilde{v}<V$ such that $\lambda(\tilde{u},\tilde{v})<0$.
It follows that $r_{\sup}=0$ for $\mathcal{B}^+\cap \{v=V\}
\cap\{ u\ge U_1\}$, and thus, by periodicity, $r_{\sup}=0$ identically on $\mathcal{B}^+$.
We have shown thus the second paragraph of the statement of the proposition.

Next, we turn to show the statement of the third paragraph. Without loss
of generality, let us consider $\mathcal{B}^+$.
First note that $\mathcal{N}^i_x$ is a connected (possibly empty)
half-open segment for all $x\in\mathcal{B}^+_1$.
For this, let $\hat{\mathcal{N}}^i_x\subset \{v=V\}$, where $V=v(x)$,
and suppose $y=(\tilde{u},V)\in\hat{\mathcal{N}}^i_x$ such that $r_{\inf} (y)=0$.
Let $(\tilde{u}_i,V_i)\to (\tilde{u},V)$ such that $r(\tilde{u}_i,V_i)\to 0$.
In view of the fact that $r\ge c$ on $\mathcal{\tilde{S}}$, we have by Raychaudhuri that
$\nu (\tilde{u}_i,V_i)<0$ for large enough $i$. It follows
by Raychaudhuri that $\nu(u,V_i) <0$
for $u\ge \tilde{u}$ and large enough $i$, and thus that $r(u,V_i)\to 0$ for all
such $u$. Thus $r_{\inf} (u,V)=0$ for any $u\ge \tilde{u}$, $(u,V)\in\hat{\mathcal{N}}^i_x$.
The desired statement about $\mathcal{N}^i_x$ follows immediately.

Note that since Raychaudhuri implies $\lambda(u,V_i)<0$ for sufficiently large
$i$, we have that $r_{\sup} =0$ on $\hat{\mathcal{N}}^i_x\setminus \mathcal{N}^i_x$.

Let $\mathcal{B}^+_{s,1}$, $\mathcal{B}^+_s$ 
be defined as in the statement of the Proposition. 

First we show that $r_{\sup}=0$ on $\mathcal{B}^+_{s,1}$.
Let $(\tilde{u},\tilde{v})\in\mathcal{B}^+_{s,1}$, and let $(\tilde{u}_i,\tilde{v}_i)\to
(\tilde{u},\tilde{v})$ with $r(\tilde{u}_i,\tilde{v}_i)\to 0$. We may choose $(\tilde{u}_i,\tilde{v}_i)$
so that $\tilde{u}_i<\tilde{u}$, $\tilde{v}_i<\tilde{v}$. 
By Raychaudhuri, we have
that $\nu(\tilde{u}_i,\tilde{v}_i)<0$, $\lambda(\tilde{u}_i,\tilde{v}_i)<0$ for large enough
$i$. By repeated use of Raychaudhuri as in the arguments above, one obtains
that there exists $\tilde{u}'<\tilde{u}$, $\tilde{v}'<\tilde{v}$ with $\lambda<0$, $\nu<0$
for $u\ge \tilde{u}'$, $v\ge \tilde{v}'$. It follows in particular that 
$r_{\sup}=0$ on $\mathcal{B}^+_{s,1}$, 
and in view of the previous statements, on the whole of $\mathcal{B}^+_s$.
Also, from this is follows that if $x\in\mathcal{B}^+_{s,1}$, then $r_{\sup}=0$
on $\hat{\mathcal{N}}^i_x$. Thus, if $\mathcal{N}^i_x\ne\emptyset$, then
the interior of $\hat{\mathcal{N}}^i_x$ is contained in $\mathcal{B}_s$.
The future endpoint of $\hat{\mathcal{N}}^i_x$ is contained in $\mathcal{B}_s$
iff it is \emph{not} the future endpoint of a $\mathcal{N}^i_y$.

We proceed to show that $\mathcal{B}^+_s$ is open.
Note that if $x \in \mathcal{B}^+_{s,1}$, then there
exists an open $\mathcal{U}\subset\mathbb R^{1+1}$ containing $x$ such that
$\mathcal{U}\cap (\mathcal{B}^+_1\setminus \mathcal{B}^+_{s,1})=\emptyset$.
For this, let  $\tilde{u}'$, $\tilde{v}'$ be as in the previous paragraph, and
consider $\mathcal{U}=\{u>\tilde{u}'\}\cap\{ v>\tilde{v}'\}$, and
let $(\hat{u},\hat{v}) \in \mathcal{U}\cap \mathcal{B}^+_1$.
Without loss of generality, say $\hat{v}>\tilde{v}$. 
We have from $(\ref{newconst})$, that for $u>\tilde{u}'$, $\hat{v}>\tilde{v}$,
the inequality 
$\kappa(u,\hat{v})\le \kappa(\tilde{u}',\hat{v} )$ holds, and thus
\begin{eqnarray*}
\int_{\tilde{u}'}^{\hat{u}}\Omega^2 (u,\hat{v}) du&=
&\int_{\tilde{v}'}^{\hat{v}}-4\nu\kappa (u,\hat{v}) du\\
&\le&\kappa (\tilde{u}',\hat{v}) \int_{\tilde{v}'}^{\hat{v}}(-4\nu)du\\
&\le&\kappa (\tilde{u}',\hat{v}) 4r(\tilde{u}',\hat{v})<\infty.
\end{eqnarray*}
By Theorem~\ref{extenthe},
it follows that since $\int_{\tilde{u}'}^{\hat{u}}\Omega^2 (u,\hat{v}) du \neq\infty$ and
$r(u, \hat{v}) \le r(\tilde{u}',\hat{v})$, we have
$(\hat{u},\hat{v})\in \mathcal{B}^+_{s,1}$.  

It follows that 
\begin{eqnarray*}
\mathcal{B}^+\cap \mathcal{U}&=&\cup_{x\in\mathcal{B}^+_{s,1}}
\{x\cup \hat{\mathcal{N}}^1_x\cup \hat{\mathcal{N}}^2_x\}\cap \mathcal{U}\\
&\subset& \mathcal{B}^+_s
\end{eqnarray*}
where the latter inclusion follows from the fact that $\mathcal{N}^i_x=\emptyset$
for $x\in \mathcal{B}^+_{s,1}$, and the openness of $\mathcal{U}$.
Thus $\mathcal{B}^+_{s,1}\subset {\rm int} \mathcal{B}^+_s$.

To show then that $\mathcal{B}^+_s$ is open, we have reduced to showing
that $\hat{\mathcal{N}}^i_x\cap \mathcal{B}^+_s\subset {\rm int}\mathcal{B}_s$. 
If $z$ is an interior point of $\hat{\mathcal{N}}^i_x$, there is nothing to show,
in view of the connectedness of $\hat{\mathcal{B}}^i_x$. 
Thus, it suffices to consider the case where $z$ is a future endpoint
of $\hat{\mathcal{N}}^i_x$. Again, if $z$ is also a future endpoint
of $\hat{\mathcal{N}}^i_y$, there is nothing to show. Thus, we may assume
this not to be the case. Let $z=(\tilde{u},\tilde{v})$ and $\mathcal{N}^i_x\subset
\{v=\tilde{v}\}$. By Raychaudhuri, we deduce that there exists $\tilde{u}'<\tilde{u}$,
$\tilde{v}'<\tilde{v}$, such that, defining $\mathcal{U}=\{u> \tilde{u}'\}\cap
\{v>\tilde{v}'\}$, we have that $\nu<0$, $\lambda<0$ on $\tilde{\mathcal{Q}}\cap\mathcal{U}$.
We show as above that $\mathcal{U}\cap (\mathcal{B}^+_1\setminus\mathcal{B}^+_{1,s})=
\emptyset$, and thus, $\mathcal{U}\cap \mathcal{B}^+=\mathcal{U}\cap \mathcal{B}^+_s$,
i.e.~$z\in {\rm int}\mathcal{B}_s$.
We have shown that $\mathcal{B}_s$ is open.

The statement about $\mathcal{B}_H^+$ in the fifth paragraph of
the Proposition directly follows from
Theorem~\ref{extenthe} and the following fact:

\begin{lemma}
\label{al}
If $r\to \infty$ along
$\mathcal{H}^i_x$, then $\mathcal{H}^i_x$ has infinite affine length.
\end{lemma}
\begin{proof}
Let $\mathcal{H}^i_x\subset \{u=U\}$, and define a coordinate $v$ such that
$\Omega^2=1$ along $\mathcal{H}^i_x$. By equation $(\ref{const1})$, we have
that $\partial_v \lambda \le 0$. It follows that the $v$-range must be infinite
if we are to have $\infty=\int_{\mathcal{H}^i_x}\lambda dv$. But this implies
$\infty=\int_{\mathcal{H}^i_x}1dv =\int_{\mathcal{H}^i_x}\Omega^2dv$, and thus,
the affine length of $\mathcal{H}^i_x$ is infinite.
\end{proof}

The statement $\mathcal{B}^\pm_\infty=\mathcal{B}^\pm=\mathcal{N}^\pm$ under
the conditions given in the sixth paragraph of the Proposition follows from
Raychaudhuri by an argument similar to the argument for $r_{\inf}=0$ given previously.

The statement $\mathcal{B}^\pm_\infty \subset \mathcal{B}^\pm_H$ in the final 
paragraph follows from Lemma~\ref{al}.

The final decomposition of $\mathcal{B}^\pm$ now follows from the 
statements proven previously.
\end{proof}

We also have the following:
\begin{proposition}
\label{alternatifprop}
In the notation of the previous proposition,
let $p \in \mathcal{B}^\pm_1$ with $0<r_{\rm inf}(p)\le r_{\rm sup}(p)<\infty$,
and let $\tilde{\mathcal{S}}$ be the lift of a Cauchy surface.
Then 
\[
{\rm Vol}(J^\mp(p)\cap \tilde{\mathcal{Q}}\cap J^\pm(\tilde{\mathcal{S}}))=\infty,
\]
where ${\rm Vol}$ can be taken here to refer either to the Lorentzian quotient,
or the $4$-dimensional spacetime.
\end{proposition}
 \begin{proof}
 This follows immediately from the results of Section~\ref{alternatif}.
 \end{proof}
 
\section{$k\le0$, $\Lambda\ge 0$}
\label{easiersec}
\subsection{Characterization of $\mathcal{B}^{\pm}$}

\begin{proposition}
\label{eitherf0}
Suppose $k\le0$, $\Lambda\ge0$. Then, either $k=0=\Lambda=f$ and the spacetime is
 flat, or,
after possibly reversing the time orientation,
we have $\lambda>0$, $\nu>0$ everywhere.
\end{proposition}

\begin{proof}
Let $\tilde{\mathcal{S}}$ be the lift of a Cauchy surface for $\mathcal{Q}$.
Suppose there is a point where say $\lambda>0$, $\nu<0$ along $\mathcal{S}$.
Consider the connected component $I$ of the set $\{\lambda>0,\nu<0\}$,
containing this point. It is a nonempty open subset of $\tilde{\mathcal{S}}$.
Since $r$ clearly strictly increases along $I$, it is clear by the periodicity
of $r$ along $\tilde{\mathcal{S}}$, that $I\neq\tilde{\mathcal{S}}$. The closure
of $I$ thus has two endpoints $p$, $q$, in $\tilde{\mathcal{S}}$. At those points
we have $k-\frac{2m}r=0$, i.e.~$2m=rk$. In the case $k=-1$, this is a contradiction,
because, if $p$ denotes the left endpoint, $r(q)>r(p)$ but $m(q)\ge m(p)$. 
In the case $k=0$, this contradicts the non-emptyness of $I$, for one
obtains that $2m=0$ identically along $I$, and thus, we cannot have
$\lambda>0$, $\nu<0$.

Thus, we have shown that, after choosing appropriately the time orientation,
we have, $\lambda\ge 0$, $\nu\ge0$ along $\tilde{\mathcal{S}}$. By
Raychaudhuri, we have in fact $\lambda\ge0$, $\nu\ge0$ in $J^-(\tilde{\mathcal{S}})$.
But now let $p$ be a point of $\tilde{\mathcal{S}}$
where, say, $\lambda=0$. From $(\ref{newevol2})$, and
the fact that $2m=rk$ at $p$, we have, in the case $k=-1$, or
$k=0$, $\Lambda>0$ that $\partial_u\lambda>0$.
This would imply that there are points in $J^-(\tilde{\mathcal{S}})$ where $\lambda$ becomes
negative, a contradiction. Thus $\lambda>0$, $\nu>0$ along $\tilde{\mathcal{S}}$. 
By a continuity argument, it follows that under this choice of
time orientation, $\lambda>0$, $\nu>0$ in all of $\tilde{\mathcal{Q}}$.

If $k=0$, $\Lambda=0$, and $\lambda=0$ say at some $p=(0,0)$ 
on $\tilde{\mathcal{S}}$, then since from $(\ref{evol1})$ and the fact that $\lambda\ge 0$
in $J^-(\tilde{\mathcal{S}})$, we must have that $\lambda=0$ identically on $\{v=0\}\cap\{ u\le0\}$
and thus 
$T_{uv}=0$ on $\{v=0\}\cap\{ u\le0\}$. It follows from the definition of
the energy momentum tensor that $f$ must vanish identically on 
$\{v=0\}\cap\{u\le 0\}$, and, thus, by the Vlasov equation that $f$ must vanish identically
on $\tilde{\mathcal{S}}\cap \{u_1\le u\le 0\}$ for some $u_1<0$. 
Also, by $(\ref{const2})$, it follows 
that $\lambda=0$ in $\{u_1\le u\le 0\}\cap \tilde{\mathcal{S}}$  for
some $u_1 <0$. By  continuity, one obtains that $\lambda=0$, $f=0$ identically on
$\tilde{S}$. From this, one obtains that the
spacetime is in fact flat.
\end{proof}

The above result was in fact proved in~\cite{crushing}. A proof has been included here
to make the paper more self-contained and to introduce the reader to
some of the techniques used later.

\begin{proposition}
Suppose $k\le0$, $\Lambda\ge0$. 
If the time orientation is such that
$\lambda>0$, $\nu>0$, then either $\mathcal{B}^+=\mathcal{B}^+_\infty=\mathcal{N}^+$,
or 
\[
\mathcal{B}^+=\mathcal{B}^+_\infty\cup
\bigcup_{x\in\mathcal{B}^+_\infty}\mathcal{N}^1_x\cup\mathcal{N}^2_x,
\]
with $r_{\inf}=\infty$ identically on $\mathcal{B}^+$. Moreover, for this choice
of time orientation,
 either $\mathcal{B}^-=\mathcal{B}^-_s$, or $\mathcal{B}^-=\mathcal{N}^-$.
\end{proposition}

\begin{proof}
We prove only the statement about $\mathcal{B}^+$, as the second statement will
follow as a special case of the results of Section~\ref{pastev}.

Clearly, from the signs $\lambda>0$, $\nu>0$,
we have a lower bound $r\ge r_0$ in $\tilde{\mathcal{Q}}^+= J^+(\tilde{\mathcal{S}})$.
In particular, $\mathcal{B}^+_s=\emptyset$.

Consider the case first that $\mathcal{B}^+\ne \mathcal{N}^+$.
Let $(u_1,v_1)\in\mathcal{B}_1^{+}$ such that $r\le R$ for $[u_0,u_1)\times\{v_1\}$.
We will show that 
\begin{equation}
\label{toshowhere}
\int_{u_0}^{u_1}\Omega^2(u,v_1)du<\infty.
\end{equation}
For this, it suffices to obtain pointwise bounds on $\Omega^2$.
Integrating twice $(\ref{newevol2})$, it follows--in view of the signs--that 
we need only bound
\[
-\int_{v_0}^{v_1}\int_{u_0}^{u_1}2r^{-3}\kappa\nu mdudv <\infty.
\]
Again, since, if $m\ge 0$, our bounds on $r$ imply that $-m(k-\mu)^{-1}$ is bounded,
it suffices to show that
\[
\int_{v_0}^{v_1}\int_{u_0}^{u_1}\lambda\nu dudv<\infty
\]
and for this, in view again of the bounds on $r$, it suffices to show
\[
\int_{u_0}^{u_1}\sup_{v_0\le v\le v_1}|\nu(u,v)|du<\infty.
\]
Integrating $(\ref{newevol1})$ backwards in time,
in view of the fact that
\[
\int_{u_0}^{u_1}\nu(u,v_1)du<\infty
\]
by assumption, and the fact that $-m(k-\mu)^{-1}$ is bounded when $m\ge0$,
we obtain the above bound, and thus, $(\ref{toshowhere})$.

It follows that for $x\in\mathcal{B}^{\pm}_1$, then $r\to \infty$ along $\mathcal{H}^1_x$,
$\mathcal{H}^2_x$. 
From the signs $\lambda>0$, $\nu>0$, we obtain easily that
$r_{\inf}(x)=\infty$ for all $x\in\mathcal{B}^{\pm}_1$, and thus
that $\mathcal{B}^+_1=\mathcal{B}^+_\infty=\mathcal{B}^+_H$.
The decomposition follows from Proposition~\ref{reform}. Note that from $\nu>0$, $\lambda>0$,
it follows easily that $r_{\inf}(x)=\infty$ for
$\mathcal{N}_x^i$.

Finally, consider the case $\mathcal{B}^+=\mathcal{N}^+$. It is clear by the inequalities
$\lambda>0$, $\nu>0$ that either $r_{\rm inf}=\infty$ or $r\le R$ uniformly.  By 
monotonicity, we have
$\partial_u\kappa\ge0$, and by the periodicity of initial data,
it follows that $\int_{v'}^{v^*}\kappa (u^*,v)dv=\infty$, where
$(u^*,v^*)\in\mathcal{B}^+$. One sees easily from $(\ref{newevol1})$, that,
under the assumption that $r\le R$ 
\[
\int_{u'}^{u^*}\nu (u,v)\to\infty
\]
as $v\to v^*$, for any $u'<u^*$. Integrating in $u$ gives $r\to \infty$ which
contradicts $r\le R$. We thus have $r_{\rm inf}=\infty$.
By Proposition~\ref{reform}, we obtain $\mathcal{B}^+=\mathcal{N}^+=\mathcal{B}^+_\infty$.
\end{proof}
Parts of the above Proposition were in fact proved in~\cite{hrr:onthe} and~\cite{tr}.
Finally, we have

\begin{proposition}
\label{acausprop}
Suppose $k\le0$, $\Lambda\ge0$, and the time orientation
is such that $\lambda>0$, $\nu>0$. Then
$\mathcal{B}^+_\infty=\mathcal{B}^+=\mathcal{N}^+$ iff $\Lambda=0$. Moreover, if $\Lambda>0$,
$\mathcal{B}^+=\mathcal{B}^+_\infty$ is acausal.
\end{proposition}

\begin{proof}
Suppose first that $\Lambda=0$. Suppose $\mathcal{B}^+\ne\mathcal{N}^+$. Then
there exists a
$p\in\mathcal{B}^+_1$ ``preceded'' by
 a $\mathcal{D}\subset\tilde{\mathcal{Q}}$.

Let $p=(u_1,\infty)$ and
define coordinates on $\mathcal{D}$ such that
\[
\mathcal{D}=[u_0,u_1]\times[v_0,\infty]\setminus
\{(u_1,\infty)\},
\]
and $\Omega^2(\cdot,v_0)=1$, $\Omega^2(u_1,\cdot)=1$.

Dividing equation $(\ref{evol1})$ by $r$, we obtain the equation
\[
\partial_u\partial_v\log r = 2r^{-3}m\kappa\nu+4\pi T_{uv} -\frac{\lambda\nu}{r^2}.
\]
Integrating in $\mathcal{D}$ we obtain
that
\[
+\infty=\lim_{v\to \infty}\int_{u_0}^{u_1}\int_{v_0}^{v}\partial_u\partial_v\log r=\lim_{v\to \infty}
\int_{u_0}^{u_1}\int_{v_0}^{v}
 2r^{-3}m\kappa\nu+4\pi T_{uv} -\frac{\lambda\nu}{r^2}.
\]
On the other hand
\begin{eqnarray*}
\log\Omega (u_0,v)&=&-\int_{u_0}^{u_1}\int_{v_0}^{v}\partial_u\partial_v\log \Omega\\
&=&\int_{u_0}^{u_1}\int_{v_0}^v
  2r^{-3}m\kappa\nu+4\pi T_{uv} -\frac{\lambda\nu}{r^2} +\frac{\lambda\nu}{r^2}
-4\pi \kappa\nu g ^{AB}T_{AB}\\
&\ge&\int_{u_0}^{u_1}\int_{v_0}^v
  2r^{-3}m\kappa\nu+4\pi T_{uv} -\frac{\lambda\nu}{r^2}
\end{eqnarray*}
and thus
\[
\log\Omega^2(u_0,v)\to\infty
\]
as $v\to\infty$, in particular, $u=u_0$ has infinite affine length.
By the previous proposition, it follows that $r\to\infty$. But this contradicts
the fact that $(u_0,\infty)\in\tilde{\mathcal{Q}}$.

Now assume that $\Lambda>0$.
Let $p= (u_1,v_2)\in \mathcal{B}^+$, with $\{u_1\}\times[v_1,v_2)\in\tilde{\mathcal{Q}}$,
and
 choose moreover $(u_1,v_1)$ such that $r(u_1,v_1)$ is sufficiently large.
Then, for any $u_2>u_1$, we have
\begin{equation}
\label{sonraiCin}
-\frac{k\Omega^2}{4r}-r^{-1}\partial_ur\partial_vr+4\pi T_{uv}+\frac14r\Omega^2\Lambda
\ge\frac14r\Lambda\Omega^2-r^{-1}\partial_u r\partial_v r
\end{equation}
in $[u_1,u_2]\times[v_1,v_2)\cap\tilde{\mathcal{Q}}$.
Thus we have
\[
\partial_v\nu\ge \frac14r\Lambda\Omega^2-r^{-1}(\partial_v r)\nu
\ge \bar{c}r\lambda-r^{-1}(\partial_vr)\nu,
\]
for some constant $\bar{c}>0$,
where in the last inequality we have used the Raychaudhuri equation.
Multiplying by $r$, we obtain
\[
\partial_v(r\nu)\ge \bar{c}r^2\lambda
\]
and thus
\begin{equation}
\label{difeq}
\partial_ur^2\ge \frac23\bar{c}r^3-\frac23\bar{c}R^3
\end{equation}
in $[u_1,u_2]\times[v_1,v_2)\cap\tilde{\mathcal{Q}}$,
where
\[
R=\sup_{u_1\le u\le u_2} r(u,v_1).
\]
But as $r(u_1,v)\to\infty$, the blow up time of equation
$(\ref{difeq})$ in $u-u_1$ goes to $0$. In particular, if $p$ denotes the
point on $u=u_1$ which intersects $\mathcal{B}^+$, then there is no future-directed
constant-$v$ component of $\mathcal{B}^+$ through $p$.

One argues similarly for constant-$u$ components.  This means that $\mathcal{N}^i_x=\emptyset$.
It follows that
$\mathcal{B}^+$ is acausal--in particular $\mathcal{B}^+\ne\mathcal{N}^+$--and 
 $\mathcal{B}^+=\mathcal{B}^+_\infty$ by the previous proposition.
\end{proof}

\section{The past evolution of antitrapped data}
\label{pastev}
\begin{proposition} 
\label{pastevprop}
Let $k$, $\Lambda$ be arbitrary, and let $\lambda>0$,
$\nu>0$ on $\tilde{\mathcal{S}}$. Then $\mathcal{B}^{-}_H=\emptyset$, and
the Penrose diagram is as in the statement
of Theorem~\ref{introthe15}.
\end{proposition}
\begin{proof}
Note that by the Raychaudhuri equations $(\ref{const1})$, $(\ref{const2})$, it follows that
$\nu>0$ throughout the past of $\mathcal{S}$.
Suppose $p\in\mathcal{B}^-_H$.  Let $p=(u_1,v_1)$ in some system of
global null coordintates. Since 
\[
\partial_{(-u)}\frac{-\lambda}{k-\mu}\le 0,
\]
\[
\partial_{(-v)}\frac{-\nu}{k-\mu}\le 0,
\]
we have that for $(u_0,v_0)\in\tilde{\mathcal{Q}}$ with $u_0>u_1$, $v_0>v_1$
\begin{eqnarray*}
\int_{u_1}^{u_0}\Omega^2(u,v_1) du&=&\int_{u_1}^{u_0}4\frac{-\lambda}{k-\mu}(u,v_1) \nu du\\
				  &\le&\sup_{u_1\le u\le u_0}\frac{-\lambda}{k-\mu}(u,v_1)
						\int_{u_1}^{u_0}\nu(u,v_1) du\\
				&\le&\sup_{u_1\le u\le u_0}\frac{-\lambda}{k-\mu}(u,v_0)
						\int_{u_1}^{u_0}\nu(u,v_1) du\\
				&\le& C.
\end{eqnarray*}
A similar inequality holds for
\[
\int_{v_1}^{v_0}\Omega^2(u_1,v)dv.
\]
This contradicts the statement that $p\in\mathcal{B}^-_H$.

Thus we have shown $\mathcal{B}^-_H=\emptyset$. The structure of the Penrose diagram
now follows from Proposition~\ref{reform}.
\end{proof}

We also have the following, which will be particularly important in Section~\ref{SCCsec}
for the case $k\ge 0$.
\begin{proposition}
\label{otherprop}
Suppose $k\ge0$, $\Lambda\le 0$,
and $\nu>0$, $\lambda>0$ on $\mathcal{S}$. Then there exists
an $\epsilon>0$ such that $m\ge k\min_{\mathcal{S}}r/2+\epsilon$ in $J^-(\mathcal{S})$.
In particular, for $(u,v)$ with $r(u,v)\le \min_{\mathcal{S}}r$, we have
$k-\mu\le-2\epsilon(\inf_{\mathcal{S}}r)^{-1}$.

For general, $k$ and $\Lambda$, if on $\mathcal{S}$ we have $\nu<0$, $\lambda<0$
and
\[
m\ge  \max\{k,0\}\inf_{\mathcal{S}}r/2 +\max\{-\Lambda,0\}
\sup_{\mathcal{S}}r^3 +\epsilon,
\]
for some $\epsilon\ge 0$,
then
\[
m\ge  \max\{k,0\}\inf_{\mathcal{S}}r/2 +\epsilon
\]
in $J^-(\mathcal{S})$. 
In particular, for $(u,v)$ with $r(u,v)\le \min_{\mathcal{S}}r$, we have
$k-\mu\le-2\epsilon(\inf_{\mathcal{S}}r)^{-1}$.
\end{proposition}
\begin{proof}
For the first statement:
Clearly, we must have by compactness and the condition $k-\mu<0$
that $m\ge k\min_{\mathcal{S}}r/2+\epsilon$ on $\mathcal{S}$.
Consider the vector field on $J^-(\tilde{\mathcal{S}})$ defined by
\[
T= -(\nu^{-1}\partial_u + \lambda^{-1}\partial_v).
\]
In view of the previous proposition, this vector field is well defined,
past pointing timelike, and does
not depend on the choice of null coordinates.
We have
\begin{eqnarray*}
T m 	&=& -8\pi r^2\Omega^{-2} (2T_{uv} -T_{uu}\lambda\nu^{-1}-T_{uu}\nu\lambda^{-1})
				- r^2\Lambda \\
	&=&-16\pi r  m (2T_{uv} \lambda^{-1}\nu^{-1} - T_{uu}\nu^{-2} -T_{vv}\lambda^{-2})
				- r^2\Lambda\\
	&\ge& -r^2\Lambda\\
	&\ge& 0,
\end{eqnarray*}
by assumption. Thus, $m\ge \epsilon+k\min_{S}r/2$ in $J^-(\tilde{\mathcal{S}})$, since
from $Tr=-1$ we easily see that all integral curves of $T$ cross $\tilde{\mathcal{S}}$.

The second statement follows by repeating the above computation in the general case,
keeping track of the effect of the $-r^2\Lambda$ term.
\end{proof}

\section{The case $k=1$, $\Lambda\ge 0$}
In this section, let us suppose that $k=1$, $\Lambda\ge 0$. For convenience,
without loss of generality, we shall discuss only future evolution.

\subsection{$\Lambda=0$}
\label{k1L0}
We show in this section
\begin{proposition}
If $\Lambda=0$, then $\mathcal{B}^{\pm}_h=\emptyset$.
\end{proposition}
\begin{proof}
By the results of~\cite{burnett}, it follows that in the case
$\Lambda=0$, $r$ is uniformly bounded above.
Moreover it can be shown that the total volume of spacetime is bounded.
To see this, first note that it follows from the results of \cite{br:emh}
and \cite{oh:gpmc} that the spacetime can be covered by a constant mean
curvature foliation, where the mean curvature $\tau$ of the leaves runs
from $-\infty$ to $\infty$. Let $h(\tau)$ be the induced metric of the
leaf of mean curvature $\tau$ and let $\alpha$ be the lapse function of
the foliation. Then the spacetime volume can be bounded by
$\int_{-\infty}^\infty \bar\alpha (\tau) {\rm Vol}(h(\tau)) d\tau$
where $\bar\alpha(\tau)$ denotes the maximum of $\alpha$ on the leaf of
mean curvature $\tau$. The restriction of this integral to the interval
$[-1,1]$ is obviously finite and so it remains to bound its restriction
to the set $(-\infty,-1]\cup [1,\infty)$. As shown in~\cite{crushing}
$\alpha\le 3/\tau^2$ and it was proved in~\cite{br:emh} 
that ${\rm Vol}(h(\tau))$ is bounded. Putting these facts
together shows that the spacetime volume is finite.

Let $p\in\mathcal{B}^\pm_h$. We have that $r_{\inf}(p)>0$. On the other hand,
by the uniform bound on $r$, $r_{\sup}(p)<\infty$. 
Proposition~\ref{alternatifprop} thus applies.
We obtain
\[
{\rm Vol}(J^\mp(p)\cap \tilde{\mathcal{Q}}\cap J^\pm(\tilde{\mathcal{S}}))=\infty,
\]
where we can interpret the volume as volume upstairs. But the
region $J^\mp(p)\cap \tilde{\mathcal{Q}}\cap J^\pm(\tilde{\mathcal{S}})$ is covered
by compactness by finitely many fundamental domains of $\mathcal{Q}$. This
contradicts the fact that the volume of each of these was shown above to be finite.
Thus $\mathcal{B}^\pm_h=\emptyset$.
\end{proof}

\subsection{$\Lambda>0$}
Let us suppose that $x\in\mathcal{B}_H^+$. Fix say $\mathcal{H}=\mathcal{H}^1_x$,
let $\mathcal{N}$ denote $\mathcal{N}^1_x$, 
and let $r_+$ denote the limit of $r$ along $\mathcal{H}$. 
We call $\mathcal{H}$ \emph{nonextremal} if
\begin{equation}
\label{nondegcond}
r_+\neq\frac1{\sqrt{\Lambda}}.
\end{equation}
We will show in this section
\begin{theorem}
If $r_+>\frac1{\sqrt{\Lambda}}$, or else, if $r_+<\frac1{\sqrt\Lambda}$
and 
\begin{equation}
\label{revisedassump}
1-\mu\ge0, \qquad -\nu (u,v) \ge e^{\alpha\int_{v'}^v{\frac{\lambda dv}{1-\mu}}},
\end{equation}
for all sufficiently late affine 
advanced time $v\ge v'$ along $\mathcal{H}$, for some $\alpha>0$,
then $\mathcal{N}=\emptyset$.
\end{theorem}

\begin{proof}
We consider separately the two cases:
\subsubsection{$r_+<\frac1{\sqrt{\Lambda}}$}
\label{lessthan}
Let us choose a coordinate
system such that
\[
\{0\}\times[0,\infty)\subset\mathcal{H},
\] 
with $(\ref{revisedassump})$ holding,
and such that $\kappa(0,v)=1$.
We see easily that the $v$ range of this coordinate system is indeed necessarily infinite
as claimed,
and, in view of $(\ref{revisedassump})$,
\begin{equation}
\label{sesuvt}
\log \Omega^2(0,v) \ge \alpha v.
\end{equation}

We will show that for $U>0$, we cannot have $r\ge r_0>0$
in
\[
\mathcal{D}=[0,U]\times [V,\infty)\cap\mathcal{Q}.
\]

So we will suppose, for the sake of contradiction,
that in fact, $r\ge r_0>0$ in the above set.

\begin{lemma}
We can select $U, V$ such that
$\frac{\mu}{r} > \tilde{c}$ uniformly in $\mathcal{D}$,
and $\nu< 0$, with $\tilde{c}$ arbitrarily
close to $r_+^{-1}$, and thus there exists a constant $c>0$ such that
\begin{equation}
\label{arrangement}
\frac{\mu}r-r\Lambda >c>0.
\end{equation}
\end{lemma}
\begin{proof}
Note first that $\mu(0,v)\to 1$. For otherwise, by monotonicity, there would exist
an $\epsilon>0$ such that $\mu(0,v)\le 1-\epsilon$ for
all $v\ge V$. In this case,
we have $1=\kappa(0,v)= \lambda/(1-\mu)(0,v)\le c\lambda(0,v)$ for some $c>0$.
This gives a contradiction upon integration in $v$.

Thus, since $r(0,v)\to r_+$, given an arbitrary 
$\tilde{c}<r_+^{-1}$, we can choose $V$ sufficiently large so
that $\mu/r(0,v)>\tilde{c}$  for $v\ge V$.

Now, we can have chosen $V$ such that $\nu(0,v)<0$ for
$v\ge V$, in view of $(\ref{revisedassump})$. By Raychaudhuri,
we then have $\nu<0$ in $\mathcal{D}$. Given an $\epsilon>0$,
in view of the fact that $r\to r_+$ along $\mathcal{H}^+$ and $\lambda(0,v)\ge 0$,
we have $r(0,v)\le r_+$ for $v\ge V$,
and thus, since $\nu<0$, $r\le r_+$ on $\mathcal{D}$.

Thus, since $\mu\ge 1$, in the region $\{\lambda\le 0\}\cap \mathcal{D}$,
choosing $\epsilon>0$ sufficiently small 
we have that $\mu/r >\tilde{c}$ in this region.

It suffices to consider then the region $\{\lambda>0\}\cap \mathcal{D}$.
By Raychaudhuri, this region is foliated by possibly empty connected null curves
emanating from $[0,U]\times\{V\}$. 

Since $m(0,v)\to r_+/2$, as $v\to\infty$, we may choose $U$ sufficiently small and
$V$ sufficiently large such that $m(u,V)\ge (r_+/2)(1-\epsilon)$.
Since $\partial_v m \ge 0$ in $\mathcal{D}\cap \{\lambda \ge 0\}$, 
by the remark on the foliation of this region by null curves, it follows
that $m(u,v)\ge (r_+/2)(1-\epsilon)$ in $\mathcal{D}\cap\{\lambda\ge 0\}$,
and thus, since $r\le r_+$ in $\mathcal{D}$, we have that $\mu\ge 1-\epsilon$
in $\mathcal{D}\cap\{\lambda\ge 0\}$. Choosing $\epsilon>0$ sufficiently
small, we again ensure $\mu/r>\tilde{c}$ in this region. Thus,
$\mu/r>\tilde{c}$ in all of $\mathcal{D}$.

The inequality $(\ref{arrangement})$
now follows immediately.
\end{proof}

Note that in view of the mass shell relation written
\[
4\kappa(-\nu)p^up^v=1+r^2\gamma_{AB}p^Ap^B,
\]
we have that, 
\[
(-\nu)p^u + 2\kappa p^v\ge \sqrt{1+r^2\gamma_{AB}p^Ap^B}.
\]
Similarly to $(\ref{comput})$ we compute:
\begin{eqnarray}
\label{computalt}
T_{uv}	& = &	\int_0^\infty\int_{-\infty}^{\infty}
\int_{-\infty}^{\infty}r^2 p_up_v f\frac{dp^u}{p^u}\sqrt{\gamma} dp^A dp^B\\
	\nonumber
	& = &	(g_{uv})^2
			\int_0^\infty\int_{-\infty}^{\infty}
			\int_{-\infty}^{\infty}
				r^2 p^up^v 
			f\frac{dp^u}{p^u}\sqrt{\gamma} dp^A dp^B\\
	\nonumber
	& = &	-\frac12 g_{uv}\int_0^\infty\int_{-\infty}^{\infty}
			\int_{-\infty}^{\infty}
				r^2\Omega^2 p^u p^v  f\frac{dp^u}{p^u}\sqrt{\gamma} dp^A dp^B\\
	\nonumber
	& = &	\frac14\Omega^2 \int_0^\infty\int_{-\infty}^{\infty}\int_{-\infty}^{\infty}
				r^2 (1+r^2\gamma_{AB}p^A p^B)  f\frac{dp^u}{p^u}\sqrt{\gamma} dp^A dp^B\\
	\nonumber
	&\le&	\frac14\Omega^2 \sqrt{1+Xr_0^{-2}} \int_0^\infty\int_{-\infty}^{\infty}\int_{-\infty}^{\infty}
			r^2 ((-\nu)p^u + 2\kappa p^v) f \frac{dp^u}{p^u}\sqrt{\gamma} dp^A dp^B\\
	\nonumber
	&=&	\sqrt{1+Xr_0^{-2}}(-\nu)\frac14\Omega^2 N^u+
	\sqrt{1+Xr_0^{-2}}\frac12   \kappa \Omega^2 N^v.
\end{eqnarray}
Note that given $\epsilon>0$, we can choose $U$ sufficiently small and $V$ sufficiently
large so that the flux of particle current through $0\times[V,\infty]\cup [0,U]\times\{V\}$
is less than $\epsilon$. By conservation of particle current it follows that
\begin{equation}
\label{fbnd1}
\int_0^u N^v\Omega^2 r^2(\bar{u},v) d\bar{u} \le \epsilon
\end{equation}
\begin{equation}
\label{fbnd2}
\int_V^v  N^u\Omega^2 r^2(u,\bar{v}) d\bar{v} \le \epsilon
\end{equation}
for all $(u,v)\in \mathcal{D}$.

Integrating $(\ref{newevol1})$ backwards $v$ in view of $(\ref{arrangement})$,
 we obtain
\begin{eqnarray*}
\sup_{V\le \bar{v}\le v}|\nu(u,\bar{v})|&\le& \left(|\nu(u,v)|+\int_V^v 
2\pi\sqrt{1+Xr_0^{-2}}\kappa \Omega^2N^v
dv\right)\cdot\\
&&\hbox{}\cdot e^{-\int_V^v\pi r\sqrt{1+Xr_0^{-2}}\Omega^2 N^u dv }\\
	&\le&	C\left((-\nu)(u,v)+\int\kappa r^2 \Omega^2 N^v dv\right),
\end{eqnarray*}
where we have used
$(\ref{fbnd2})$.
Thus we have
\begin{eqnarray}
\label{kaitoe3ns}
\nonumber
\int_0^u \sup_{V\le\bar{v}\le v}|\nu(u,\bar{v})|du&\le&	
C\left(  \int_0^u |\nu|(u,v)du+ \epsilon\int_V^v\sup_u \kappa \right)\\
&\le& Cr_+ + C\epsilon (v-V).
\end{eqnarray}
The constant $C$ can be chosen independently of $\epsilon$.

Recall from the proof of the lemma that 
$\mathcal{D}\cap\{\lambda \ge0\}$ is foliated
by constant-$u$ segments emanating from $[0,U]\times\{V\}$.
Supposing $U,V$ are such that $r_+\ge (1-\epsilon)r_+$
on $[0,U]\times\{V\}$, it follows that 
on a $\{u\}\times [V,v]\subset\{\lambda\ge0\}$ we
have $\int_V^v r^{-2} \lambda(u,v) \le 
\epsilon(1-\epsilon)^{-1} r_+^{-1}$.
Thus, from $(\ref{kaitoe3ns})$, we have
\begin{equation}
\label{arrangement2}
\int_0^u\int_V^v 2r^{-2} \partial_ur\partial_vr \ge -\epsilon(1-\epsilon)^{-1} 
r_+^{-1}(Cr_+ + C\epsilon (v-V))
\end{equation}
holds for $(u,v)\in\mathcal{D}$.

On the other hand, integrating $(\ref{evol2})$, in view of  
$(\ref{sesuvt})$,
$(\ref{computalt})$, the inequality $\kappa\le 1$ in $\mathcal{D}$,
and the inequalities $(\ref{kaitoe3ns})$  and $(\ref{arrangement2})$,
we have
\begin{eqnarray*}
\log \Omega^2(u,v)	&\ge& \alpha v -\int_0^u\int_V^v T_{uv}-
				\epsilon(1-\epsilon)^{-1} 
r_+^{-1}(Cr_+ + C\epsilon (v-V))	\\	
		&\ge& \alpha v -\frac14 r_0^{-2}\sqrt{1+Xr_0^{-2}}
			\int_0^u\sup_{V\le \bar{v}\le v} |\nu(\bar{u},\bar{v})| d\bar{u} \int_V^v N^u\Omega^2 
			r^2(u,\bar{v}) d\bar{v}\\
		&&\hbox{}
		-\frac12 r_0^{-2}\sqrt{1+Xr_0^{-2}}\int_V^v d\bar{v} \int_0^uN^v\Omega^2 
			r^2(\bar{u},v) d\bar{u}\\
		&&\hbox{}-\epsilon(1-\epsilon)^{-1} 
r_+^{-1}(Cr_+ + C\epsilon (v-V))\\
		&\ge& \alpha v- \frac14 r_0^{-2}\sqrt{1+Xr_0^{-2}}\big((Cr_++C\epsilon(v-V))\epsilon
					+2\epsilon (v-V)\big)\\
		&&\hbox{}-\epsilon(1-\epsilon)^{-1} 
r_+^{-1}(Cr_+ + C\epsilon (v-V)).
\end{eqnarray*}
It follows that by choosing $V$ sufficiently large, and $\epsilon$ sufficiently small,
we may obtain
\[
\log\Omega^2 \ge \tilde{\alpha} v
\]
for any $\tilde{\alpha}<\alpha $. In particular, 
$\Omega^2\ge 1$ in all of $\mathcal{D}$.

We can revisit $(\ref{computalt})$ and estimate now
\[
\int_0^u\int_V^v rT_{uv}dudv\le \epsilon\tilde{c}\int_0^u\int_V^v\Omega^2dudv,
\]
for a constant $\tilde{c}$ independent of $\epsilon$.
Finally, integrating $(\ref{newevol1})$ in $u$ and $v$, we estimate
in view of $(\ref{arrangement})$ and the above inequality
\begin{eqnarray*}
\int_0^u\nu 
&\le& -c'\int_0^u\int_V^v \Omega^2dudv+\int_0^u\int_V^v rT_{uv}dudv\\ 
&\le& (-c'+\epsilon \tilde{c})\int_0^u\int_V^v \Omega^2dudv.
\end{eqnarray*}
For sufficiently small $\epsilon>0$,
the right hand side above goes to $-\infty$ as $v\to\infty$, and this is a contradiction.

\subsubsection{$r_+>\frac{1}{\sqrt\Lambda}$}
\label{cosmohor}
Clearly, it follows that $\nu\to\infty$ along 
$\mathcal{H}$. Choose coordinates such that $\Omega^2=1$
along $\mathcal{H}$, and such that $\mathcal{H}$ is $u=0$,
choose $(0,0)$ along $\mathcal{H}$ such
that $\nu>\delta>0$, and $r(0,0)>1/\sqrt{\Lambda}$  and choose $(U,0)$ such that
$\nu>\delta/2 $ along $[0,U]\times\{0\}$. 
Define
\[
\mathcal{D}=[0,U]\times [0,\infty)\cap\mathcal{Q},
\]
and define
\[
\mathcal{D}'=\{(u,v)\in\mathcal{D}:\nu(u^*,v^*)>0, (u^*,v^*)\in
J^-(u,v)\cap\mathcal{D}\}.
\]

Clearly, since $\lambda(0,v)\ge0$, and $\nu>0$ in $\mathcal{D}'$,
we have
\begin{equation}
\label{rbdd'}
r\ge r(0,0)>1/\sqrt{\Lambda}
\end{equation}
in $\mathcal{D}'$. 
Now, integrating $(\ref{newevol1})$, we have
\begin{equation}
\label{tobeimpr}
\nu(u,v) \ge \nu(u,0)e^{\int{\frac{\mu}{r}\frac{\lambda}{1-\mu}+
\Lambda r}}.
\end{equation}
We can choose $\mu_0$ close to $1$, $\mu_0>1$, such that for
$\mu\le \mu_0$, the second term
in the integrand dominates, in view of the inequality
$(\ref{rbdd'})$.
Thus,
\begin{eqnarray*}
\nu(u,v)&\ge& \nu(u,0)e^{\int{\frac{\mu_0}{1-\mu_0}\frac{\lambda}r}}\\
	&\ge& \delta r^{\frac{\mu_0}{1-\mu_0}}.
\end{eqnarray*}
The above estimate shows that $\mathcal{D}'$ is closed
in $\mathcal{D}$, and thus, since it is clearly open,
by connectedness we have $\mathcal{D}=\mathcal{D}'$.
In particular $(\ref{rbdd'})$ holds throughout $\mathcal{D}$.

The fact that the affine length of $\mathcal{H}$ is infinite
implies that
\[
\int_0^\infty\frac{(-\lambda)}{1-\mu}(0,v)=\infty.
\]
The fact that $\nu>0$ implies that $\frac{(-\lambda)}{1-\mu}$
is nondecreasing in $u$. Let us suppose that
\begin{equation}
\label{letsuppose}
\{u_1\}\times [0,\infty)\subset\mathcal{D}
\end{equation}
and
\begin{equation}
\label{letsuppose2}
r(u_1,v)\le R<\infty.
\end{equation}
Clearly, this must be true also then for all $(u,v)\in\mathcal{D}$
with $u\le u_1$.
From $(\ref{tobeimpr})$, we can in fact obtain for such $(u,v)$
\begin{eqnarray*}
\nu(u,v)&\ge& \nu(u,0)e^{\int{\frac{\mu_0}{1-\mu_0}\frac{\lambda}r
		+\epsilon\frac{-\lambda}{1-\mu_0}}}\\
	&\ge& \delta r^{\frac{\mu_0}{1-\mu_0}}+\delta 
		e^{\int_0^v\frac{(-\lambda)}{1-\mu}(0,v)}\\
	&\ge&  \delta r(u_1,\infty)^{\frac{\mu_0}{1-\mu_0}}+\delta 
		e^{\int_0^v\frac{(-\lambda)}{1-\mu}(0,v)}.
\end{eqnarray*}
Thus $\nu(u,v)\to\infty$, as $v\to\infty$ uniformly in $u$
for $u\in[0,u_1]$. Integrating in $u$, we contradict 
$(\ref{letsuppose2})$.

We have shown that $\mathcal{N}=\emptyset$.
\end{proof}

We note that the results of Section~\ref{cosmohor} apply
not only to $x\in\mathcal{B}^+_H$, but to $x\in
\mathcal{N}_y$. Thus we have $r_{\sup}<\infty$ on interior points of $\mathcal{N}_y$.

Moreover, from the above argument we also easily retrieve the following result of~\cite{tsb}
\begin{proposition}
\label{toxreiaz}
Suppose $r> \frac{1}{\sqrt{\Lambda}}$, $\lambda>0$, $\nu>0$ on $\mathcal{S}$. 
Then $\mathcal{B}^+
=\mathcal{B}^+_\infty$.
\end{proposition}

\subsubsection{The structure of the boundary revisited}
The results of the Section~\ref{cosmohor} give immediately the following
refinement of Proposition~\ref{reform}:
\begin{proposition}
$\mathcal{B}^{\pm}_\infty$ is an acausal, open subset of $\mathcal{B}^\pm$.
We have the decomposition
\[
\mathcal{B}^\pm=\mathcal{B}^\pm_s\cup \mathcal{B}^\pm_\infty\cup
\bigcup_{x\in\mathcal{B}^\pm_h}{(\{x\}\cup \mathcal{N}^1_x\cup\mathcal{N}^2_x)}
\]
where $r_{\sup}<\infty$ on interior points of $\mathcal{N}^i_x$.
\end{proposition}

\subsubsection{The finiteness theorem}

Note that by the results of Sections~\ref{lessthan} and~\ref{cosmohor}, one deduces easily that 
the set
\begin{equation}
\label{thenonextreme}
\{x\in\mathcal{B}_h^\pm:\mathcal{H}^i_x {\rm\ satisfies\ }(\ref{case1}){\rm\ or\ }(\ref{case2})\}
\end{equation}
is a set of isolated points.
In fact, in the above we may weaken $(\ref{case2})$ by not requiring the first inequality to
be strict, and dropping the third inequality.

The proof of Theorem~\ref{finitetheorem}   is practically immediate.
Suppose the set $\mathcal{B}_h^\pm$ coincides with $(\ref{thenonextreme})$,
and that $\mathcal{B}_h^\pm\cap\mathcal{F}$ is infinite, where $\mathcal{F}$ is
a fundamental domain for $\mathcal{Q}$ in $\tilde{\mathcal{Q}}$. 
Note in this case we have
\[
\mathcal{B}^\pm=\mathcal{B}^\pm_h\cup \mathcal{B}^\pm_s\cup \mathcal{B}^\pm_\infty,
\]
where the union is disjoint.
Let $x_i\in\mathcal{B}_h^\pm$ be a sequence. By compactness, there exists
a convergent subsequence to a point of $x\in \mathcal{B}^\pm$. Since 
$\mathcal{B}^\pm_\infty$ and $\mathcal{B}^\pm_s$ are open
subsets of $\mathcal{B}^\pm$, disjoint from $\mathcal{B}_h^\pm$, then
$x\in\mathcal{B}_h^\pm$. But this contradicts the statement that $(\ref{thenonextreme})$
is discrete.

\section{Black holes in asymptotically flat spacetimes}
\label{bhs}
Spherically symmetric asymptotically flat solutions of
the Einstein-Vlasov system have been discussed in~\cite{dr:ep},
where an extension principle in the regular region, away from the 
centre, was proven. The extension principle of~\cite{dr:ep} can
in fact be reproved
easily from the results of Section~\ref{localest}. 
But, in fact, we can now
say much more. The relevant global estimate is provided by:
\begin{proposition}
Let $\mathcal{Q}^+$ denote the future evolution of the data considered
in~\cite{dr:ep}, and let $\mathcal{I}^+\subset\{u=U\}$. For $u_1<u_2<U$,
then if $r(u,v)\ge r_0>0$ on $[u_1,u_2)\times\{v\}$, then
\begin{equation}
\label{noother}
\int_{u_1}^{u_2}\Omega^2(u,v)du<\infty.
\end{equation}
\end{proposition}
\begin{proof}
Recall that $\nu<0$, $\kappa>0$ in $\mathcal{Q}^+$.
By equation $(\ref{newconst})$, it follows that $\partial_u\kappa\le 0$.
Thus, we have
\begin{equation}
\label{kappabnd}
\int_{u_1}^{u_2}\kappa(u,v)dv<\infty,
\end{equation}
in view of our assumptions on initial data.
On the other hand, since 
\[
r_0\le r(u,v)\le \sup_{\mathcal{S}\cap \{u\le u_2\}}r=R<\infty,
\]
we also have 
\begin{equation}
\label{lambdabnd}
\int_{u_1}^{u_2}|\lambda|(u,v)dv<\infty,
\end{equation}
as $\lambda$ can change sign at most once.
Since $\Omega^2=-4\nu\kappa $, to obtain $(\ref{noother})$, it suffices to obtain
a pointwise bound for $\nu$. 
In view of the fact that $\nu<0$, we have
\[
|\nu|(u,v)\le |\nu(u_1,v)|e^{\int_{u_1}^u\frac{\mu\lambda}{r(1-\mu)}(u,v)dv}.
\]
But now, we partition the integrand into the set where $\mu\ge 2$, and
$\mu\le 2$, and estimate the integral over the former set via $(\ref{lambdabnd})$,
and estimate the integral over the latter set via $(\ref{kappabnd})$.
Thus we obtain
\[
|\nu(u,v)|<C|\nu(u_1,v)|,
\]
as desired.
\end{proof}
We can now apply Theorem~\ref{extenthe} and this yields the Penrose
diagram of Theorem~\ref{afth}.
Let $M_f$ denote the final Bondi mass, and let $r_+$ denote the asymptotic
area radius of the event horizon. Recall from~\cite{trapped} that
$1\le 2M_f r_+^{-1}$. 
\begin{proposition}
There exists a constant $\delta_0>1$ such that if
\begin{equation}
\label{eav}
2M_f r_+^{-1} < \delta_0
\end{equation}
then $(\ref{revisedassump})$   holds on $\mathcal{H}^+$.
\end{proposition}

\begin{proof}
Consider coordinates such that $\log \Omega^2 =0$ on some ingoing null ray $v=0$,
and $\kappa=1$ on some late retarded time $U$, such that
\begin{equation}
\label{havearranged}
1-\epsilon\le 2mr_+^{-1} \le \delta_0
\end{equation}
for $v\ge 0$, $u\ge U$.
Let the event horizon $\mathcal{H}^+$ correspond to $u=U'>U$.
 
We note first that, from $(\ref{pum})$, we have the inequality
\begin{equation}
\label{havethein}
2\pi \kappa^{-1} r^2 T_{uv} \le -\partial_u m .
\end{equation}
On the other hand, from $(\ref{newconst})$, we have
\begin{equation}
\label{k<=1}
\kappa \le 1,
\end{equation}
and thus
integrating in $u$, in view of $(\ref{havearranged})$, we obtain that
\[
\int_{U}^{U'} 2\pi r^2 T_{uv} \le \frac{r_+}2 (\delta_0 -1+\epsilon)
\]
whence
\begin{equation}
\label{evawhence}
\int_{0}^{v}\int_{U}^{U'} {2\pi r^2 T_{uv}} \le \frac{r_+}2 (\delta_0 -1+\epsilon)v.
\end{equation}

On the other hand, we may reexpress $(\ref{havethein})$ as
\[
 2\pi (1-\mu) (-\nu)^{-1}r^2T_{uu}\le  -\partial_u m
\]
from which, 
rewriting as 
\[
4\pi (-\nu)^{-1}rT_{uu} \le 2(r-2m)^{-1}(-\partial_u m)
\]
we obtain using
$(\ref{newconst})$ 
that 
\[
\kappa \ge e^{-\frac{\delta_0-1+\epsilon}{\delta_0} }
\]
in the region where $r \ge 2\delta_0 r_+$.
Let $v=0$ be chosen late enough so that, along
$\{U\}\times[0,\infty)$, the inequality $r\ge 2\delta_0r_+$ holds, in fact
so that $r\ge \epsilon^{-1/2}r_+$ holds.
Let $(u,v)\in[U,U']\times[0,\infty)$ be such that $r(u,v)\le 2\delta_0r_+$. 
Denoting by $\mathcal{X}(u,v)=[U,u]\times [0,v]\cap \{r\ge 2\delta_0r_+\}=
[U,U']\times[0,v]\cap\{r\ge 2\delta_0r_+\}$, we have
\begin{eqnarray}
\label{oxiwhence}
\nonumber
\int_0^v\int_{U}^u{mr^3 \kappa(-\nu)}dudv &\ge& \int \int_{\mathcal{X}(u,v)} 
\frac{r_+}2(1-\epsilon)
r^{-3}(-\nu) 
e^{-\frac{\delta_0-1+\epsilon}{\delta_0} }du dv\\
\nonumber
&=& \frac{r_+}4(1-\epsilon)(2\delta_0)^{-2}(r_+^{-2}-r(U,v)^{-2})
	e^{-\frac{\delta_0-1+\epsilon}{\delta_0} }\int_0^v dv\\
\nonumber
&=&\frac{r_+}4(1-\epsilon)(2\delta_0)^{-2}(r_+^{-2}-r(U,v)^{-2})
	e^{-\frac{\delta_0-1+\epsilon}{\delta_0} } v\\
&\ge&\frac{1}{16}r_+^{-1} (1-\epsilon)^2 \delta_0^{-2}
	e^{-\frac{\delta_0-1+\epsilon}{\delta_0} } v.
\end{eqnarray}
Note that we have a lower bound $-\nu\ge e^{4c}$ on $u=U$, for some $c>0$, whence we have
a lower bound $\log\Omega^2 \ge c$ on this curve.
Integrating $(\ref{newevol2})$, in view of the bound
\[
4\pi \kappa(-\nu) g^{AB}T_{AB} \le 4\pi T_{uv},
\]
we obtain, for $(u,v)\in[U,U']\times[0,\infty)$ with $r(u,v)\le 2\delta_0r_+$,
\begin{eqnarray*}
\log \Omega^2(u,v) &\ge& c + \int_0^v\int_{U'}^u 2mr^{-3}\kappa(-\nu) dudv\\
&&\hbox{} - r_+^{-2}(1 -\epsilon)^{-2} 2\int_0^v \int_{U'}^U 4\pi r^2 T_{uv}\\
&\ge& c +\frac{1}{8}r_+^{-1} (1-\epsilon)^2 \delta_0^{-2}
	e^{-\frac{\delta_0-1+\epsilon}{\delta_0} } v\\
&&\hbox{}	-r_+^{-2}(1 -\epsilon)^{-2}  r_+(\delta_0 -1+\epsilon)v,
\end{eqnarray*}
where we have used $(\ref{evawhence})$, $(\ref{oxiwhence})$, and 
the inequality $r\ge r_+(1-\epsilon)$, which follows from $(\ref{havearranged})$
and $\mu\le 1$
in $[U,U']\times[0,v]$.
Choosing $\epsilon$ sufficiently small, we see that
for  $\delta_0$
satisfying 
\[
\frac18 \delta_0^{-2} e^{-1+\delta_0^{-1}} - (\delta_0-1) >0,
\]
we have for $r(u,v)\le 2\delta_0r_+$, 
\[
\log 4+\log (-\nu)(u,v) + \log \kappa(u,v) =\log\Omega^2 (u,v)\ge   c+ a_0v.
\]
We obtain
\[
\log (-\nu)(u,v) \ge c-\log 4 +a_0  v
\]
whence, in view also of $(\ref{k<=1})$,
 $(\ref{revisedassump})$ follows for an $\alpha>0$, when integrated along the $\mathcal{H}^+$.
\end{proof}

The statement $\mathcal{CH}^+=\emptyset$  of Theorem~\ref{afth}
in the case $(\ref{eav})$ follows immediately from
Section~\ref{lessthan}. 

Finally, since it follows now that $\mathcal{B}_s\ne\emptyset$, there exist by
the Raychaudhuri equation points in $\mathcal{A}=\{\lambda=0\}$,
i.e.~$\mathcal{A}\ne\emptyset$. The statement $(\ref{property})$ follows
from easy monotonicity arguments, in view of the fact that
$i^+\in\overline{\mathcal{B}_s}$.

\section{Strong cosmic censorship}
\label{SCCsec}
We prove in this section that the various inextendibility
statements quoted in the theorems of the introduction hold.

\subsection{The generic condition}
First we introduce the generic condition
\begin{assumption}
\label{genassump}
There exists a $W>0$ with the following property: 
Let $\mathcal{V}\subset (\pi_1\circ \pi)^{-1}(\mathcal{S})$ be open
such that $\mathcal{V}\cap
\{r^4\gamma_{AB}p^Ap^B<W\}\ne\emptyset$. Then $f$ does not vanish
identically in $\mathcal{V}\cap
\{r^4\gamma_{AB}p^Ap^B<W\}$.
\end{assumption}
We will invoke the above assumption when necessary in the statements of the propositions to 
follow.

\subsection{The extendibility theorems}
Suppose $(\mathcal{M},g)$ is
extendible as a manifold with $C^2$ Lorentzian metric, and let $\hat\gamma$ be a causal
geodesic leaving $\mathcal{M}$. One easily sees that
\begin{equation}
\label{withgeo}
\overline{\pi_1(\hat\gamma)}\cap\mathcal{B}^\pm\ne\emptyset.
\end{equation}
On the other hand, by the results of the previous sections, we
have a decomposition of $\mathcal{B}^{\pm}$. To derive a contradiction,
it suffices to show that the intersection of $\overline{\pi_1(\hat\gamma)}$
with the various possible components of $\mathcal{B}^\pm$ we have described
is necessarily empty,
at least for data satisfying the generic condition above (or weaker conditions).
We proceed to prove such statements in the sections that follow.

We may rephrase the above in a more convenient way (at the
expense of some additional notation), as follows: Let us denote by
\[
\mathcal{Z}^\pm\subset \mathcal{B}^\pm
\]
the subset of points $p$ such that there exists a causal geodesic $\hat\gamma$
exiting the spacetime as above so that 
\[
p\in\overline{\widetilde{\pi_1(\hat\gamma)}}\cap\mathcal{B}^\pm.
\]
In view of $(\ref{withgeo})$, 
to prove strong cosmic censorship it suffices to show that
for generic initial data, $\mathcal{Z}^\pm=\emptyset$. In the sections that follow,
we will restrict  $\mathcal{Z}^\pm$ further and further.

\subsubsection{The reduction to radial null geodesic inextendiblity or $r=0$}
Let 
\[
\mathcal{Z}^\pm_{radnull}\subset \mathcal{Z}^\pm
\]
denote the subset consisting of all $p\in\mathcal{Z}^\pm$ where the $\hat\gamma$ of the definition 
of $\mathcal{Z}^\pm$ can be taken to be a \emph{radial} null geodesic.

We will show in this section that
\begin{proposition}
\label{scc1}
Let $(\mathcal{M},g)$ be a maximal development as considered in the statement of
Theorem~\ref{introthe},~\ref{introthe15},~\ref{introthe2}, or~\ref{afth},
let $\mathcal{B}^\pm$ be as in Section~\ref{sssec}.
Then
\[
\mathcal{Z}^\pm\subset \overline{\mathcal{Z}^\pm_{radnull}} \cup \mathcal{B}^\pm_s
\]
where this union is not necessarily disjoint.
\end{proposition}

\begin{proof}
Without loss of generality, let us talk about the future.
Let ${\cal H}^+$
denote the future boundary of $\cal M$ in an extension ${\cal M}'$, and let
$q$ be an arbitrary point of $\mathcal{Z}^+$. 
Let $p\in \mathcal{H}^+$ be a point such that a $\hat\gamma$ corresponding
to $q$ in the definition of $\mathcal{Z}^+$ crosses $p$.

Our task is to show that $q\in \overline{\mathcal{Z}^+_{radnull}}\cup \mathcal{B}^+_s$.
We will show equivalently that if $q\not\in\overline{\mathcal{Z}^+_{radnull}}$, then
$q\in\mathcal{B}^+_s$.

Recall that $\mathcal{H}^+$ is differentiable on a dense subset~\cite{chrusciel}.
Let $p_i\to p$ be a sequence of such points where $\mathcal{H}^+$ is regular,
 and let $q_i\in\mathcal{B}^+$
be corresponding points. Since the complement of
$\overline{\mathcal{Z}^+_{radnull}}$ is open, we can arrange this sequence such that
\begin{equation}
\label{tocontradicthere}
q_i\not\in \overline{\mathcal{Z}^+_{radnull}}.
\end{equation}

Now for each $p_i$ we may associate planes $O_i$, $T_i$ as follows: Choosing a causal curve
$\gamma_i$ exiting the spacetime at $p_i$, and choosing a sequence of points $p_{ij}\to p_i$
along $\gamma_i$,
we may draw convergent subsequences from the sequence of orthogonal planes
$O_{ij}$, $T_{ij}$, where these denote the planes orthogonal and tangential, respectively,
to the symmetric surfaces at $p_{ij}$.

A priori, the planes $O_i$ are either null or timelike.  We claim that
they are in fact necessarily null, and their null generator $K_i$ is necessarily
tangential to $\mathcal{H}^+_{p_i}$. For otherwise, there would exist a null 
geodesic tangent to $O_i$ entering the spacetime. By a continuity argument and conservation
of angular momentum, this
is easily seen to be a null radial geodesic, and this contradicts $(\ref{tocontradicthere})$.

Now, one can extract a subsequence $O_i$ converging to a necessarily null plane $O$ at $p$. 
Let the corresponding $T_i$ converge to $T$. Since $T_i$ and $O_i$ are orthogonal,
$T_i$ is also null, and there exists a null vector $K \in O\cap T$.
 
Since $\mathcal{H}^+$ is achronal at $p$, there exist timelike geodesics entering $\mathcal{M}$
at $p$. Let $\gamma$ be such a geodesic. We have $g(K,\dot \gamma)\ne 0$. Let $K_j$ be
a sequence of vectors tangential to $\Sigma$ along $\gamma$ such that $K_j\to K$.
We have that 
\[
r^4\gamma_{AB}p^Ap^B \ge r^2 g(K_j, \dot\gamma) (g(K_j,K_j))^{-1/2}.
\]
By conservation of angular momentum, since $g(K_i,K_i)\to 0$ we must then have
$r(p_i)\to 0$, i.e.~$q\in\mathcal{B}^+_s$, as desired.
\end{proof}

\subsubsection{Inextendibility across $r=\infty$ and ``horizon'' points}
\begin{proposition}
Let $(\mathcal{M},g)$ be as in Proposition~\ref{scc1}. We have
\[
\mathcal{Z}^\pm_{radnull}\cap(\mathcal{B}^\pm_\infty \cup \mathcal{B}^\pm_h)=\emptyset,
\]
\[
\mathcal{Z}^+_{radnull}\cap(\mathcal{I}^+\cup i^+)=\emptyset.
\]
\end{proposition}
\begin{proof}
Null radial geodesics whose projections meet $\mathcal{B}^\pm_h$ or $i^+$ have already
 been shown
to have infinite affine length in view of the extension principle. 
For null radial geodesics whose projections meet $\mathcal{B}^\pm_\infty$
or $\mathcal{I}^+$,
the infiniteness of their affine length follows immediately
by the Raychaudhuri equation: For, without loss of generality, 
let $u=u_0$ be a null curve terminating
at $\mathcal{B}^+$. Reparametrize $v$ so that $\Omega^2=1$
along this curve. Raychaudhuri gives that $\partial^2_v r\le 0$. In particular,
$r$ cannot go to infinity in finite affine time. Thus, the curve has infinite
affine length.
\end{proof}

\subsubsection{The cases $r=0$}
\begin{theorem}
\label{scc15}
Let $(\mathcal{M},g)$ be as in Proposition~\ref{scc1}.
If $k=1$, or if $k=0$, $\Lambda<0$, then 
\begin{equation}
\label{stokevtro}
\mathcal{Z}^\pm\cap\mathcal{B}^\pm_s=\emptyset.
\end{equation}
If $k=0$, $\Lambda=0$ then if $f$ does not vanish identically,
$(\ref{stokevtro})$ holds. Finally, 
if $k< 0$, or if $k=0$, $\Lambda >0$, and  Assumption~\ref{genassump} is satisfied, then
again
$(\ref{stokevtro})$ holds.
\end{theorem}

\begin{proof}
If $k=1$ and 
$\overline{\widetilde{\pi_1(\hat\gamma)}}\cap\mathcal{B}^+_s\ne\emptyset$, 
then by Appendix~\ref{curvexp},
the Kretschmann scalar blows up along $\hat\gamma$ as $\mathcal{H}^+$ is approached.
But this contradicts the statement that $(\mathcal{M}',g')$ is $C^2$.
The relation $(\ref{stokevtro})$ then follows in this case.

In the case $k=0$, $\Lambda=0$, then unless $f=0$ identically,
the assumption of Proposition~\ref{otherprop} holds, and thus $m\ge \epsilon$ in $J^-(\mathcal{S})$.
One can apply again Appendix~\ref{curvexp} to obtain $(\ref{stokevtro})$.
In the case $k=0$, $\Lambda<0$, then Proposition~\ref{otherprop}  holds in view of the assumptions
of the only theorem applicable to this case, i.e.~Theorem~\ref{introthe15}, and
thus we again obtain $(\ref{stokevtro})$.

Suppose we are in the
 case $k<0$, and suppose 
\begin{equation}
\label{suppose...}
\overline{\pi_1(\hat\gamma)}\cap
\mathcal{B}^+_s\ne\emptyset.
\end{equation}
Let $\mathcal{H}^+_0$ denote the subset of $\mathcal{H}^+$ consisting of
all points corresponding to
 geodesics $\hat{\gamma}$ satisfying $(\ref{suppose...})$. In view of the non-emptyness
assumption, and what we have shown about the structure of the Penrose diagram, and the previous
propositions, we have in this
case that 
\begin{equation}
\label{suppose2...}
\emptyset\neq\mathcal{H}^+_0=\mathcal{H}^+.
\end{equation}
We will need the following

\begin{proposition}
\label{Riccivanish}
If $k=0, \Lambda>0$, or if $k<0$, then
assuming the non-emptyness $(\ref{suppose2...})$, it follows that
$\mathcal{H}^+$ is a $C^3$ null hypersurface on an open
dense subset, on which moreover
$Ric(\bar{K},\bar{K})=0$,
where $\bar K$ denotes any null vector in the direction of the null generator
of $T\mathcal{H}^+$.
\end{proposition}
\begin{proof}
In the case $k=0$, $\Lambda>0$, this follows from general results about
$T^2$-symmetric spacetimes proven in~\cite{dr3}. In the
case $k=-1$, this follows from Theorem~\ref{rigid}.
\end{proof}

On the other hand we have the following
\begin{proposition}
\label{positRicci}
Suppose Assumption~\ref{genassump}  is  satisfied, and $k\le0$. Then 
\begin{equation}
\label{statement}
{\rm Ric}(\bar{K},\bar{K})>0
\end{equation}
on a dense open subset of regular points of $\mathcal{H}^+$, where $\bar{K}$ denotes
a null generator.
\end{proposition}

\begin{proof}
In the case $k=0$, this follows from the results of~\cite{dr3} for general $T^2$-symmetric
spacetimes.

In what follows we assume $k<0$.
Let $\mathcal{W}\subset \mathcal{H}^+_1\cup\mathcal{H}^+_2$ be an arbitrary open set,
where $\mathcal{H}^+_1$, $\mathcal{H}^+_2$  are as in  Section~\ref{Rigidsection}, 
let $p\in\mathcal{W}$, 
let  $K$, $L$ be as in the null frame of Section~\ref{framesec}, and
consider the unit timelike geodesic $\gamma_c$ through $p$ 
with tangent vector $T_c= \frac12 c L + \frac12 c^{-1} K$. 
We will show that if $c$ is sufficiently small, this
geodesic will have angular momentum less than $W/4$.

Choose Killing fields $X$, $Y$ and $Z$ as in~\cite{dr2} in a neighborhood of $p$,
and extend these to Killing fields along each $\gamma_c$. Let $T_c$ denote
the tangent vector field of $\gamma_c$, and let $q_c\in\mathcal{S}$
denote the point where $\gamma_c$ intesects a Cauchy surface $\mathcal{S}$.
Now by the remark on Killing tensors in Section~\ref{vlasovsec},
\begin{equation}
\label{alavis;}
r^4\gamma_{AB}p^Ap^B(q_c)= g(X,T_c)^2+g(Y,T_c)^2-g(Z,T_c)^2.
\end{equation}
On the other hand, by the Killing equation
\[
T_c g(X,T_c)=0, T_cg(Y,T_c) =0, T_cg(Z,T_c)=0.
\]
For any choice of $X$, $Y$, $Z$ at $p$, we have 
that these vectors are in the orthogonal space of $K$, and thus
\[
g(X,T_c), g(Y,T_c), g(Z,T_c)\to 0
\]
as $c\to0$.
It follows from $(\ref{alavis;})$ and conservation of angular momentum
that we can ensure in particular
\[
r^4 \gamma_{AB}p^Ap^B \le  W/4
\]
holds along $\gamma_{c}$ for small enough $c$.

Let us choose such a geodesic $\gamma_{c}$ and let us denote it
in what follows by $\hat{\gamma}$.
For an arbitrary neighborhood $\mathcal{V}$ of $\hat{\gamma}'(p)$ in the mass shell
$P\cap \pi^{-1}(\mathcal{S})$,
we know by Assumption~\ref{genassump} that there exists an open subset $\tilde{\mathcal{V}}\subset \mathcal{V}$ such that
$f>0$ on $\tilde{\mathcal{V}}$. 
By the continuity properties of geodesic flow and the fact that $\hat\gamma$ intersects
$\mathcal{H}^+$ transversely, we may choose $\mathcal{V}$ so that the geodesics
with tangent vectors in $\mathcal{\tilde{V}}$ intersect $\mathcal{H}^+$ precisely in an open
set $\tilde{\mathcal{W}}\subset \mathcal{H}^+_1\cup\mathcal{H}^+_2$ with $\mathcal{W}\cap\tilde{\mathcal{W}}\ne
\emptyset$.

Let $\tilde{p}\in  \tilde{\mathcal{W}}$.
By continuity of geodesic flow and transversality, one sees that
$f$ extends continuously to $P\cap \pi^{-1}(\tilde{\mathcal{W}})$. Let $V$ be a null
vector field in a neighborhood of $\tilde{p}$ such that $V(\tilde{p})=\bar{K}$ is the null
generator of $\mathcal{H}^+$ at $\tilde{p}$.
Since $f$ is constant on geodesics, by construction
we have that $f>0$ at some point of $P\cap \pi^{-1}(\tilde{p})$. By continuity,
$f>0$ on an open set of $P\cap \pi^{-1}(\tilde{p})$. In particular, the
integral defined by $(\ref{EMdef})$ is strictly positive at $q$ when contracted
twice with $\tilde{V}$.

Take now a sequence of points $\tilde{p}_i\to \tilde{p}$, with $\tilde{p}_i\in
\mathcal{M}$. By the $C^2$ property of the extension, 
\begin{equation}
\label{thelimit}
{\rm Ric}(V,V)(\tilde{p}_i)\to
{\rm Ric}(V,V)(\tilde{p})={\rm Ric}(\bar{K},\bar{K}).
\end{equation}
On the other hand, by Fatou's lemma, the right hand side of $(\ref{EMdef})$
contracted twice with $\bar{K}=V(\tilde{p})$ is less than or equal to the limit of
its value at $\tilde{p}_i$ contracted twice with $V(\tilde{p}_i)$. The former we have
just shown to be strictly positive, while the latter equals ${\rm Ric}(\bar{K},\bar{K})$
in view of $(\ref{thelimit})$ and $(\ref{Eeq})$.
We have thus shown $(\ref{statement})$ for  all $\tilde{p}\in \tilde{\mathcal{W}}$.
Since $\tilde{\mathcal{W}}$ is open, $\tilde{\mathcal{W}}\cap \mathcal{W}\neq\emptyset$,
and $\mathcal{W}$ was an arbitrary open subset of $\mathcal{H}^+_1\cup\mathcal{H}^+_2$,
we have that $(\ref{statement})$ holds on a dense open subset of 
$\mathcal{H}^+_1\cup\mathcal{H}^+_2$, and thus, on a dense open subset of $\mathcal{H}^+$.
\end{proof}

The theorem now follows by contradiction. 
\end{proof}
\subsubsection{The case $\mathcal{B}^{\pm}=\mathcal{N}^{\pm}$, $\infty>r_{\pm}>0$}
\label{thelastcases}

\begin{theorem} 
\label{togevikot}
Let $(\mathcal{M},g)$ be as in the statement of Theorems~\ref{introthe}--\ref{introthe2}, and 
assume $\mathcal{B}^\pm=\mathcal{N}^\pm$,
with $\infty>r_\pm>0$. Then if $k\le 0$, or if $k=1$, $\Lambda\le 0$, then
either
$f= 0$ identically,
or the spacetime is null radial geodesically inextendible beyond $\mathcal{N}^\pm$, i.e.
\begin{equation}
\label{stokevtro2}
\mathcal{Z}_{radnull}\cap\mathcal{N}^\pm=\emptyset.
\end{equation}
If $k=1$, $\Lambda>0$ and $f$ does not vanish identically, then
$(\ref{stokevtro2})$ holds if there exists a point $p\in\mathcal{Q}$
such that 
\begin{equation}
\label{dekaviki}
r(p)\ge r_{\pm}.
\end{equation}
\end{theorem}

\begin{proof}
Without loss of generality, we can restrict to the case of $\mathcal{B}^-$.

Assume the spacetime to be extendible, and assume that there
exists moreover a null radial geodesic crossing into a nontrivial extension 
of $\mathcal{M}$, i.e.~assume $(\ref{stokevtro})$ does \emph{not} hold.

Choose a fundamental domain $\mathcal{F}'$ for $\mathcal{Q}$ bounded by two null curves, 
and consider
the union $\mathcal{F}=\mathcal{F}'\cup \tau(\mathcal{F}')\cap \tau^{-1}(\mathcal{F}')$
where $\tau$ generates the deck transformations.
The set $\mathcal{F}$ is a connected set itself bounded by two null curves: 
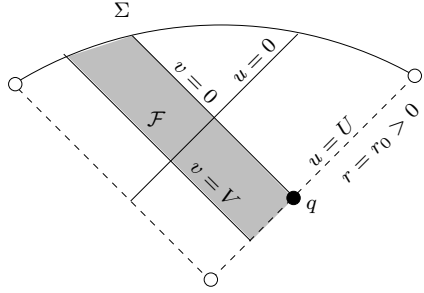
\[
\input{past4.pstex_t}
\]
Moreover, let these have been chosen so that the future boundary
of $\mathcal{F}$ corresponds to the lift of the projection
of a null radial geodesic passing into the extension,
i.e.~such that $q\in\mathcal{Z}_{radnull}$.

Choose new coordinates (not respecting periodicity) such that $\Omega^2=1$ on 
the ray through $q$
depicted, and so that this becomes $v=0$, and also $\Omega^2=1$ on 
some conjugate ray in the spacetime,
so that this becomes $u=0$. The other ray 
depicted (isometric to $v=0$) corresponds to $v=V>-\infty$, 
and the null boundary $u=U<-\infty$. The latter inequality is strict because
$v=0$ must be affine incomplete.
The original fundamental domain $\mathcal{F}'$ corresponds to $v_1\le v\le v_0$
for some $V< v_1<v_0< 0$.

Note that, in the case $k\le 0$ 
 we have, that if $f$ does not vanish identically,
  \begin{equation}
 \label{lpositivity}
 \lambda>0,
 \end{equation} 
 \begin{equation}
 \label{npositivity}
 \nu>0
 \end{equation}
 in these coordinates. For $\Lambda\ge0$, this 
 follows in view of $(\ref{flux})$ from Propostion~\ref{eitherf0}.
 For $\Lambda<0$, this follows by fiat, as the only applicable theorem is
 Theorem~\ref{introthe15}.

If $k=1$, $\Lambda \ge 0$, it is not necessarily the case that $(\ref{lpositivity})$,
$(\ref{npositivity})$ hold. Let us assume, however, $(\ref{dekaviki})$.
If $\nu (p)\le 0$, then $\nu \le 0$ on $\{v= v(p)\}\cap \{u\le u(p)\}$. Thus, $r\ge r_-$ 
along $v=v(p)$.  If $r< r_-$ along some other $v=v_c$ for all $u\le u_c$, then
by considering the intersection of a constant-$u$ ray  sufficiently close to $u=U$
with $v=v_c+V$,  with $v_{p}$, and with $v_c-V$, in view of periodicity,
one contradicts the statement
that on such a ray $\lambda$ can change sign at most once.
Thus, we have $r\ge r_-$ everywhere. From Raychaudhuri, this implies
that one can select the point $(0,0)$ sufficiently far in the past so that
$\nu \ge 0$, $\lambda\ge 0$ in $(U,0]\times [V,0]$.

Now, in the above argument, had we assumed that  $r(p)>r_-$,
we would have obtained $\nu>0$, $\lambda >0$. Thus, $(0,0)$ can
be chosen so that either
$(\ref{lpositivity})$ and $(\ref{npositivity})$ hold or $r=r_-$  identically in 
$(U,0]\times [V,0]$.

In the latter case,
from the equations $(\ref{const1})$ and $(\ref{const2})$, one obtains that
$T_{uu}=0=T_{vv}$ and thus $f=0$ identically
in $\pi^{-1}((U,0]\times [V,0])$. Now consider any unit timelike geodesic in $\mathcal{M}$
from $\mathcal{S}$. By global hyperbolicity, the projection of this geodesic to
$\mathcal{Q}$ must intersect the projection of
$(U,0]\times [V,0]$ to $\mathcal{Q}$. Thus, in view of the previous statements
and the Vlasov equation, we have that $f=0$ along  this geodesic.
It follows that $f$ vanishes identically.

In what follows, we assume that $f$ does not vanish identically.
We have thus reduced the Theorem to the case where $(\ref{lpositivity})$ and
$(\ref{npositivity})$ hold in $J^-(0,0)$, or the case where
$k=1$, $\Lambda=0$ and these inequalities
do not hold.

Let us note first that the particle flux is uniformly bounded
along any null ray in $\mathcal{F}$, and approaches the 
initial flux through $\mathcal{F}'\cap\Sigma$ as one computes it on constant-$u$
rays in $\mathcal{F}'$,  as $u\to U$.

To see this, first note that $T_{uv}$, $T_{vv}$, $T_{uu}$ are uniformly bounded in these
coordinates along $v=0$, since these can  be related the components of curvature
in a parallely propagated null frame on a null geodesic entering a $C^2$ extension.
It follows that $N^v$ is bounded pointwise, and thus
the flux through $v=0$ is bounded, i.e.
\begin{equation}
\label{yeniyeniyIldIz}
\int_{U}^u r^2\Omega^2N^v(\bar{u},0) d\bar{u} <\infty
\end{equation}
for any $u$ with $(u,0) \in J^-(\Sigma)$. 

By periodicity we have
\begin{equation}
\label{yeniyIldIz}
\int_{U}^u r^2\Omega^2N^v(\bar{u},V) d\bar{u} <\infty.
\end{equation}
But now one can bound uniformly the flux
through any constant-$u$ curve in $\mathcal{F}$ by conservation of particle current.
The last part of the claim of the previous to the previous paragraph 
follows by noting on the one hand that the flux through $\{v=v_0\}\cap \{u\ge u'\}$
equals that through $\{v=v_1\}\cap \{u\ge u''(u')\}$ for any $u'\ge U$, by periodicity,
and on the other that as $u'\to U$, we have $u'' \to U$, and, the flux
through $\{v=v_0\}\cap \{U<u\le u'\}$ and through $\{v=v_1\}\cap\{U<u\le u''\}$ both go to $0$,
in view of the uniform boundedness.

It follows that since $f$ does not vanish identity, the initial flux is
non-zero, and thus by the above
\begin{equation}
\label{flux}
\lim_{u\to U}\int_{v_1}^{v_0} r^2\Omega^{2}N^u(u,v)dv =\delta_0 > 0.
\end{equation}

Let us assume first the former case, i.e.~the case where $(\ref{lpositivity})$
and $(\ref{npositivity})$ hold in $J^-((0,0))$.

We derive an upper bound for $\Omega^2$ in $\mathcal{F}\cap \{u\le 0\}$
as follows:
Set first $R=r(0,0)=\sup_{J^{-}(0,0)}r$, and $M=\frac{R}2(k+1)$.
If $m\ge M$, we have 
\[
k-\mu\le -1.
\]
Note also by $(\ref{newconst})$, $(\ref{npositivity})$
that $0> \int_v^0 \kappa (u,\bar{v})\ge \int_v^0 \kappa (0,\bar{v})\doteq -K$.
We estimate
\begin{eqnarray}
\label{nubo}
\nonumber
\nu(u,0)	&\ge&	\nu(u,v)e^{\int_v^0 2r^{-2}m\kappa-r\kappa\Lambda d\bar{v}}\\
\nonumber
		&\ge&	\nu(u,v)e^{\int_v^0 2r^{-2}M\kappa-2r^{-2}\lambda -r\kappa\Lambda d\bar{v}}\\
\nonumber
		&\ge&	\nu(u,v)e^{\int_v^0 2r^{-2}M\kappa-2r^{-2}\lambda d\bar{v}-r\kappa
								\max\{-\Lambda,0\}}\\
\nonumber
		&\ge&	\nu(u,v)e^{\int_v^0 (2Mr^{-2}+ r \max\{-\Lambda,0\})(u,\bar{v})
					\kappa(0,\bar{v})-2r^{-2}\lambda (u,\bar{v})d\bar{v}}\\
\nonumber
		&\ge&	\nu(u,v)e^{(-2Mr_-^{-2}-R\max\{-\Lambda,0\})K-2(R^{-1}-r_-^{-1})}\\
		&\doteq&	c_1\nu(u,v).
\end{eqnarray}
Thus, integrating
twice $(\ref{newevol2})$, we
have
\begin{eqnarray*}
\log\Omega(u,v)&=&\int_u^0\int_v^0-4\pi T_{uv}-2r^{-3}\kappa\nu m +4\pi\kappa\nu g^{AB}T_{AB}\\
			&\le&\int_u^0\int_v^0-2r^{-3}\kappa\nu m\\
			&\le&\int_u^0\int_v^0-2r^{-3}\kappa\nu M+\int_u^0\int_v^0 2r^{-3}\lambda\nu\\
			&\le&\int_u^0\int_v^0-2Mr^{-3}\nu(u,v)\kappa(0,v)
						+\int_u^0\int_v^0 2r^{-3}\lambda\nu(u,v)\\
			&\le&M(r_-^{-2}-R^{-2})\int_v^0-\kappa(0,v)dv
						+\int_u^0\int_v^0 2r^{-3}\lambda\nu(u,v)\\
			&\le&M(r_-^{-2}-R^{-2})K+c_1^{-1}\int_u^0\int_v^0 2r^{-3}\lambda(u,v)\nu(u,0)\\
			&\le&M(r_-^{-2}-R^{-2})K+c_1^{-1}(R_-^{-2}-r_-^{-2})R_-
\end{eqnarray*}
so
\begin{equation}
\label{Omep}
\Omega^2 \le C_1.
\end{equation}

But now, we can obtain in addition a \emph{lower} bound
for $\Omega^{2}$: First note that, by the mass-shell relation $(\ref{MSR})$
and the angular momentum bound $(\ref{am0})$,
we have on the one hand an estimate
\[
\Omega^2 p^up^v = 1+r^2\gamma_{AB}p^Ap^B \le 1+Xr_-^{-2},
\]
on the support of $f$.
Thus, if either
$p^u\ge 1$, or $\Omega^2 p^v \ge 1$
then
\begin{equation}
\label{isxuei...}
\Omega^2 p^up^v \le 1+r^2\gamma_{AB}p^Ap^B \le (1+Xr_-^{-2})(p^u +\Omega^2 p^v).
\end{equation}
Otherwise,
if both $p^u\le 1$ and $\Omega^2p^v\le 1$, we have
\[
\Omega^2 p^up^v \le\frac12( (p^u)^2+\Omega^4(p^v)^2)\le \frac12(p^u+\Omega^2 p^v).
\]
The above bound is better than $(\ref{isxuei...})$. It follows that $(\ref{isxuei...})$
holds on the whole support of $f$.

Now we may compute
\begin{eqnarray}
\label{computALT}
T_{uv}	& = &	\int_0^\infty\int_{-\infty}^{\infty}
\int_{-\infty}^{\infty}r^2 p_up_v f\frac{dp^u}{p^u}\sqrt{\gamma} dp^A dp^B\\
	\nonumber
	& = &	(g_{uv})^2
			\int_0^\infty\int_{-\infty}^{\infty}
			\int_{-\infty}^{\infty}
				r^2 p^up^v 
			f\frac{dp^u}{p^u}\sqrt{\gamma} dp^A dp^B\\
	\nonumber
	& = &	-\frac12 g_{uv}\int_0^\infty\int_{-\infty}^{\infty}
			\int_{-\infty}^{\infty}
				r^2 \Omega^2 p^u p^v  f\frac{dp^u}{p^u}\sqrt{\gamma} dp^A dp^B\\
	\nonumber
	& \le &	\frac14(1+Xr_-^{-2})\Omega^2 \int_0^\infty\int_{-\infty}^{\infty}\int_{-\infty}^{\infty}
				r^2 (p^u +\Omega ^2 p^v)  f\frac{dp^u}{p^u}\sqrt{\gamma} dp^A dp^B\\
	\nonumber
	&\le&	\frac14\Omega^2 (1+Xr_0^{-2}) \int_0^\infty\int_{-\infty}^{\infty}\int_{-\infty}^{\infty}
			r^2 p^u f \frac{dp^u}{p^u}\sqrt{\gamma} dp^A dp^B\\
\nonumber
			&&\hbox{}+\frac14\Omega^4
			(1+Xr_0^{-2}) \int_0^\infty\int_{-\infty}^{\infty}\int_{-\infty}^{\infty}
			r^2 p^u f \frac{dp^u}{p^u}\sqrt{\gamma} dp^A dp^B \\
	\nonumber
	&=&	(1+Xr_0^{-2})\frac14\Omega^2 N^u+
	(1+Xr_0^{-2})\frac14\Omega^4 N^v.
\end{eqnarray}

Integrating twice
$(\ref{evol2})$, 
we have
\begin{eqnarray*}
\log\Omega(u,v)&=&\int_u^0\int_v^0-4\pi T_{uv}+\frac14 kr^{-2} \Omega^2+r^{-2}\lambda\nu 
						+4\pi\kappa\nu g^{AB}T_{AB}\\
			&\ge&\int_u^0\int_v^0-8\pi T_{uv}+\int_u^0\int_v^0r^{-2}(\max\{-k,0\})\kappa\nu.
\end{eqnarray*}
The second term on the right hand side has already been bounded below by a negative constant
in the context of the derivation of $(\ref{Omep})$. 
On the other hand, since, by uniform boundedness of the flux through $\mathcal{F}$,
there exists a constant $B$ such that, for
all $(u,v)\in (U,0)\times[V,0]$,
\[
\int_v^0r^2\Omega^2N^u (u,\bar{v})d\bar{v}\le B,
\]
\[
\int_u^0r^2\Omega^2N^v(\bar{u},v)d\bar{u}\le B,
\]
we estimate, using $(\ref{computALT})$ and $(\ref{Omep})$,
\begin{eqnarray*}
\int_u^0\int_v^0 8\pi T_{uv}	&\le&	\frac14(1+Xr_0^{-2})\int_u^0\int_v^0\Omega^2N^u+
										\Omega^4N^v\\
						&\le&	\frac14(1+Xr_0^{-2})r_-^{-2}\int_u^0
								\int_v^0 r^2\Omega^2 N^u d\bar{u}d\bar{v}\\
						&&\hbox{}
										+\frac14(1+Xr_0^{-2})r_-^{-2}C_1\int_u^0\int_v^0
												r^2N^vd\bar{u}d\bar{v}\\
						&\le&	\frac14(1+Xr_0^{-2})r_-^{-2} (|U|B +C_1|V|B).
\end{eqnarray*}
Thus, we have
\begin{equation}
\label{Omepallo}
\Omega^2 \ge c_2>0.
\end{equation}

Now since
\[
\lim_{u\to U} \int_{v_0}^0 \lambda(u,v)=0,
\]
and 
\[
\lim_{u\to U}\int_{V}^{v_1}\lambda(u,v)=0,
\]
there must exist for every $u>U$, a $(u,\tilde{v}_1(u))$, $(u, \tilde{v}_0(u)$),
where $V\le \tilde{v}_1(u)\le v_1$, $v_0\le \tilde{v}_0(u)\le 0$, such that
\begin{equation}
\label{veostropos}
\lim_{u\to U} \lambda(u,\tilde{v}_1(u))=
\lim_{u\to U} \lambda(u,\tilde{v}_0(u))=0.
\end{equation}

Integrating now equation $(\ref{const1})$,
we have
\[
\int_{\tilde{v}_1(u)}^{\tilde{v}_0(u)} -4\pi rT_{vv}\Omega^{-2} (u,\bar{v})d\bar{v}
=\Omega^{-2}\lambda(u,\tilde{v}_0(u))-\Omega^{-2}\lambda(u,\tilde{v}_1(u)).
\]
In view of the bounds $(\ref{veostropos})$ and $(\ref{Omepallo})$,
it follows that the right hand side of the above tends to $0$.
We thus have
\begin{eqnarray}
\label{claim}
\nonumber
\lim_{u\to U}\int_{v_1}^{v_0} r^2\Omega^{-2}T_{vv}(u,v)dv&\le&
\lim_{u\to U}\int_{\tilde{v}_1}^{\tilde{v}_0} r^2\Omega^{-2}T_{vv}(u,v)dv\\
&\le& R\lim_{u\to U}\int_{\tilde{v}_1}^{\tilde{v}_0} r\Omega^{-2}T_{vv}(u,v)dv
\to 0.
\end{eqnarray}

Note the trivial fact that
if $p^u\ge \delta $, we have
$p^u\le \delta^{-1} (p^u)^2$.
In view of this and the angular momentum bound $(\ref{am0})$, we compute
\begin{eqnarray}
\label{ieqhelp}
\nonumber
N^u	& = &	 \int_0^\infty\int_{-\infty}^{\infty}\int_{-\infty}^{\infty}
			r^2 p^u f \frac{dp^u}{p^u}\sqrt{\gamma} dp^A dp^B\\
				\nonumber
	& \le&	
			\delta^{-1}\int_0^\infty\int_{-\infty}^{\infty}
			\int_{-\infty}^{\infty}
				r^2 (p^u)^2
			f\frac{dp^u}{p^u}\sqrt{\gamma} dp^A dp^B\\
			\nonumber
			&&\hbox{}+\int_0^\delta\int_{-\infty}^{\infty}\int_{-\infty}^{\infty}
			r^2 p^u f \frac{dp^u}{p^u}\sqrt{\gamma} dp^A dp^B \\
	\nonumber
	& = &	\delta^{-1}(g^{uv})^2\int_0^\infty\int_{-\infty}^{\infty}
			\int_{-\infty}^{\infty}
				r^2 (p_v)^2  f\frac{dp^u}{p^u}\sqrt{\gamma} dp^A dp^B\\
			\nonumber
			&&\hbox{}+\int_0^\delta\int_{-\infty}^{\infty}\int_{-\infty}^{\infty}
			r^2 p^u f \frac{dp^u}{p^u}\sqrt{\gamma} dp^A dp^B \\
	& \le &	\frac14\Omega^{-4} \delta^{-1} T_{vv}+ F Xr_-^{2}\delta.
\end{eqnarray}

Thus,  from $(\ref{claim})$ and $(\ref{Omep})$,
\begin{eqnarray*}
\lim_{u\to U}\int_{v_1}^{v_0} r^2\Omega^{2}N^u(u,v)dv		&\le& \lim_{u\to U}
												\int_{v_1}^{v_0}\left(\frac14\delta^{-1}
													r^2\Omega^{-2}T_{vv}
												+\delta FXr_-^2r^2\Omega^2
												\right) (u,\bar{v})d\bar{v}\\
										&\le&\frac14\delta^{-1}\lim_{u\to U}
												\int_{v_1}^{v_0}r^2\Omega^{-2}T_{vv}
												(u,\bar{v})d\bar{v}
											+ \delta FXr_-^2 C_1R^2|V| \\
										&=& \delta FXr_-^2 C_1R^2|V|.
\end{eqnarray*}
Choosing $\delta<\delta_0 F^{-1}X^{-1}r_-^{-2}C_1^{-1}R^{-2}|V|^{-1}$,
we contradict $(\ref{flux})$. 

We now turn to the case of $k=0$, $\Lambda=0$. In view of
the previous, 
we may assume $\lambda<0$, $\nu< 0$ in $[U,0]\times(V,0]$.
By the bound on spacetime volume of Section~\ref{k1L0}, we have
\[
\int _U^0\int_V^0 \Omega^2 du dv \le C.
\]
By the pigeonhole principle, we may have chosen $v=0$ such that
\[
\int_U^0 \Omega^2 du <\infty,
\]
and thus, $U>-\infty$.

We again obtain an upper bound on $\Omega^2$
by
integrating equation $(\ref{evol1})$:
\begin{eqnarray*}
\log \Omega (u,v)&=&\int_v^0\int_u^0 -4\pi T_{uv} -\frac1r\lambda\nu -\frac{\Omega^2}{4r}\\
			&\le&\int_v^0\int_u^0-\frac{\Omega^2}{4r}\\
			&\le&\frac14 r^{-1}(0,0) \int_v^0\int_u^0\Omega^2\\
			&\le&\frac14 r^{-1}(0,0)C.
\end{eqnarray*}
We may estimate
now $\int_u^0\int_v^0 T_{uv}$ as before, and thus, we obtain
again $(\ref{Omepallo})$. 
We continue as before.
\end{proof}

In various special cases, we can in fact prove more.
First we show the following:
\begin{proposition}
\label{stoparel9ov}
Suppose there exists a Cauchy surface
with $\lambda>0$, $\nu>0$, $\mathcal{B}^-=\mathcal{H}^-$, 
and $\int -\frac{\lambda}{1-\nu} dv <\infty$, $\int -\frac{\nu}{1-\mu}du <\infty$
on a  radial null constant-$u$ and constant-$v$ geodesic, respectively, meeting $\mathcal{B}^-$.
 Then $f=0$ or  $r_-=0$.
\end{proposition}
\begin{proof}
Suppose $r_->0$. We will show that $f=0$.

Let $\mathcal{S}'$ denote the lift of the Cauchy surface of the statement
of the Proposition. We have $\lambda>0$, $\nu>0$ in $J^-(\mathcal{S}')$.

Choose two conjugate null rays emanating from a point $p$, and choose
coordinates on $J^-(p)$ as follows: Let the two rays be $v=0$, $u=0$,
respectively, and let $\Omega^2=1$ on these rays.

Note that since $\lambda>0$, $\nu>0$, it follows that the $u$-range
and $v$-range of these coordinates in the past is finite. 
Let $\mathcal{B}^-$ correspond to $u=U>-\infty$, $v=V>-\infty$.
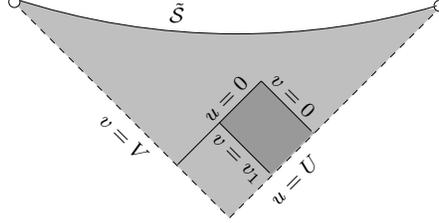
\[
\input{exclude2.pstex_t}
\]

Denoting $R=\sup_{\tilde{S}}r$, we have in particular that $r\le R$ in $J^-(p)$.
By $(\ref{newconst})$ and the periodicity, if follows that $\int -\frac{\lambda}{1-\mu} dv<\infty$
on \emph{any} constant-$u$ ray meeting $\mathcal{B}^-$.
(Similarly, it follows by our assumption that $\int -\frac{\nu}{1-\mu}du<\infty$
for any constant-$v$ curve meeting $\mathcal{B}^-$.)
 In particular, for $u=0$.
Thus,
for any $v_1<0$, we have
\[
\int_{v_1}^{0} -\frac{\lambda}{1-\mu}(u_2,v) dv \le
\int_{V}^0 -\kappa (u_2, v)dv \doteq K<\infty
\]
and consequently, from $(\ref{newconst})$, it follows that
\[
\int_{U}^{0}\int_{v_1}^{0} -4\pi r\kappa\nu^{-1}T_{uu}dudv\le K.
\]
In particular, there exists a $v_0$ such that
\[
\int_{U}^{0} -4\pi r\kappa \nu^{-1}T_{uu} (u,v_0)du  \le |v_1|^{-1}K.
\]

As in $(\ref{ieqhelp})$, we bound
\[
N^v
 \le 	\frac14\Omega^{-4} \delta^{-1} T_{uu}+ F Xr_-^{-2} \delta.
\]

Thus, setting $\delta=  -\frac14 \Omega^{-2} \kappa^{-1}\nu= \Omega^{-4} \nu^2 $, we have
\[
\frac14 \delta^{-1} \Omega^{-2} = -\kappa \nu^{-1},
\]
and thus, for $u>U$, 
\begin{eqnarray*}
\int_{u}^{0} r^2\Omega^{2}N^v(\bar{u},v_0)d\bar{u}		&\le& 
												\int_{u}^{0}\left(\frac14\delta^{-1}
													r^2\Omega^{-2}T_{uu}
												+\delta FXr_-^{-2}r^2\Omega^2
												\right) (\bar{u},v_0)d\bar{u}\\
										&\le&
												\int_{u}^{0}r^2(-\kappa\nu^{-1})												T_{uu}
												(\bar{u},v_0)d\bar{u}\\
										&&\hbox{}
											+  FXr_-^{-2} R^2\int_{u}^{0}
												\Omega^{-4}\nu^2(\bar{u},v_0)
														d\bar{u} \\
										&\le& (4\pi)^{-1} R |v_1|^{-1}K+
										 FXr_-^{-2} R^2\int_{u}^{0}
												\Omega^{-4}\nu^2(\bar{u},v_0)
														d\bar{u}.
\end{eqnarray*}
If we can uniformly bound
\begin{equation}
\label{cub}
\int_{u}^{0}\Omega^{-4}\nu^2(\bar{u},v_0)d\bar{u}
\end{equation}
then we will have that the flux of particles through $v=v_0$ is finite. One can then
repeat the argument of Theorem~\ref{togevikot} to show that $f=0$ identically.

Thus, the proposition is reduced to showing $(\ref{cub})$ is uniformly bounded
as $u\to U$.

We derive an upper bound for $\Omega^2$ in $J^-(p)$
as follows. Set  $M=\frac{R}2(k+1)$.
If $m\ge M$, we have 
\[
k-\mu\le -1.
\]
Recall by $(\ref{newconst})$
that $0> \int_v^0 \kappa (u,\bar{v})\ge \int_v^0 \kappa (0,\bar{v})\ge  -K$.
We estimate
\[
\nu(u,0)	\ge	c_1\nu(u,v).
\]
as in $(\ref{nubo})$ of the proof of Theorem~\ref{togevikot}.
Thus, integrating
twice $(\ref{newevol2})$, in view of $\Omega^2(0,\tilde{v})=1$, $\Omega^2(0,\tilde{u})$, we
have for $u\le0$, $v\le0$,
we obtain again $(\ref{Omep})$.

Now, integrating twice $(\ref{evol1})$, we obtain
\[
\int_u^0 \int_v^0 \partial_u\partial_v r d\bar{u} d\bar{v}=
			\int_u^0 \int_v^0  \frac{-k\Omega^2}{4r} -\frac1r\partial_u r\partial_vr +4\pi r T_{uv}
					+\frac14 r\Omega^2\Lambda d\bar{u} d\bar{v}
\]
and thus
\begin{eqnarray*}
\int_u^0 \int_v^0 4\pi r T_{uv} &=& \int_u^0 \int_v^0 \partial_u\partial_v r+
						 \frac{k\Omega^2}{4r} +\frac1r\partial_u r\partial_vr
							-\frac14 r\Omega^2 \Lambda d\bar{u}d\bar{v}\\
				&\le& \int_u^0 \int_v^0 \partial_u\partial_v r+
						 \frac{k\Omega^2}{4r} -\frac14 r\Omega^2 \Lambda d\bar{u}d\bar{v}\\
				&&\hbox{}+c_1^{-1} \int_u^0\int_v^0 \nu(u,0)r^{-1} \lambda(u,v) d\bar{u}d\bar{v} \\
			 	&\le& 2R +\max\{k,0\} UVC_1(4r_-)^{-1}\\
				&&\hbox{}
						+ 4^{-1}RUVC_1 \max\{-\Lambda,0\}
						+c_1^{-1} R\log (R/r_-).
\end{eqnarray*}
Integrating again equation $(\ref{evol2})$, we obtain
a lower bound
\[
\tilde{c}_1\le \Omega^2,
\]					
thus, we have 
\begin{equation}
\label{Omeuplo}
\tilde{c}_1\le \Omega^2\le C_1.
\end{equation}		

In particular, to show $(\ref{cub})$ is uniformly bounded, it suffices
to uniformly bound
\begin{equation}
\label{cub2}
\int_{u}^{0}\nu^2(\bar{u},v_0)d\bar{u}.
\end{equation}

Multiplying $(\ref{newevol1})$ by $\nu$, we obtain
\[
\partial_v \nu^2 = r^{-2}m\kappa\nu^2 + 8\pi rT_{uv}\nu  -2r\kappa\Lambda \nu^2,
\]
we have that for $v\le v_0$.
\begin{eqnarray*}
\nu^2 (u,v_0)&\le& \left(\nu^2( u, v) +\int _{v}^{v_0} 4\pi r T_{uv}\nu d\bar{v}\right)e^{ 
\int_v^{v_0}r^{-2}m\kappa 
+r\kappa\max\{-\Lambda,0\}(u,\bar{v})d\bar{v}}\\
		&\le&C\nu^2(u,v) +4\pi CR \int_v^{v_0} T_{uv} \nu d\bar{v}.
\end{eqnarray*}
On the other hand, as $v\to V$, for any $u_1, u_2$ we have $\int_{u_1}^{u_2} 
\nu(u,v)\to0$, while we have in addition the monotonicity
$\partial_u(\Omega^{-2} \partial_ur )\le 0$. It follows that
\[
\lim_{v \to V} \nu(u, v)= 0.
\]
Consequently, to show the uniform boundedness of $(\ref{cub2})$, it suffices
to show the uniform boundedness of
\begin{equation}
\label{paliitsuf}
\int_u^{0}\int_v^{v_0} T_{uv}\nu,
\end{equation}
for all $u>U$, $v>V$.

Now, for any $\delta_1>0$, $\delta_2>0$, we claim that
\begin{equation}
\label{weclaim}
T_{uv} \le    C\left( \delta_1 T_{uu} +\delta_2  T_{vv}
				+ \int _{\delta_2^{1/2}}^{\delta_1^{-1/2}} p_v^{-1} dp^v \right),
\end{equation}
where $C$ depends on $F$, $X$, and $r_-$. 
To see this:
In view of the mass-shell relation, and our bounds $(\ref{Omeuplo})$ on $\Omega^2$,
the multiple of $f$ in the integral defining the left hand side is on the order of 
$1$.\footnote{Here we consider $(p^v)^{-1}dp^v$ as the volume form.}
We have that $1  \le  \delta_1 (p^v)^2$
if $p^v \ge \sqrt{\delta_1^{-1}}$
while similarly, $1\le \delta_2 (p^u)^2$ for
$p^u \ge \sqrt{\delta_2^{-1}}$ and thus
 $p^v\sim (p^u)^{-1} \le \sqrt{ \delta_2}$. Inequality $(\ref{weclaim})$ now follows
 easily from these considerations.

Now applying the above with 
$\delta_1=-\kappa\nu^{-2}$, $\delta_2= -(k-\mu)^{-1}\lambda^{-1}$,
in view also of $(\ref{Omeuplo})$,
we obtain
\begin{eqnarray*}
T_{uv}\nu 	&\le& C\left( \kappa\nu^{-1} T_{uu}- \frac{\nu}{k-\mu} \lambda^{-1}T_{vv} +
\nu \int_{\nu^{-1/2}\lambda^{-1}}^{\nu^{3/2}} (p^v)^{-1} dp^v\right)\\
&\le& C\left( \kappa\nu^{-1} T_{uu}- \frac{\nu}{k-\mu} \lambda^{-1}T_{vv} +
\max\{0, \nu\log \nu^2\lambda\} \right)\\
&\le& C\left( \kappa\nu^{-1} T_{uu}- \frac{\nu}{k-\mu} \lambda^{-1}T_{vv} +
(2\max\{0, \nu\log \nu\}+\max\{0,\nu \log \lambda \})\right).
\end{eqnarray*}

We have already noted a uniform bound
on 
\[
\int_u^0\int_v^{v_0}  \kappa\nu^{-1} T_{uu}d\bar{u}d\bar{v} .
\]
The fact that 
\[
\int_u^0\int_v^{v_0 }- \frac{\nu}{k-\mu} \lambda^{-1}T_{vv} d\bar{u}d\bar{v}
\]
is uniformly bounded follows by interchanging $u$ and $v$ in that argument.
Finally, 
\begin{eqnarray*}
\int_u^0\int_v^{v_0 }\max\{0,\nu \log \lambda \} d\bar{u} d\bar{v}
&\le& \int_u^0\int_v^{v_0 } \nu \lambda d\bar{u} d\bar{v}\\
&\le&c_1^{-1} \int_u^0\int_v^{v_0} \lambda(\bar{u},\bar{v})\nu(\bar{u},0) d\bar{u}d\bar{v}\\
&\le&c_1^{-1}R^2.
\end{eqnarray*}

Thus, to bound $(\ref{paliitsuf})$ uniformly, it sufficies
to bound
\begin{equation}
\label{newsuf}
\int_u^0\int_v^{v_0} \max\{0,\nu\log\nu\}(\bar{u},\bar{v})d\bar{u}d\bar{v}.
\end{equation}
We will bound in fact 
\begin{equation}
\label{simdi}
\int_u^0 \max \{0,\nu\log\nu\}(\bar{u},\bar{v})
\end{equation}
uniformly in $u$ and $\bar{v}$.

We note that 
\begin{eqnarray*}
\partial_v (\nu \log \nu )	&=&   2r^{-2}m \kappa \nu + 4\pi rT_{uv} - r\kappa \nu \Lambda\\
					&&\hbox{}		+\nu \log \nu 2r^{-2}m\kappa + 4\pi r\log \nu T_{uv} -
						r\kappa\nu \log \nu \Lambda,
\end{eqnarray*}
and thus
\begin{eqnarray*}
\nu\log\nu (\bar{u},\bar{v}) &\le& \left(\max\{0,\nu\log\nu (\bar{u}, v)\} +
						\int_v^{\bar{v}}\big( 2r^{-2} \max\{-m,0\}\kappa\nu +4\pi rT_{uv}\right.\\
						&&\hbox{}\left.-r\kappa\nu\max\{
							\Lambda,0\}
						+4\pi r\max\{0,\log\nu \}T_{uv}\big) d\tilde{v}\right)
							e^ {\int_{v}^{\bar{v}}2r^{-2}m\kappa  +r\kappa\Lambda}\\
					&\le& C \left(\max\{0, \nu\log\nu (\bar{u}, v)\} +
						\int 2r^{-2} \max\{-m,0\}
						\kappa\nu \right.\\
					&&\hbox{}\left.+4\pi rT_{uv}-r\kappa\nu\max\{
							\Lambda,0\}
						+4\pi r\max\{0,\log\nu\} T_{uv}\right).
\end{eqnarray*}	
Since as we remarked earlier, in any region $u_1\le u\le u_2$, we have
$\nu(u,v)\to 0$ as $v\to V$, and we have bounded
\[
\int_u^0\int_v^0 2r^{-2} \max\{-m,0\}\kappa\nu+ 4\pi r T_{uv}-r\kappa\nu\max\{
							\Lambda,0\}d\bar{u}d\bar{v},
\]
we see that to bound $(\ref{simdi})$, it suffices to bound
\[
\int_v^0 \int_u^0\max\{0, \log\nu\} T_{uv}.
\]

For this, we return to $(\ref{weclaim})$, and now choose 
$\delta_1 = -\kappa \nu^{-1} (\log \nu)^{-1}$,
$\delta_2 = -(k-\mu)^{-1}\lambda^{-1}\nu  (\log\nu)^{-1}$.
We obtain
\begin{eqnarray*}
\max\{0,T_{uv}\log\nu\} 	&\le& C\left( \kappa\nu^{-1} T_{uu}- \frac{\nu}{k-\mu} \lambda^{-1}T_{vv} +
\max\{0,\log\nu\} \right.\\
&&\hbox{}\left.+
\int_{\lambda^{-1}\nu^{-1}(\log\nu)^{-1/2}}^{\nu\sqrt{\log\nu }} (p^v)^{-1} dp^v\right)\\
&\le& C\left( \kappa\nu^{-1} T_{uu}- \frac{\nu}{k-\mu} \lambda^{-1}T_{vv}\right.\\
&&\hbox{}\left. +
\max\{0, \log \nu \max\{0, \log (\nu^2\lambda \max\{0, \log\nu\} )\}\} \right).
\end{eqnarray*}

So it suffices to bound
\begin{equation}
\label{telikosuf}
\int\int\max\{0, \log \nu \max\{0, \log (\nu^2\lambda \log\nu)\}\} d\bar{u}d\bar{v}.
\end{equation}
But
\begin{align*}
\max\{0, &\log \nu \max\{0, \log (\nu^2\lambda \max\{0, \log\nu\})\}\}\\
\le& \max\{0, (\log\nu)^2\}+
	(\max\{0,\log\nu)(\max\{0,\log\lambda)\\
&\hbox{}+ \max\{0,\log \nu \} \max\{0,\log\max\{0,\log \nu\}\}\\
\le& \nu+\nu\lambda+\nu.
\end{align*}
Since $\int\int \nu\le VR$, and we have already bounded the spacetime
integral of the middle term, we have shown
$(\ref{telikosuf})$. 
The proposition is thus proven.
\end{proof}

\begin{proposition}
\label{stomellov}
Suppose $\Lambda\ge 0$, 
$\lambda>0$, $\nu>0$, and $\mathcal{B}^+=\mathcal{H}^+$, with $r_+<\infty$.
Then if $f$ does not identically vanish, then
there cannot exist both a future incomplete radial null geodesic of constant $v$ and $u$.
\end{proposition}
\begin{proof}
Suppose the spacetime contains a null radial future incomplete
geodesic of constant $v$ and of constant $u$.

Let $u=u'$ be a null radial incomplete geodesic. We have
$\int\Omega^2(u',v)dv <\infty$. 
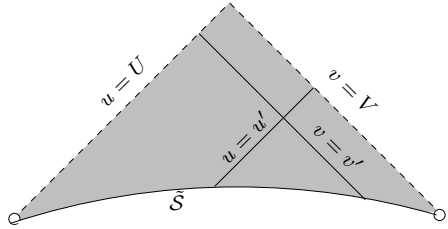
\[
\input{exclude3.pstex_t}
\]
Choose a coordinate $v$ such that $\Omega^2=1$ on $u=u'$.
The $v$-range of this coordinate is thus finite. Let $v=V<\infty$
correspond to $\mathcal{B}^+$. 

From $(\ref{const1})$, we obtain that for any $v_0<V$, 
\[
\int_{v_0}^V -4\pi rT_{vv}(u',v)<\infty.
\]

In view
of $\Omega^2=1$ we have 
\[
N^u \le C(T_{vv} + 1)
\]
along $u=u'$, where $C$ depends on $F$, $X$, and a lower bound on $r$
along $\tilde{\mathcal{S}}$.

In view of the upper and lower bounds on $r$, and the finiteness of the
range of $v$, we obtain
that there exists a $B<\infty$ such that for all $v_0$ with $(u',v_0)\in J^+(
\tilde{\mathcal{S}})$, 
\[
\int_{v_0}^V r^2\Omega^2N^u(u',v)dv\le B.
\]

Consider now a null radial incomplete geodesic of constant $v$, 
say $v=v'$.
Repeating the same argument as before, we obtain
\[
\int_{u_0} ^U r^2\Omega^2N^u(\bar{u},v')d\bar{u}\le \tilde{B},
\]
where $u=U$ corresponds to the constant $v$-component of $\mathcal{B}^+$.
Now let $u'_i$ denote a series of translates of $u'$ by the deck transformations
of $\mathcal{Q}$, with $u'_i\to U$. We have
by periodicity that for any $v$ with $(u_i,v)\in J^+(\tilde{\mathcal{S}})$,
\[
\int_v^ {v'}r^2\Omega^2N^u(u'_i,\bar{v})d\bar{v}\le \tilde{B}.
\]
Thus, by conservation of matter, it follows that
the flux of particles through $\tilde{S}\cap \{v\le  v'\}$ is bounded by $B+\tilde{B}$.
By periodicity, however, this flux must either vanish or be infinite.
This is a contradiction, unless the flux vanishes, in which 
case $f=0$ identically. Thus, $f=0$ identically.
\end{proof}

The previous two propositions in particular give the following
\begin{corollary}
In the case $k=1$, $\Lambda=0$, we have that if $\mathcal{B}^\pm=\mathcal{N}^\pm$,
and $f$ does not identically vanish, 
then $r_{\pm}=0$.
\end{corollary}
\begin{proof}
If say $\mathcal{B}^\pm=\mathcal{N}^\pm$, then, after changing the time orientation,
in view of the assumption that $f$ does not identically vanish, then we have
either $\mathcal{B}^+=\mathcal{H}^+$ and $\lambda>0$, $\nu>0$ everywhere,
or $\mathcal{B}^-=\mathcal{H}^-$ and $\lambda>0$, $\nu>0$ on a
certain Cauchy surface.

Consider the former case. By our volume estimate
\[
\int\int \Omega^2dudv<\infty
\]
in any fundamental domain,
it follows that there exists a $u=u'$ such that
\[
\int\Omega^2 (u',v)dv <\infty
\]
and a $v=v'$ such that
\[
\int\Omega^2(u,v') du<\infty.
\]
Apply Proposition~\ref{stomellov} to yield a contradiction unless $f=0$ identically.
It follows that this case cannot occur.

We must then be in the latter case. Let $\mathcal{S}'$ denote a Cauchy surface,
and let $r_- \le r_0\le \inf_{\mathcal{S}'}r$. $\{r\ge r_0\}\cap J^-(\mathcal{S}')$ is compact,
whereas for $r\le r_0$ we have the result of 
Proposition~\ref{otherprop}. Thus we have the uniform bound
$1-\mu \le-\epsilon<0$ in $J^-(\mathcal{S}')$. Consequently, 
$\int-\frac{\lambda}{1-\mu}dv \le \epsilon^{-1}\int\lambda\le \epsilon^{-1}R$,
and similarly for $\int-\frac{\nu}{1-\mu}du$, and thus
the conditions
of Proposition~\ref{stoparel9ov} are satisfied. If $f$ does not vanish
identically, it follows that $r_-=0$.
\end{proof}

The proof
of the following two corollaries follows immediately from Proposition~\ref{otherprop},
as in the above proof:
\begin{corollary}
In the case $k=0$, $\Lambda=0$, then if $f$ does not vanish identically,
and $\mathcal{B}^-=\mathcal{N}^-$, then $r_-=0$.
\end{corollary}

\begin{corollary} 
Let $k$, $\Lambda$ be arbitrary, 
and suppose the initial data are antitrapped.
If
\[
m>  \max\{k,0\}\inf_{\mathcal{S}}r/2 +\max\{-\Lambda,0\}
\sup_{\mathcal{S}}r^3,
\]
on the initial hypersurface $\mathcal{S}$, then, if $f$ does not vanish identically
and $\mathcal{B}^-=\mathcal{N}^-$, it follows that $r_-=0$.
\end{corollary}

Finally, 
as consolation for excluding the case where $(\ref{dekaviki})$ does not hold, we offer
the following: 
\begin{proposition}
Let $(\mathcal{M},g)$ be such that $k=1$, $\Lambda>0$,
$f$ does not vanish identically, and $\mathcal{B}^+=\mathcal{N}^+$ with $r\le r_+$.
Then $(\mathcal{M},g)$ is past inextendible with $r_-=0$. 
\end{proposition}
\begin{proof}
For this, it suffices to remark that $\lambda>0$, $\nu>0$ must hold in a neighborhood
of $\mathcal{B}^+$, and thus, by Raychaudhuri, on $\mathcal{Q}$. But this means
that Proposition~\ref{pastevprop}  applies, and thus, as is shown immediately
following in Section~\ref{PIAT}, Theorem~\ref{introthe15}.
\end{proof}

\subsubsection{Putting it all together}
\label{PIAT}
The preceeding propositions show that under the genericity assumptions of the theorems
of the introduction, 
\[
\mathcal{Z}^\pm_{radnull} \subset 
\bigcup_{x\in\mathcal{B}_h} \mathcal{N}^1_x\cup\mathcal{N}^2_x,
\]
or
\[
\mathcal{Z}^+_{radnull}\subset (\mathcal{B}_0\setminus\overline{\Gamma}) \cup \mathcal{CH}^+,
\]
as applicable,
while
\[
\mathcal{Z}^\pm\cap \mathcal{B}_s=\emptyset.
\]
Thus, if $ \mathcal{N}^1_x\cup\mathcal{N}^2_x=\emptyset$ for all $x\in
\mathcal{B}_h$ or $ (\mathcal{B}_0\setminus\overline{\Gamma})
\cup \mathcal{CH}^+=\emptyset$ as the case may be,
then by Proposition~\ref{scc1} we have that
\[
\mathcal{Z}^\pm=\emptyset.
\]
Otherwise, by the same proposition,
\[
\mathcal{Z}^\pm\subset
\cup_{x\in\mathcal{B}_h}(\overline{\mathcal{N}^1_x}\cup\overline{\mathcal{N}^2_x}),
\]
or
\[
\mathcal{Z}^+\subset \overline{\mathcal{B}_0\setminus\overline{\Gamma}} \cup 
\overline{\mathcal{CH}^+}.
\]
These relations yield the full statements of the theorems of the introduction. 
$\Box$

\section{Local rigidity of $\mathcal{H}^+$ in the case $k<0$}
\label{Rigidsection}
In this section we shall prove a general local rigidity theorem for
Cauchy horizons in hyperbolic symmetric spacetimes 
\begin{theorem}
\label{rigid}
Let $(\mathcal{M},g)$ be a globally hyperbolic spacetime with Cauchy surface
$\mathcal{C}$, where the symmetric surfaces $\Sigma$ are hyperbolic spaces, or compact quotients thereof. Assume
$(\mathcal{M},g)$ is future extendible, let $(\tilde{\mathcal{M}},\tilde{g})$ denote
a $C^2$ extension, and let $\mathcal{H}^+$ 
denote an open subset of the Cauchy horizon of $\mathcal{C}$ in $\tilde{\mathcal{M}}$
where $r$ extends continuously to $0$. 
Then there exists a dense subset $\mathcal{S}\subset\mathcal{H}^+$, 
at which $T_p\mathcal{H}^+$ is a hyperplane whose orthogonal complement is spanned by
a Killing vector field $K$, depending on $p\in\mathcal{S}$, such that $K(p)$ is null,
\[
{\rm Ric}(K,K)(p)\le 0.
\]
Moreover, if $\nabla_KK\ne0$, then 
\[
{\rm Ric}(K,K)=0
\]
\end{theorem}

\begin{proof}
Let $X$, $Y$, and $Z$ be the (locally defined) Killing fields of~\cite{dr2} on $\mathcal{M}$
corresponding
to hyperbolic symmetry.
Note that
\begin{equation}
\label{commut1}
[X,Y]=Z,
\end{equation}
\begin{equation}
\label{commut2}
[Y,Z]=-X,
\end{equation}
\begin{equation}
\label{commut3}
[Z,X]=-Y.
\end{equation}
By the results of~\cite{dr2}, $X$, $Y$, and $Z$ extend to $C^2$ Killing fields
on $\mathcal{H}^+$ in a neighborhood of some $p\in \mathcal{H}^+$. 
(This can be interpreted to mean the following:
We can extend $X$, $Y$, and $Z$ to the extension $\tilde{\mathcal{M}}$ in a neighborhood
of $p$ as $C^2$ vector fields, not necessarily
Killing.) 

The integral curves of $X$, $Y$, and $Z$ through points of $p$ must stay on $\mathcal{H}^+$.
For otherwise, one could find integral curves emanating from points of $\mathcal{M}$
which leave $\mathcal{M}$. This contradicts the fact that $r$ is constant along such 
integral curves. 
In particular, at the set $S$ of
differentiable points of $\mathcal{H}^+$, $X$, $Y$, and $Z$ are tangent
to $\mathcal{H}^+$. By the results of~\cite{chrusciel}, the set $S$ of such points is
dense in $\mathcal{H}^+$.

Let us define 
\[
\mathcal{H}^+_1= {\rm int}(\{p\in\mathcal{H}: \dim({\rm Span} (X,Y,Z))=1)\})
\]
and 
\[
\mathcal{H}^+_2=\{p\in\mathcal{H}: \dim({\rm Span} (X,Y, Z))=2\}.
\]
These definitions are easily seen to be independent of the choice of local
Killing vector fields. The interior is taken with respect to the topology
of $\mathcal{H}^+$. These sets are manifestly open.

\subsection{Regularity on a dense open subset}
The proof of Proposition~\ref{Riccivanish} will be accomplished with the
help of several Lemmas.

\begin{lemma}
\label{Nprevlem}
The sets $\mathcal{H}^+_i$, $i=1, 2$, are $C^3$ null hypersurfaces, and
\begin{equation}
\label{Ndecomp}
\mathcal{H}^+=\overline{\mathcal{H}^+_1\cup\mathcal{H}^+_2}. 
\end{equation}
For each point $p$ of $\mathcal{H}^+_1\cup\mathcal{H}^+_2$, there exists a Killing field $K$ such
that $K(p)$ is future pointing and null, and
the identity
\begin{equation}
\label{gvwstne3}
\nabla_KK(p)=\kappa K(p)
\end{equation}
holds
for some $\kappa\le 0$.
\end{lemma}
Clearly, it follows from the above lemma that
 $\mathcal{H}^+_1$ (but not necessarily $\mathcal{H}^+_2$)
is locally a Killing horizon,
i.e.~for $p\in \mathcal{H}^+_1$, $K$ can be chosen so that
$K$ is null on $\mathcal{H}^+_1$ in a neighborhood of $p$.

\begin{proof}
Since $\dim({\rm Span}(X,Y,Z))\le 2$, to show $(\ref{Ndecomp})$ it suffices to
show that 
\begin{equation}
\label{vem}
{\rm int}(\{p\in\mathcal{H}^+: X=Y=Z=0)\})=\emptyset.
\end{equation}
Recall from before that the set $S$ of differentiable points of $\mathcal{H}^+$ is dense
in $\mathcal{H}^+$.
Thus, since the left hand side of $(\ref{vem})$ is manifestly an open set, it suffices
to show
\[
{\rm int}(\{p\in\mathcal{H}^+: X=Y=Z=0)\})\cap S=\emptyset.
\]
We will in fact show that 
\begin{equation}
\label{ifws}
{\rm int}(\{p\in\mathcal{H}^+: V=0)\})\cap S=\emptyset
\end{equation}
for any nontrivial Killing field $V$ in the span of $X$, $Y$, and $Z$.

Let $V$ be thus such a Killing field, and  let $p$ be a point in the set on the left
hand side of $(\ref{ifws})$. 
Since $p\in S$, we have that $\mathcal{H}^+$ is a differentiable null hypersurface
at $p$, and we can choose $\hat{K}$ a null generator at $p$. Let now $\hat{E}_1$,
$\hat{E}_2$,
$\hat{K}$, $\hat{L}$ denote a null frame, where $\hat{E}_1$ and
$\hat{E}_2$ denote unit vectors tangent to $\mathcal{H}^+$ at $p$.

By the Killing equation, we have that
\[
g(\nabla_{\hat L}V,{\hat L})=0,
\]
\[
g(\nabla_{\hat L}V,\hat{E}_i)+g(\nabla_{\hat{E}_i}V,\hat{L})=0,
\]
\[
g(\nabla_{\hat{K}} V, \hat{L})+g(\nabla_{\hat L}V, \hat{K})=0.
\]
In view of the fact that $\nabla_{\hat K}V=0$, $\nabla_{\hat{E}_i}V=0$, as these directions
are tangential to $\mathcal{H}^+$ and, by assumption,
$V=0$ in a neighborhood in $\mathcal{H}^+$ of $p$,
we obtain
$\nabla_LV=0$ and thus
$\nabla V=0$. By continuity of $\nabla V$, we have that $V=0, \nabla  V=0$ for all points
of $\mathcal{H}^+$ in a neighborhood of $p$.

But if indeed $V=0$, $\nabla V=0$ along $\mathcal{H}^+$ in a neighborhood of $p$,
then, from the well known relation
\begin{equation}
\label{WK}
\nabla_\alpha\nabla_\beta V_\gamma=R_{\alpha\beta\gamma\delta}V^\delta
\end{equation}
which holds for any Killing vector field $V$, by considering a family
of timelike geodesics transverse to $\mathcal{H}^-$ and integrating, it follows
immediately that $V$ must vanish identically in a neighborhood of $p$
in $\mathcal{M}$. This is a contradiction.

Thus, we have established $(\ref{ifws})$, and as a consequence, $(\ref{Ndecomp})$.
We proceed to show the remaining statements of the proposition.

By $(\ref{ifws})$, the set of points where $X\ne 0, Y\ne 0$ is open
and dense in $\mathcal{H}^+$.
Let $\tilde{S}=S\cap \{X\ne0\}\cap \{Y\ne0\}$.
This set is again dense.

Let $p\in \tilde{S}\cap \mathcal{H}^+_1$.
As $X$ does not vanish at $p$,
we may complete $X$ to a $C^2$ frame $X$, $V_1$, $V_2$, $V_3$ for the tangent bundle
in a neighborhood of $p$. We may then write
\[
Y=\alpha X + \beta_1V_1+\beta_2V_2+\beta_3V_3
\] 
where $\alpha$, $\beta_1$, $\beta_2$, $\beta_3$ are $C^2$ functions.
Since $p\in \mathcal{H}^+_1$, we have
\begin{equation}
\label{periergo}
\beta_1=\beta_2=\beta_3=0
\end{equation}
 along $\mathcal{H}^+$ in some neighborhood
of $p$.
Then
\begin{eqnarray*}
0	&=&	g(\nabla_XY,X)\\
	&=&	g(\nabla_X(\alpha X)X),X)+g(\nabla_X(\beta_1V_1+\beta_2V_2+\beta_3V_3),X)\\
	&=& (X\alpha) g(X,X)+\sum_i\beta_ig(\nabla_X V_i,X)+\sum_i X\beta_i g(V_i,X),
\end{eqnarray*}
where we have used the Killing equation for $X$ and $Y$. Since $\beta_i=0$ for
all points of $\mathcal{H}^+$ in this neighborhood, and $X\beta_i(q)=0$ for all
$q\in\tilde{S}$ in this neighborhood,
we have that for all $q\in\tilde{S}$, either
\begin{equation}
\label{nauto}
g(X,X)=0,
\end{equation}
or
\begin{equation}
\label{ntoallo}
\nabla_X\alpha=0.
\end{equation}
By the continuity of the functions $g(X,X)$, and $\nabla_X\alpha$, 
it follows that either $(\ref{nauto})$
or $(\ref{ntoallo})$ holds
for \emph{all} points in a neighborhood in $\mathcal{H}^+_1$ of $p$. In the 
case $(\ref{ntoallo})$, we note that $(\ref{commut1})$ implies that
$Z=0$ in this neighborhood, a contradiction, in view of $(\ref{ifws})$.

We thus must have
$(\ref{nauto})$ in a neighborhood of $p$ in $\mathcal{H}^+_1$. 
Since integral curves of $X$ must stay on $\mathcal{H}^+_1$, and $X$
is null, it follows by the characterization of~\cite{chrusciel} that $\mathcal{H}^+_1$
is differentiable. But this means that $\mathcal{H}^+_1$ is tangent at every point
to the $C^2$ orthogonal distribution of $X$. Thus $\mathcal{H}^+_1$ is
a $C^3$ null hypersurface, which is locally a Killing horizon around every point.

In the case of $\mathcal{H}^+_2$, let us first show that the span of
the Killing vectors at points of $\mathcal{H}^+_2$ is a null plane.

To see this we use the equality of $(\ref{suppose2...})$. 
Note first that there exists a dense subset $\hat{S}$ of $\mathcal{H}^+_2$
with the property that for each point $p\in\hat{S}$, 
there exists a non-radial null geodesic $\hat{\gamma}$ crossing $p$.
Since the span of the Killing fields constitutes a $C^2$ distribution of the tangent
bundle over $\hat\gamma$, if this distribution is a spacelike two plane at $p$, then the associated 
projection map is well defined and regular in a neighborhood of $p$.
Thus $\gamma_{AB}p^Ap^B\not\to\infty$ along $\hat\gamma$. 
On the other hand $r\to 0$ by $(\ref{suppose2...})$. Thus, since the quantity
$r^4\gamma_{AB}p^Ap^B$ is constant on $\hat\gamma$, it follows that it must vanish.
But this implies that $\hat{\gamma}$ is radial, a contradiction.
Thus, the distribution must be null at $p$.
By continuity, it follows that this distribution is null for all points of $\mathcal{H}^+_2$.

Let $V$ be a Killing field which is
null at $p$. Since $Vg(V,V)=0$, it follows that the integral curves of $V$ are null.
On the other hand, these integral curves must remain on $\mathcal{H}^+_2$.
It follows from~\cite{chrusciel} that $\mathcal{H}^+_2$ is a differentiable null
hypersurface. 

We may now show that $\mathcal{H}^+$ is $C^2$ as follows. Let $p\in\mathcal{H}^+_2$,
and let $K$ be a Killing field such that $K(p)$ is null. Let $E$ be any other Killing field
such that $E(p)$ is spacelike. Define the vector field $\tilde{K}$ in a nieghborhood of $p$ by
\[
\tilde{K}=K-g(K,E)(g(E,E))^{-1}E.
\]
Clearly, $\tilde{K}$ is null along $\mathcal{H}^+_2$ in a neighborhood of $p$,
is $C^2$,
and tangent to its null generator. 
It follows that $\mathcal{H}^+_2$ is tangent to the $C^2$ orthogonal 
distribution of $\tilde{K}$, and thus is a $C^3$ null hypersurface.

We proceed to show $(\ref{gvwstne3})$. We have already established
that for $p\in\mathcal{H}^+_1\cup\mathcal{H}^+_2$, there exists a Killing field $K$ such
that $K(p)$ is (after reversing the sign, if necessary) future directed, null.
Let us extend $\hat{K}=K(p)$ to a null frame $\hat{E}_1$, $\hat{E}_2$, $\hat{L}$ at $p$,
where $\hat{L}$ is future directed and $g(\hat{K},\hat{L})=-2$, and where
$\hat{E}_1$, $\hat{E}_2$ are tangent to $\mathcal{H}^+$ at $p$.
By the Killing property
we have
\[
g(\nabla_KK,K) = 0.
\]
On the other hand, since the function 
$g(K,K)$ restricted to $\mathcal{H}^+$ has a local minimum at $p$,
we have
\[
0=(\hat{E}_1 g(K,K))_p=2g(\nabla_{\hat{E}_1}K,K)=-2g(\nabla_KK,\hat{E}_1),
\]
\[
0=(\hat{E}_1 g(K,K))_p=2g(\nabla_{\hat{E}_1}K,K)=-2g(\nabla_KK,\hat{E}_1).
\]
Thus,
\[
\nabla_KK = -\frac12 g(\nabla_KK,\hat{L})K =\frac12 g(\nabla_{\hat{L}}K,K)K =\frac14
(\hat{L}g(K,K)) K,
\]
i.e. $(\ref{gvwstne3})$ holds for
\[
\kappa\doteq \frac14\hat{L}g(K,K)\le0,
\]
where the latter inequality holds since $\hat{L}$ is future pointing and $g(K,K)>0$ in 
$\mathcal{M}$.
\end{proof}

\subsection{Frames}
\label{framesec}
\begin{lemma}
\label{difficult}
Let $p\in\mathcal{H}^+_2$, and let
$K$, $\kappa$
be as in the previous lemma.
In some neighborhood $\mathcal{U}_p$,
$K$ can be completed to a $C^2$ frame $K$, $L$, $E_1$, $E_2$
for the tangent bundle over $\mathcal{U}_p$,
such that
at $p$, the vectors $K(p), L(p), E_1(p), E_2(p)$ constitute
a null frame, the vector field $E_1$ is Killing in $\mathcal{M}\cup\mathcal{H}^+\cap
\mathcal{U}_p$, and $E_2$ is
orthogonal to $K$ and $E_1$.
If $\kappa<0$, then either 
\begin{equation}
\label{gvwstne32}
g(\nabla_KE_1(p), E_1(p))=\kappa,
\end{equation}
\begin{equation}
\label{yeniyIldIz*}
E_1E_1g(K,K)(p)=2\kappa^2,
\end{equation}
or
\begin{equation}
\label{evallakt}
g(\nabla_K E_1(p), E_1(p))=0,
\end{equation}
in which case
\begin{equation}
\label{yeniyIldIz*evallakt}
E_1E_1g(K,K)(p)=0.
\end{equation}
Finally, for all $\kappa\le 0$, 
\begin{equation}
\label{yeniyildiz**}
E_2E_2 g(K,K)(p)=0, E_2 g(K,E_2)(p)=0.
\end{equation}
\end{lemma}
\begin{proof}
Let $E_1$ be any Killing vector with $0\ne E_1(p)\ne K(p)$, and let $E_2$ be 
a $C^2$ section of the $C^2$ distribution orthogonal to that generated by $E_1$ and $K$,
such that in addition $g(E_2(p),E_2(p))=1$. Finally let $L$ be an arbitrary $C^2$ extension
of a null vector with $g(E_1,L)(p)=0$, $g(E_2,L)(p)=0$, $g(K,L)(p)=-2$.

To show $(\ref{gvwstne32})$, consider the linear map $\phi$
from the span of $K(p)$ and $E_1(p)$ to itself defined as follows.
For $V=c_1K(p)+c_2E_1(p)$ consider the Killing field
$c_1K+c_2E_1$ and let 
\[
\phi(V) \doteq \nabla_K(c_1K+c_2E_1)(p).
\]

Now applying twist-free condition for $V$ (See Appendix~\ref{TF}) we have
\[
g(\nabla_K V,E_2)=g(dV,E_2\wedge K)=0.
\]
The first step uses the 
Killing property and the second the fact that $dV$ is proportional to $V$.
On the other hand, since $V$ is Killing we have, 
\[
g(\nabla_K V,K)=0.
\]
Writing 
\[
\phi (E_1(p)) =  h_1 E_1(p)+   h_2 K(p),
\]
and noting that by what we have just shown, 
\begin{equation}
\label{phiK}
\phi(K(p)) = \kappa K(p).
\end{equation}
it follows that 
the characteristic polynomial $p(\lambda)$ of $\phi$ is
\begin{equation}
\label{charpol}
p(\lambda)= (\lambda- h_1)(\lambda - \kappa).
\end{equation}
In particular, $h_1$ does not depend on the choice of Killing field $E_1$.

More is in fact true. It turns out that, except possibly in the cases $(\ref{evallakt})$
or $\kappa=0$, the entire
map defined above, which \emph{a priori} depends
on the choice of the Killing vectors which at $q$ point in the direction
$K$, $E_1$, actually depends only on the vector $K(q)$.

Let $X, Y, Z$ be Killing vector fields as before, and without loss of generality,
write
$K(p)=k Y$ for some $k\ne 0$. (We can arrange this by rotating $X$ and $Y$.)
We deduce from
\[
g(X,X)(p)+g(Y,Y)(p)-g(Z,Z) (p)= r^2(p) =0 
\]
 that $g(X,X)(p)=g(Z,Z)(p)$.  Since $g(Y,Y)\ge 0$ on $\mathcal{H}^+$ and
$=0$ at $p$, it follows that $p$ is a critical point of this function restricted
to $\mathcal{H}^+$. 
We thus have in particular that $Xg(Y,Y)(p)=0$ and thus
\[
Xg(X,X)(p)=X g(Z,Z)(p).
\]
On the other hand, by the Killing equation $Xg(X,X)=0$, so
we have
\[
X g(Z,Z)(p)=0.
\]
Since, again by the Killing equation, 
\[
Zg(Z,Z)(p)=0,
\]
it follows that, if $Z(p)$ and $X(p)$ are linearly independent, then
\[
Kg(Z,Z)(p) = c_1 X g(Z,Z)+c_2 Z g(Z,Z)=0.
\]
Similarly one shows $Kg(X,X)(p)=0$, and since $Kg(K,K)(p)=0$, it follows
that $Kg(E_1,E_1)(p)=0$ from which 
$(\ref{evallakt})$ follows immediately.

In what follows then, let us assume that $Z(p)$ and $X(p)$
are not linearly independent.
In this case, we must have then $Z(p)=\mp X(p)$, since $Z$ and
$X$ have the same length.

Now given our Killing vector $E_1$, then
any other Killing vector $\hat{E}_1$ such that $\hat{E}_1(p)=E_1(p)$
must be of the form $\hat{E}_1=E_1+ c(X\pm Z)$.
We compute:
\[
\nabla_K E_{1} = \nabla_{E_{1}} K + [E_1, K],
\]
whereas
\begin{eqnarray*}
\nabla_K \hat{E}_1 &=& \nabla_{E_{1}}K + [\hat{E}_1,K]\\
				&=&\nabla_{E_{1}}K + [E_1+c(X\pm Z), kY]\\
				&=& \nabla_K E_{1} + c(Z\pm X)
\end{eqnarray*}
and thus $\nabla_K E_1(p)=\nabla_K \hat{E}_1(p)$.

Claim. If $\kappa\ne0$, then either $K$ is the only eigenvector of
$\phi$ or every vector is an eigenvector.

Proof. Assume the contrary. Let $c_1, c_2$ be chosen so that
$c_1X+c_2Y $ is the unique unit eigenvector not equal to $K$.  Note that $c_1\ne 0$
since $kY(p)= K(p)$.  (Note also that $\nabla_K Y(p) = \kappa Y(p)$.) 
Since the direction $K$ is preserved by symmetries fixing $p$,
we must have that the direction $c_1 X+ c_2 Y$ be preserved by such symmetries.
We shall see in what follows that this is not the case.

Explicitly:
consider then the local group of transformations generated by
$X\pm Z$. Let us call this group $\phi_t$.  This fixes $p$. 
We have that
\[
((\phi_t)_*K)(p) = c_0 K(p)
\]
since there is a unique null direction in the span of the Killing fields,
while
\begin{eqnarray*}
((\phi_t)_* (c_1 X+c_2 Y))(p)  &=& (c_1 X +c_2 Y - [ X\pm Z,c_1 X+c_2 Y])(p)\\
			&=&( c_1X+c_2Y -c_2 [X,Y] \mp c_1[Z,X] \mp c_2 [Z,Y])(p)\\
			&=&(c_1X+c_2 Y -c_2 Z \pm c_1 Y \mp c_2 X)(p)\\
			&=&(c_1X+c_2 Y \pm c_1 Y -c_2(X\pm Z))p\\
			&=&(c_1X+c_2 Y \pm c_1 Y)(p)\\
			&\ne& (c_1 X+c_2 Y)(p).
\end{eqnarray*}
Thus, we compute from the above:
\begin{eqnarray*}
\nabla_ {(\phi_t)_*K} ((\phi_t)_* (c_1 X+c_2 Y))(p) 
 &=& c_0 \nabla_K (c_1X+c_2 Y\pm c_1 Y)(p) \\
 &=&  c_0 h_1 (c_1 X+c_2 Y)(q)\pm c_0c_1 \nabla_K Y(p)\\
&=&  c_0 h_1 (c_1 X+c_2 Y)(q)\pm c_0c_1 \kappa Y(p)
 \end{eqnarray*}
 while, since isometries of $\mathcal{M}$ preserve the connection,
 we have
 \begin{eqnarray*}
 \nabla_ {(\phi_t)_*K} ((\phi_t)_* (c_1 X+c_2 Y))(p) 
 &=& \nabla_K (c_1 X+c_2 Y)(p)\\
 & =&  h_1 (c_1 X+c_2 Y)(p).
\end{eqnarray*}
Since $\kappa\ne 0$, this is a contradiction.
 
Given this, it follows that the characteristic polynomial $(\ref{charpol})$ 
of $\phi$ must
have a double root, i.e.~$h_1= \kappa$. This gives $(\ref{gvwstne32})$ immediately.

We turn to showing $(\ref{yeniyIldIz*})$
or $(\ref{yeniyIldIz*evallakt})$ for $\kappa<0$, and
 $(\ref{yeniyildiz**})$ for all values of $\kappa$.
   
Since $\mathcal{H}^+$ coincides with the set
\[
\{g(K,K)g(E_1,E_1)-g(K,E_1)^2 =0\}
\]
in a neighborhood of $p$, and the vector field $E_1$ is tangent
to $\mathcal{H}^+$ we have
\begin{equation}
\label{willexp1}
E_1E_1(g(K,K)g(E_1,E_1)-g(K,E_1)^2)=0.
\end{equation}

Expanding $(\ref{willexp1})$ and evaluating at $p$ we obtain
\[
(E_1E_1g(K,K))(p)=2(E_1g(K,E_1))^2(p),
\]
where we use in particular that $E_1g(E_1,E_1)=0$ identically, since $E_1$ is Killing.
On the other hand, 
\begin{eqnarray*}
E_1g(K,E_1)	&=&	g(\nabla_{E_1}K,E_1)+g(K,\nabla_{E_1}E_1)\\
			&=&	g(K,\nabla_{E_1}E_1)\\
			&=&	-g(E_1,\nabla_K E_1),
\end{eqnarray*}
where we have used the Killing properties for $K$ and $E_1$.
Identities $(\ref{yeniyIldIz*})$
or $(\ref{yeniyIldIz*evallakt})$ now follow from $(\ref{gvwstne32})$ or $(\ref{evallakt})$, respectively.

We finally turn to showing $(\ref{yeniyildiz**})$. First note that $E_2$ is also tangent to
$\mathcal{H}^+$, for instance since $g(E_2,\tilde{K})=0$, and thus we have
\begin{equation}
\label{willexp2}
E_2E_2(g(K,K)g(E_1,E_1)-g(K,E_1)^2)=0.
\end{equation}

In addition, note that by the twist-free (See Appendix~\ref{TF}) 
and Killing properties of $K$ and $E_1$ we derive
\[
g(\nabla_{E_2} K,E_1)=g(dK,E_1\wedge E_2)=0,
\]
\[
g(K,\nabla_{E_2}E_1)=g(dE_1,K\wedge E_2)=0.
\]

Expanding $(\ref{willexp2})$ and evaluating at $q$ we obtain
\begin{eqnarray*}
0	&=&	E_2E_2 g(K,K)(p) +E_2g(K,K)E_2g(E_1,E_1) - 2(E_2 g(K,E_1))^2\\
	&=&	E_2E_2 g(K,K)(p) - 2(E_2 g(K,E_1))^2\\
	&=&	E_2E_2 g(K,K)(p) -2(g(\nabla_{E_2}K,E_1)+g(K,\nabla_{E_2}E_1))^2\\
	&=&  E_2E_2 g(K,K)(p) -2(g(K,\nabla_{E_2}E_1))^2\\
	&=&  E_2E_2 g(K,K)(p) -2(g(E_2,\nabla_{K }E_1))^2\\
	&=&  E_2E_2 g(K,K)(p) -2(g(E_2,\nabla_{E_1}K+[K,E_1]))^2\\
	&=&	E_2E_2 g(K,K)(p) -2(g(E_2,\nabla_{E_1}K)+g(E_2,[K,E_1]))^2\\
	&=& E_2E_2 g(K,K)(p),
\end{eqnarray*}
where we have used the Killing and twist free properties of $E_1$, $K$,
 as well as the orthogonality
of $E_2$ with the Lie algebra spanned by $K$ and $E_1$.
The above gives $(\ref{yeniyildiz**})$.
\end{proof}

\begin{lemma} 
\label{killinghor}
Let $K$ be as in Lemma~\ref{Nprevlem}, and let
$p\in\mathcal{H}^+_1$. 
In some neighborhood $\mathcal{U}_p$,
$K$ can be completed to a $C^2$ frame $K$, $L$, $E_1$, $E_2$
for the tangent bundle over $\mathcal{U}_p$,
such that
at $p$, the vectors $K(p), L(p), E_1(p), E_2(p)$ constitute
a null frame, $E_1$ and $E_2$ are tangent to
$\mathcal{H}^+_1$ near $p$,
and
\begin{equation}
\label{evallakt-ap}
g(\nabla_K E_1(p), E_1(p))=0,
\end{equation}
\begin{equation}
\label{yeniyIldIz*evallakt-ap}
E_1E_1g(K,K)(p)=0,
\end{equation} 
\begin{equation}
\label{yeniyildiz**-ap}
E_2E_2 g(K,K)(p)=0, E_2 g(K,E_2)(p)=0.
\end{equation}
\end{lemma}
\begin{proof} 
Let $E_1$, $E_2$ be $C^2$ sections of the $C^2$ orthogonal bundle
to $K$ in a neighborhood $\mathcal{U}_p$, such that $E_1(p)$ and $E_2(p)$ are orthonormal. 
Clearly, $E_1$ and $E_2$ are tangent to $\mathcal{H}^+$.
Complete
$E_1$, $E_2$, $K$ to a frame for the tangent bundle by adding a $C^2$ vector
field $L$, such
that $E_1(p)$, $E_2(p)$, $K(p)$, $L(p)$ constitutes a null frame at $p$. 

Note that $E_1(E_1g(K,K))(q)$ and $E_2(E_2g(K,K))$ are well defined and vanish, on account of the fact
that the $C^3$ integral curves of $E_1$ and $E_2$ lie on $\mathcal{H}^+$ which is precisely
the set where $g(K,K)=0$. This gives $(\ref{yeniyIldIz*evallakt-ap})$ and
the first equality of $(\ref{yeniyildiz**-ap})$. Similarly, we compute
that
\[
g(K,\nabla_{E_1}E_1)(q)= E_1g(K,E_1)(q)-g(\nabla_{E_1}K,E_1)(q)=0,
\]
\[
g(K,\nabla_{E_2}E_2)(q)= E_1g(K,E_2)(q)-g(\nabla_{E_2}K,E_2)(q)=0,
\]
by the Killing property of $K$ together with the fact that $g(K,E_1)=0$, $g(K,E_2)=0$
along $\mathcal{H}^+$.
On the other hand, $g(\nabla_{E_1}K,E_2)=0$ by the twist-free property,
and certainly $\nabla_{E_1}g(K,K)=0=\nabla_{E_2} g(K,K)$. 
This gives the remaining statements.
\end{proof}

\subsection{The rigidity computation}
We now turn to complete 
the proof of Proposition~\ref{Riccivanish}.
The open dense set will be $\mathcal{H}^1\cup\mathcal{H}^2$.
If $p\in \mathcal{H}^1$, then we can obtain ${\rm Ric}(K(p),K(p))=0$
from Proposition  6.15 of  Heusler~\cite{unique}.
As our argument for the case $p\in\mathcal{H}^2$ is in any
case motivated by the computation
of~\cite{unique}, we will give a self-contained treatment for all $p\in\mathcal{H}^1\cup
\mathcal{H}^2$.

Note first the identity
\begin{equation}
\label{theid}
\Box g(K,K)=-2Ric (K,K) +  2 g(\nabla K,\nabla K).
\end{equation}

Let us first consider the case $\kappa\ne0$.
We evaluate $(\ref{theid})$ at a $p\in\mathcal{H}^+_1\cup\mathcal{H}^+_2$.
Let $K$, $L$, $E_1$, $E_2$ be as in Lemma~\ref{difficult}, or Lemma~\ref{killinghor}, accordingly.
In view of the properties of the frame we obtain
\begin{eqnarray*}
\Box g(K,K)(p) &=&
-2\nabla^2_{L,K}g(K,K)
+\nabla^2_{E_1,E_1}g(K,K)+\nabla^2_{E_2,E_2}g(K,K)\\
&=&-2L (K(g(K,K))+2\nabla_{{\nabla_L}K}g(K,K)
+E_1(E_1(g(K,K)))\\
&&\hbox{} + E_2(E_2(g(K,K)))
-\nabla_{\nabla_{E_1}E_1}g(K,K)-\nabla_{\nabla_{E_2}E_2}g(K,K)\\
&=&E_1(E_1(g(K,K)))+E_2(E_2(g(K,K)))+
4g(\nabla_{\nabla_L K}K,K)\\
&&\hbox{}-2g(\nabla_{\nabla_{E_1}E_1}K,K)
-2g(\nabla_{\nabla_{E_2}E_2}K,K)\\
&=&E_1(E_1(g(K,K)))+E_2(E_2(g(K,K)))-
4g(\nabla_K K,\nabla_L K)\\
&&\hbox{}+2g(\nabla_KK,\nabla_{E_1}E_1)
+2g(\nabla_KK,\nabla_{E_2}E_2)\\
&=&E_1(E_1(g(K,K)))+E_2(E_2(g(K,K)))-
8\kappa^2\\
&&\hbox{} +2g(\nabla_KK,\nabla_{E_1}E_1)
+2g(\nabla_KK,\nabla_{E_2}E_2)\\
&=&E_1(E_1(g(K,K)))+E_2(E_2(g(K,K)))-
8\kappa^2\\
&&\hbox{} +2\kappa g(K,\nabla_{E_1}E_1)
+2\kappa g(K,\nabla_{E_2}E_2).
\end{eqnarray*}
In the case of $p\in\mathcal{H}^2$, $\kappa(p)\ne 0$, we
obtain from
Lemma~\ref{difficult} and the above,
\begin{equation}
\label{PE1}
\Box g(K,K)(p) = -8\kappa^2.
\end{equation}
In the case, $p\in\mathcal{H}^2$, $\kappa(p)=0$, we obtain
from Lemma~\ref{difficult},
\begin{equation}
\label{PE2}
\Box g(K,K)(p)=E_1E_1g(K,K)\ge 0,
\end{equation}
where the inequality follows from the fact that $g(K,K)$ restricted to $\mathcal{H}^+$
has a local minimum at $p$, and $E_1$ is tangent to $\mathcal{H}^+$.
Finally, in the case of $p\in\mathcal{H}^1$, we obtain now from Lemma~\ref{killinghor},
\begin{equation}
\label{PE3}
\Box g(K,K)(p)=-8\kappa^2.
\end{equation}

On the other hand, for $p\in\mathcal{H}^+_1\cup\mathcal{H}^+_2$,
\begin{eqnarray}
\label{OTOH1}
\nonumber
2 g(\nabla K, \nabla K)(p)	 &=&-4 g(\nabla_L K,\nabla_K K)+ 2 g(\nabla_{E_1}K,\nabla_{E_1}K)
		+2 g(\nabla_{E_2}K,\nabla_{E_2}K)\\
\nonumber
&=&		-8\kappa^2+ 2(g(\nabla_{E_1}K,E_1))^2+2(g(\nabla_{E_1}K,E_2))^2\\
\nonumber
&&\hbox{}
		-4g(\nabla_{E_1}K,K)g(\nabla_{E_1}K,L)+2(g(\nabla_{E_2}K,E_2))^2 \\
\nonumber
&&	\hbox{}+2(g(\nabla_{E_2}K,E_1))^2
				-4g(\nabla_{E_2}K,K)g(\nabla_{E_2}K,L) \\
&=&-8\kappa^2,
\end{eqnarray}
where to obtain the final equality, we have used the fact that $K$ is twist-free.

Thus, in the case $p\in\mathcal{H}^+_2$, $\kappa(p)\ne0$, we
obtain from $(\ref{theid})$, $(\ref{PE1})$  and $(\ref{OTOH1})$ that
\[
0=-2{\rm Ric} (K,K)(p).
\]
so we conclude ${\rm Ric}(K, K)=0$ at $p$.
The same identity holds for $p\in\mathcal{H}^+_1$, after applying $(\ref{theid})$,
$(\ref{PE2})$ and $(\ref{OTOH1})$.

Finally, for $p\in\mathcal{H}^+_2$, $\kappa(p)=0$, 
$(\ref{theid})$, $(\ref{PE2})$ and $(\ref{OTOH1})$ gives
\[
{\rm Ric}(K,K)(p)\le 0.
\]
The proof of the Theorem is complete.
\end{proof}
Note that if
 $(\mathcal{M},g)$ satisfies the null convergence condition, then by continuity,
it follows that ${\rm Ric}(K,K)=0$ even if $\nabla_KK=0$.

\section{Open questions and conjectures}
\label{ops}
We begin with the cosmological case.
The first obvious open problem left unresolved  is the following
\begin{conjecture}
\label{leftout}
Strong cosmic censorship holds in the case $k=1$, $\Lambda>0$. 
\end{conjecture}
In the case where $(\ref{evwsis})$ holds,
this would follow from a positive resolution to
\begin{conjecture}
\label{nonext}
For generic initial data in the case $k=1$, $\Lambda>0$, all horizons satisfy $(\ref{case1})$
or $(\ref{case2})$.
\end{conjecture}
On the other hand, as the inextendibility statement is not in fact violated
by the extremal case depicted in Section~\ref{???}, one could imagine a proof of
Conjecture~\ref{leftout} that does not go through Conjecture~\ref{nonext}.

Another interesting question about horizons in the $k=1$ case is provided
by the following
\begin{question}
Let $\mathcal{F}$ denote a fundamental domain for $\mathcal{Q}$ in
$\tilde{\mathcal{Q}}$. Is $|\overline{\mathcal{F}}\cap\mathcal{B}^{\pm}_h|<\infty$?
\end{question}
In view of Theorem~\ref{finitetheorem}, this would be true at least for
 generic initial data if Conjecture~\ref{nonext} holds.
See also Appendix~\ref{homog}.

An additional open problem is try to apply the methods of this paper to the study of
the case with $\Lambda<0$, i.e.~to answer
\begin{question}
Does strong cosmic censorship hold for $\Lambda<0$?
\end{question}

We turn to the asymptotically flat case. In the absence of known
counterexamples, one might reasonably conjecture
\begin{conjecture}
For all spherically symmetric asymptotically flat initial data, $\mathcal{CH}^+=\emptyset$.
\end{conjecture}
This would follow from
\begin{conjecture}
For all spherically symmetric asymptotically flat initial data,
$(\ref{case2})$ is satisfied on $\mathcal{H}^+$.
\end{conjecture}
Of course, for applications to strong cosmic censorship it would be
sufficient to prove the above statements for generic initial data. 

As shown in Section~\ref{bhs}, $(\ref{case2})$ follows from $(\ref{veasuv9nkn})$.
Here, it is worth comparing with the case of a self-gravitating scalar field. In that case, it follows 
from~\cite{di} or~\cite{chr:mt} that for all data leading to a black hole, the equality
\begin{equation}
\label{muchstronger}
2M_fr_+^{-1}=1
\end{equation}
holds.
That is to say, in the case of a self-gravitating scalar field, there can be no
persistent atmosphere of a black hole. The field either falls into the black hole
or disperses to infinity. In the case of collisionless matter, one does not expect
this to be true,
and thus, many interesting questions arise. Some of these are summarised below in
\begin{question}
Can one formulate conditions on initial data ensuring 
$(\ref{muchstronger})$? Can one formulate
conditions on initial data (other than the trivial ones obtained
by monotonicity arguments) ensuring $(\ref{veasuv9nkn})$?
Are there restrictions on the possible values of $2M_fr_+^{-1}$,
in particular, can it be arbitrarily close to $1$ or arbitrarily large?
Can one construct open sets of initial data whose solutions
violate $(\ref{veasuv9nkn})$?
\end{question}

Turning to the cases which are ``successfully'' handled in this paper,
although our results indeed prove strong cosmic censorship
in its $C^2$ formulation, they do not resolve all interesting questions
about the maximal development. Indeed, the lesson of this paper may
be that the $C^2$ formulation of the conjecture was not appropriate
in the first place. For the fundamental questions of what happens to 
macroscopic classical observers remain unanswered: 
In the expanding direction, can they observe for all time?
Are observers living for only finite time necessarily destroyed?

The answer to the former question in the case of $k\le 0$, $\Lambda\ge 0$,
is most certainly ``Yes.'',
i.e.~we can reasonably state the following as a 
\begin{conjecture}
In the statement of Theorem~\ref{introthe}, the spacetime is future causally
geodesically complete.
\end{conjecture}
The above conjecture is known to be true in the case $\Lambda>0$,
as proved in~\cite{tr}.
The question of the fate of observers who live for only finite proper 
time is much more
delicate.  
\begin{question}
\label{con1}
In the statement of Theorems~\ref{introthe}--\ref{afth}, can one replace
$C^2$ inextendibility with $C^0$ inextendibility?\footnote{For remarks on the
appropriateness of this formulation for strong cosmic censorship, 
see Christodoulou~\cite{chr:givp}.}
\end{question}
The answer to the above may 
be related to the genericity of the second Penrose diagram
for past evolution in Theorems~\ref{introthe}--\ref{introthe15}, i.e., it 
may depend on the answer
to the following 
\begin{question}
\label{con15}
In the statement of Theorems~\ref{introthe}--\ref{introthe2}, 
is there an open set of initial
data leading to the second Penrose diagram with $\infty>r_\pm>0$?
\end{question}
It is worth noting that the considerations of Appendix~\ref{homog} indicate
that solutions with $\mathcal{B}^+=\mathcal{N}^+$ might be obtained by
fine tuning in $1$-parameter families interpolating between recollapse
and infinite expansion. Thus, these solutions
may be analogous to critical behaviour in gravitational collapse. 
It may then be of interest to study  the fine properties of the
 set of such solutions, even if the answer to 
the above question is ``No.''.

A negative answer to Question~\ref{con15} in the case of the assumptions
of Theorem~\ref{introthe2} would also exclude counterexamples to
strong cosmic censorship arising from
$\mathcal{B}^\pm=\mathcal{N}^\pm$.

Irrespectively of the answer to the above questions,
in the case where black holes can form, it is now widely thought that in more
realistic models, the boundary of the maximal 
development will have a null portion emanating from
$i^+$, i.e. $\mathcal{CH}^+\ne\emptyset$. 
Similar heuristics should apply in the cosmological case for
perturbations of Schwarzschild-de Sitter~\cite{strong}. 
In spherical symmetry, this mechanism can be produced by adding charge. 
Indeed,
the picture of a weak null singularity has been rigorously confirmed for the collapse
of a spherically symmetric scalar field coupled gravitationally with a 
Maxwell field~\cite{md:si, md:cbh}. In particular, for this model,
the answer to Question~\ref{con1} has been shown to be ``No.''. This motivates:
\begin{conjecture}
\label{con2}
Consider the setup of Theorem~\ref{introthe15} or Theorem~\ref{introthe2}
for the Einstein-Vlasov-Maxwell
system. In the former case,
then there are initial data for which
$\mathcal{N}^1_x\ne\emptyset$, even if $\mathcal{H}^1_x$ is not
extremal (see the Penrose diagram after Theorem~\ref{introthe2}),
and in the latter case, there are intial data for which
there is a non-empty null component $\mathcal{CH}^+$ to the
boundary of the maximal development, where $r$ is strictly positive in the limit:
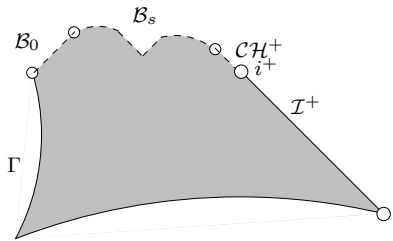
\[
\input{asymptflat3.pstex_t}
\]
Moreover, the answer to Question~\ref{con1} is ``No.'' for this system.
\end{conjecture}

Despite the fact that the above picture of
a weak null singularity, at least on a heuristic
level~\cite{pi:ih}, has been known for many years, one often still sees in
the literature the statement that singularities are thought to be
``generically spacelike''.
A good way to familiarise oneself with the real issues involved in this problem
would be to try to resolve Conjecture~\ref{con2}.

In the cosmological case, one might still hope that horizons (and the associated
issues they raise) can be excluded if one remains sufficiently close to a homogeneous solution.
That even this is not the case, in general, is shown by explicit example in 
Appendix~\ref{homog}.
Nevertheless, the following is still unanswered:
\begin{question}
\label{nq}
Are there spherically
symmetric solutions of the Einstein-Vlasov system with $\Lambda>0$,
arising from data arbitrarily close to homogeneous,
with spatial topology $\mathbb S^1\times \mathbb S^2$,  such that $\lambda>0$, $\nu>0$
initially, and
$\mathcal{B}^+_h\ne \emptyset$, $\mathcal{B}^+_\infty\ne\emptyset$,
 $\{\lambda<0\}\cap\{\nu<0\}\ne \emptyset$?
\end{question}

Finally, the nature of the set $\mathcal{A}$ of spherically symmetric marginally trapped
spheres remains to be explored. A reasonable conjecture would be
\begin{conjecture}
\label{dh}
For sufficiently large $v$, $\mathcal{A}\cap\{v\ge V\}$ is achronal.
\end{conjecture}
The solution of the above conjecture may in fact be relevant to the previous. This is in
fact the case for the Einstein-Maxwell-scalar field system, for which
Conjecture~\ref{dh} is proven in~\cite{md:cbh}. For this, the results of~\cite{di},
in particular, the decay of $T_{vv}$ along the event horizon, is essential.
The connection of radiation decay on the event horizon and the eventual achronality
of the set $\mathcal{A}$ is often overlooked in the literature on ``dynamical horizons''.
Again, a good way to familiarize oneself with the real issues involved would
be to try to resolve Conjecture~\ref{dh}.

\appendix
\section{Curvature expressions}
\label{curvexp}
In the following we collect expressions for the curvature in surface
symmetric spacetimes. The basic unknowns are the two-dimensional
Lorentzian metric on the quotient manifold and the area radius $r$.
The components of the curvature are:
\begin{eqnarray*}
R^a_{bcd}&=&K(\delta^a_cg_{bd}-\delta^a_dg_{bc}),         \\
R^a_{BcD}&=&-r\nabla^a\nabla_c r\gamma_{BD},      \\
R^A_{BCD}&=&(k-\nabla_a r\nabla^a r)(\delta^A_C\gamma_{BC}-\delta^A_D
\gamma_{BC}).
\end{eqnarray*}
Here $K$ is the Gaussian curvature of the quotient metric and $k$ is the
curvature parameter of the orbits. The explicit formula for the Gaussian
curvature in double null coordinates is
\[
K=4\Omega^{-2}(\Omega^{-1}\partial_u\partial_v\Omega-\Omega^{-2}
\partial_u\Omega\partial_v\Omega).
\]
Lower and upper case Latin
indices correspond to objects on the quotient manifold and the orbits 
respectively. The metric $\gamma_{AB}$ is equal to $r^{-2}g_{AB}$.
The Kretschmann scalar is given by
\[
R^{\alpha\beta\gamma\delta}R_{\alpha\beta\gamma\delta}
=4K^2+4r^{-4}(k-\nabla^a r\nabla_a r)^2+12r^{-2}\nabla_a\nabla_b r
\nabla^a\nabla^b r.
\]
Note that the coefficient of the last term differs from that given in~\cite{nature}. 
The non-vanishing components
of the Ricci tensor are:
\begin{eqnarray*}
R_{ab}&=&Kg_{ab}-2r^{-1}\nabla_a\nabla_b r,     \\
R_{AB}&=&(-r\nabla^a\nabla_a r+k-\nabla^a r\nabla_a r)\gamma_{AB}.
\end{eqnarray*}
Using this the following field equation can be derived:
\begin{equation}
\nabla_a\nabla_b r=\frac{1}{2r}(k-\nabla_c r\nabla^c r)g_{ab}
-4\pi r(T_{ab}-{\rm tr}Tg_{ab})-\frac12\Lambda rg_{ab}
\end{equation}
where ${\rm tr}T=g^{ab}T_{ab}$. It follows that the last term in the
expression for the Kretschmann scalar can be written as
\[
24 r^{-2}(\frac{1}{2r}(k-\nabla_c r\nabla^c r)+2\pi r{\rm tr}T-r\Lambda/2)^2
+96\pi^2(T_{ab}-\frac12 {\rm tr}T g_{ab})(T^{ab}-\frac12 {\rm tr}T g^{ab}).
\]
The only contribution to the Kretschmann scalar which is not a square is
the last one. This last term is positive provided the tensor $T^a_{b}$ is
diagonalizable. This holds if the matter model satisfies the dominant
energy condition.

\section{The twist-free condition}
\label{TF}
If $k$ is a Killing vector and we denote the one-form corresponding to
it via the metric by the same letter then $k$ is said to be \emph{twist-free}
if $dk\wedge k=0$. This is equivalent to the property that 
$dk=\eta\wedge k$ for some $\eta$. By Frobenius' theorem this is 
equivalent to the property that the orthogonal complement of $k$ is
integrable.

\begin{proposition} Consider a spacetime admitting an isometry group with
two-dimensional spacelike orbits and suppose that the planes orthogonal to
the orbits are surface-forming. Then any one of the Killing vectors 
corresponding to the group action is twist-free.
\end{proposition}

\begin{proof}
Without loss of generality we may restrict consideration
to a point where the Killing vector $k$ is non-vanishing since points
of this kind are dense. It then suffices to show that the orthogonal 
complement of $k$ is integrable. Let $\gamma$ be a curve which lies in 
an orbit and is orthogonal to the integral curves of $k$. It is evident 
that such curves exist at least locally. Let $W$ be the union of the 
integral manifolds of the orthogonal complement of the orbits passing 
through points of $\gamma$. Then $W$ is an integral manifold of the 
orthogonal complement of $k$. For it is a three-dimensional manifold 
whose tangent space is by construction orthogonal to $k$.
\end{proof}

\noindent
{\it Remark.} The above Proposition is applicable to surface symmetric spacetimes,
as can be seen from the usual coordinate form of the metric. For the
same reason it also applies to spacetimes with Gowdy symmetry. It does not 
apply to general $T^2$-symmetric spacetimes. In this last case there are 
two Killing vectors $k$ and $l$ and the fact that the spacetime does not 
have Gowdy symmetry is equivalent to the fact that one of $dk\wedge k\wedge l$
or $dl\wedge k\wedge l$ is non-zero. It follows that at least one of the
two Killing vectors must fail to be twist-free.

\section{Homogeneous solutions and their stability properties}
\label{homog}
Let us consider in this section the $k=1$, $\Lambda\ge 0$ case.
We will consider here certain stability and instability phenomena of homogeneous solutions 
of the Einstein-Vlasov system within
the spherically symmetric class.

First we note there 
exists a spherically symmetric and
homogeneous vacuum solution, where $f$ vanishes, and $r=\frac{1}{\sqrt{\Lambda}}$
identically. This is sometimes called the Nariai solution.
The Penrose diagram of the universal cover of the quotient is easily seen to be
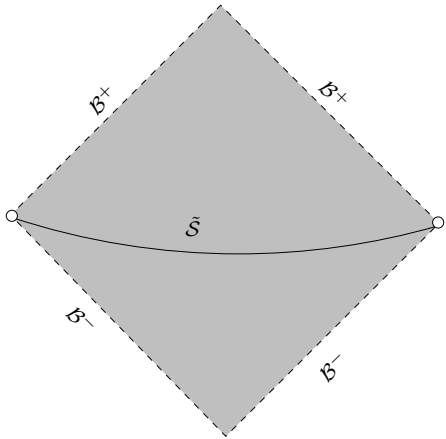
\[
\input{special.pstex_t}
\]
It is interesting to note that arbitrary small neighborhoods of
the initial data induced on $\mathcal{\tilde{S}}$ in the space of Einstein-Vlasov initial data
contain data sets with regions 
coinciding with arbitrarily large regions of 
(almost extremal) Schwarzschild-de Sitter initial data
sets.
By the domain of dependence property, the resulting solutions will have in particular horizons,
in fact, one can construct such arbitrary close solutions with arbitrary many horizons.
Thus, the Nariai solution is a homogeneous solution with the property that horizons
occur for arbitrary small spherically symmetric
perturbations.\footnote{Some form of this
fact was first enunciated in a linearised setting by~\cite{gper}.}

One might object that the above cannot really be seen as a cosmological solution.
We discuss in the sequel a method for constructing more interesting examples.

First, one can infer--from our previous results and ``soft arguments''--the existence of a homogeneous non-vacuum solution with $k=1$, $\Lambda>0$,
such that there exists a Cauchy surface with $\lambda>0$, $\nu>0$,  
and such that, either  $\mathcal{B}^+=\mathcal{N}^+$ with $\infty>r_+>0$, or
$\mathcal{B}^+=\mathcal{B}^+_h$.

This we do as follows:
Consider a one parameter family of homogeneous solutions with $f$ not vanishing identically,
arising from initial
data with constant and fixed
$r$, such that $\lambda>0$, $\nu>0$ on that surface. Let the parameter
be $\Lambda$ itself, 
the cosmological constant, and let it take all values in $[0,\infty)$. For $\Lambda=0$, 
we have shown $r_+=0$. This means that there exists a later Cauchy surface  $\mathcal{S}'$
such that $\lambda<0$, $\nu<0$. Consider the set of $\Lambda$ such that there exists such 
a later Cauchy surface  $\mathcal{S}'$. 
By Cauchy stability\footnote{Here we are perturbing both the data and the equation by
changing $\Lambda$!}, perturbations of such data in the spherically symmetric
class also have this property. 
In particular, this set of $\Lambda$ is open and nonempty.

On the other hand, consider such a homogeneous solution with
 $r> \frac{1}{\sqrt{\Lambda}}$,
$\nu>0$, $\lambda>0$
on $\mathcal{S}$. By the Proposition~\ref{toxreiaz}, it follows that $\mathcal{B}^+=\mathcal{B}_\infty$.
On the other hand, any solution such that $\mathcal{B}^+=\mathcal{B}^+_\infty$ will have a later
Cauchy surface $\mathcal{S}'$ such that $r> \frac{1}{\sqrt{\Lambda}}$,
$\lambda>0$, $\nu>0$.
By Cauchy stability, it follows that sufficently small spherically
symmetric perturbations of homogeneous
solutions with $f$ not identically $0$ and with $r_+=\infty$ also satisfy $r_+=\infty$.

In particular, the set of $\Lambda$ in our one-parameter family such that $r_+=\infty$
is an open, non-empty subset of $[0,\infty)$, 
as is the set of $\Lambda$ for which $r_+=0$. 

By connectedness of $[0,\infty)$, it follows that there exists a solution corresponding
to $\Lambda_0$ such that either $\mathcal{B}^+=\mathcal{B}^+_h$,
or $\mathcal{B}^+=\mathcal{N}^+$ and $r_+\le \frac{1}{\sqrt{\Lambda}}$.

Finally, we can now rescale the parameters $r$, $f$, $\Omega$, $\Lambda$ 
in this family in such a way
so that the rescaled solutions (except for the one with $\Lambda=0$, which we discard)
all have cosmological constant $\Lambda_0$. 
Let the parameter of the one-parameter family of homogeneous solutions thus obtained
now be denoted by $s$.
Let $\mathcal{R}_{0}$ denote the
class of solutions for which $\lambda<0$, $\nu<0$ on some late Cauchy surface $\mathcal{S}'$, 
and let
$\mathcal{E}_{0}$ denote the class of solutions for which $\mathcal{B}^+=\mathcal{B}^+_\infty$.
Let $\mathcal{R}$ denote the set of all spherically symmetric
solutions with fixed $\Lambda_0$ such that $\lambda<0$, $\nu<0$ on some late
$\mathcal{S}'$,
and let $\mathcal{E}$ denote the set of all spherically symmetric solutions for
which $\mathcal{B}^+_\infty=\mathcal{B}^+$.

Let $s_0$ denote $\sup_{\mathcal{R}_0} s$,
and consider the homogeneous solution  corresponding to $s_0$ considered
as a solution with spatial topology $\mathbb R\times \mathbb S^2$. Now 
consider arbitrary small spherically symmetric
perturbations of this solution, which are periodic in the direction of $\mathbb R$\footnote{and
thus can be quotiented to $\mathbb S^1\times \mathbb S^2$}.
That is to say, let $\mathcal{U}$ be an arbitrarily small neighborhood of spherically
symmetric solutions around the homogeneous one, topologized by closeness on a
fixed initial surface  in a suitable norm.

If $s_0\in\overline{\mathcal{E}_0}$, then $\mathcal{U}\cap\mathcal{R}_0\ne\emptyset$,
$\mathcal{U}\cap \mathcal{E}_0\ne\emptyset$.
It is clear that by an easy domain of dependence argument, one can arrange 
an arbitrarily small spherically symmetric
perturbation of the initial data corresponding to the homogeneous solution with $s=s_0$
such that
$\mathcal{B}^+_\infty\ne\emptyset$, and $\{\lambda<0\}\cap\{  \nu<0\}\ne\emptyset$.
For this, we just need to chose  a perturbation coinciding
with a homogeneous solution with $r_+=0$ on a sufficiently large
subset of initial data, and similarly, a homogeneous solution with $r_+=\infty$
on a sufficiently large subset.
For the perturbations constructed, it follows that $\mathcal{B}^+_h\ne\emptyset$.

If  $s_0\not\in\overline{\mathcal{E}_0}$, since $\mathcal{U}\cap \mathcal{R}$, $\mathcal{U}\cap \mathcal{E}$
are open, by connectedness, one argues, after possibly changing $s_0$, that
either there exists an open $\mathcal{V}\subset \mathcal{U}$ 
such that $\mathcal{V}\cap \mathcal{R}\ne \emptyset$
and $\mathcal{V}\cap \mathcal{E}\ne \emptyset$, or there exists
an open $\mathcal{V}\subset {\rm int}(\mathcal{U}\setminus(\mathcal{R}\cup \mathcal{E}))$.
In the former case, one repeats the argument of the previous paragraph.
In the latter case, every solution in $\mathcal{V}$ will have $\mathcal{B}_h\ne\emptyset$
or $\mathcal{B}^+=\mathcal{N}^+$ with $r_+\le \frac{1}{\sqrt{\Lambda}}$.

Thus, either ``global horizons'' (the case of $\mathcal{N}^+$) or
``horizon points'' (the case of $\mathcal{B}^+_h$) are relevant even in an arbitrarily small
neighborhood of an ``expanding'' homogeneous solution, and they occur for a set of solutions with
non-empty interior. In view of the remarks prior to Conjecture~\ref{con2}, one should expect
horizon points to give rise to null portions of the boundary of spacetime in more general
models, for instance, in the presence of charge. Thus, either way, it appears that
one should expect the boundary of
spacetime to contain null portions.
These remarks should be contrasted with various general scenarios that
have been put forth as to
the nature of ``generic singularities'' in general relativity, scenarios that perhaps do not
do sufficient
justice to the variety of causal structure that the global dynamics may give rise to,
even in an arbitrarily small neighborhood of a homogeneous cosmology.

Let us return to the case  $s_0\in\overline{\mathcal{E}_0}$, if this
indeed occurs.
In the case of topology $\mathbb S^1\times \mathbb S^2$,  the finiteness of 
the initial length of the $\mathbb S^1$
factor is an obstruction to the creation of arbitrarily small perturbations of the kind
described (i.e.~using the domain of dependence) in the $\mathbb R\times\mathbb S^2$
case. 
Nevertheless, given a smallness parameter $\epsilon$, then for sufficiently large
length, there would exist a perturbation of size less than $\epsilon$ with
 $\mathcal{B}^+_h\ne\emptyset $, $\mathcal{B}^+_\infty\ne\emptyset$, 
 $\{\lambda<0\}\cap\{\nu<0\}\ne \emptyset$.

It remains to be seen whether there
exist arbitrarily small perturbations of fixed expanding homogeneous
data on $\mathbb S^1\times\mathbb S^2$
such that $\mathcal{B}^+_h\ne\emptyset$, $\mathcal{B}^+_\infty\ne\emptyset$,
 $\{\lambda<0\}\cap\{\nu<0\}\ne \emptyset$; this is Question~\ref{nq}.

\end{document}